\documentclass[prl,twocolumn]{revtex4}%
\usepackage{ulem}
\usepackage{graphics}
\usepackage{subfig}
\usepackage{bm}
\usepackage{amsmath}
\usepackage{amsfonts}
\usepackage{amssymb}
\usepackage{graphicx}%
\begin{document}
\title{MESOPHASE FORMATION IN TWO-COMPONENT CYLINDRICAL BOTTLE-BRUSH POLYMERS}
\author{Igor Erukhimovich, $^{\text{1}}\;$ Panagiotis E. Theodorakis, $^{\text{2}}\;$
Wolfgang Paul, $^{\text{3}}\;$ and Kurt Binder $^{\text{2}}\;$
\\[\baselineskip]%
                   $^{\text{1}}$ \textit{ A.N. Nesmeyanov Institute of Organoelement Compound, RAS \\
                   and Moscow State University, Moscow 119992, Russia} \\
                   $^{\text{2}}$ \textit{Insitut f\"ur Physik, Johannes Gutenberg-Universit\"{a}t} \\
                   \textit {D-55099 Mainz, Staudinger Weg 7, Germany}\\%
                   $^{\text{3}}$ \textit{ Institut f\"{u}r Physik, Martin-Luther-Universit\"at Halle-Wittenberg,} \\
                   \textit {von Seckendorff-Platz 1, 06120 Halle,
                   Germany}}
\date{\today }

\begin{abstract}
When two types of side chains (A,B) are densely grafted to a (stiff) backbone
and the resulting bottle-brush polymer is in a solution under poor solvent
conditions, an incompatibility between A and B leads to microphase separation
in the resulting cylindrical brush. The possible types of ordering are
reminiscent of the ordering of block copolymers in cylindrical confinement.
Starting from this analogy, Leibler's theory of microphase separation in block
copolymer melts is generalized to derive a description of the system in the
weak segregation limit.

Also molecular dynamics simulation results of a corresponding coarse-grained
bead-spring model are presented. Using side chain lengths up to N = 50
effective monomers, the ratio of the Lennard-Jones energy parameter between
unlike monomers $(\epsilon_{AB})$ and monomers of the same kind $(\epsilon
_{AA} = \epsilon_{BB})$ is varied. Various correlation functions are analyzed
to study the conditions when (local) Janus cylinder-type ordering and when
(local) microphase separation in the direction along the cylinder axis occurs.
Both the analytical theory and the simulations give evidence for short range
order due to a tendency towards microphase separation in the axial direction,
with a wavelength proportional to the side chain gyration radius, irrespective
of temperature and grafting density, for a wide range of these parameters.

\end{abstract}
\maketitle

\section{1. INTRODUCTION}

Enabled by progress in chemical synthesis (see \cite{1,2} for reviews),
macromolecules with ``bottle-brush'' architecture where flexible side chains
are densely grafted to a backbone, have found much recent interest (e.g.
\cite{1,2,3,4,5,6,7,8,9,10,11,12,13}). Such molecules may be useful for
various applications (such as sensors, actuators, building blocks in
supramolecular structures \cite{7,8,9}), since these systems are
stimuli-responsive polymers, exhibiting large conformational changes when
external conditions vary. Apart from these synthetic bottle-brush polymers,
also biopolymers with a related architecture are abundant in nature, e.g.
proteoglycans \cite{14}. These brush-like polymers contain a protein backbone
with carbohydrate side chains, and are held responsible for a large variety of
biological functions (cell signaling, cell surface protection, joint
lubrication, etc. \cite{15,16,17}). At the same time, the interplay between
entropic effects and various enthalpic forces in these "soft" objects makes
the understanding of structure-property relationship of such bottle-brush
polymers a challenging problem of statistical thermodynamics
\cite{18,19,20,21,22,23,24,25,26,27,28,29,30,31,32,33,34,35,36,37,38,39,40,41,42,43,44,45,46,47,48,49,50,51,52,53,54,55}%
.

In the present paper, we focus on the interplay between chain conformations
and local order in binary A,B bottle-brush polymers, from the point of view of
both theoretical arguments and Molecular Dynamics Simulations. Previous work
\cite{39,41} has focused on the possibility of microphase separation in the
form of ``Janus cylinders'' \cite{8}, i.e., the cylinder splits in two halves
(in the most symmetric case, where the lengths $N_{A}, N_{B}$ of both
types of side chains and their grafting densities $\sigma_{A}, \sigma_{B}$ are
equal), such that the A-B-interface contains the cylinder axis (taken to be
along the $z$-axis henceforth). Then it was pointed out \cite{46} that due to
the quasi-one-dimensional character of this ordering, for finite cylinder
radius R, no true long-range order of Janus-cylinder-type should be expected:
rather there exists a finite correlation length over which the orientation of
the interface plane decorrelates. In addition, it was speculated \cite{47}
that also the ratio $\epsilon_{AB}/\epsilon$ between the interaction strength
$\epsilon_{AB}$ of unlike monomers and monomers of the same kind
$(\epsilon_{AA}=\epsilon_{BB}=\epsilon)$ matters: if the formation of A-B
interfaces is energetically much more unfavourable rather than the formation
of polymer-solvent interfaces, a phase separation along the axial (z)
direction of the cylinder could occur also in the form of a double cylinder
(the A-rich polymers form a separate cylinder from the B-rich ones, both
cylinders touch each other along the z-axis) \cite{47}. Also intermediate
cases (``Janus dumbbell''-like cross sections of the bottle-brush) were
suggested \cite{47} and some indication of such structures were observed
\cite{55}. However, again only finite correlation lengths of such orderings
along the z-axis can be expected. In addition, one needs to consider that for
not very large grafting densities the bottle-brush polymer under poor solvent
conditions does not form a structure that is homogeneous along the z-axis (the
direction of the backbone; note that we restrict attention to rigid backbones
only, for simplicity), but rather inhomogeneities in the form of a
"pearl-necklace"-structure form \cite{43,51,56}.

The problem which we investigate in the following sections is whether it can
be favorable to form a microphase-separated structure that is inhomogeneous
along the z-axis. Following the information from experiment that one
side-chain can be grafted per backbone monomer \cite{10}, we assume an
alternating grafting of the two types (A and B) of side chains \cite{55}. Of
course, this situation is essentially equivalent to the case where symmetrical
block copolymers \cite{rev1,rev2,rev3,rev4,57} would be anchored with their A-B junction to the
backbone. It is then also interesting to ask how such a situation compares to
the case where such block copolymers fill a cylinder (at the same density)
without constraint on the locations of the A-B junctions. Of course, the
latter problem has already been studied in other contexts
\cite{58,59,60,61,62,63,64,65,66,67,68,69,70,71,72}. Here we generalize the
treatment of Ref.~\cite{71} to take this constraint on the location of the A-B
junctions into account to formulate a weak segregation theory \cite{73,74} of
ordering in binary bottle-brushes (Sec.~2). Then we present data from
Molecular Dynamics simulations, extending the study of Ref.~\cite{55} by
considering now the variation of properties with $\epsilon_{AB}/\epsilon$
(Sec.~3). We summarize the main findings of our work in the concluding section (Sec.~4).

\section{2. Weak segregation theory of microphase separation in two-component bottle-brushes.}
\section{2.1. The condensed state of the two-component bottlebrush.}

The general idea of our theoretical approach is as follows. We consider the
case when incompatibility of both sorts of blocks with the solvent is high
enough to form a condensed state (globule), which in the case of a rigid
backbone would acquire the form of a cylinder uniformly filled by the monomers
of the side chains. The height $H$ of the cylinder would be just the length of
the rigid backbone and the radius $R$ would be related to the equilibrium
monomer density $\rho_{0}$ inside the cylinder. Such a condensed state could
be considered as an equivalent of a bottlebrush compressed up to the monomer
density $\rho_{0}$ via squeezing it into a tube of the radius $R$. To
consider the composition fluctuations in such a condensed (squeezed)
bottlebrush we employ some new ideas, which were put forward in refs
\cite{67,71} to generalize the seminal Leibler theory \cite{73} (see also refs
\cite{57,rev1,rev2,rev3,rev4}) to confined polymer systems. To begin with, we
find, based on the theory of polymer globules by I. Lifshitz et al.
\cite{LGK}, the value of $\rho_{0}$, which is to provide the minimum of the
two-component bottlebrush free energy given the values of the chi-parameters
$\chi_{ij}$ (a similar approach has been employed by Borisov et al. \cite{43}
for one-component bottlebrushes in a poor solvent). To determine the
equilibrium value of a two-component bottlebrush forming a condensed state due
to its overall incompatibility with the solvent we notice that, according to
I. Lifshitz et al \cite{LGK}, the free energy of any flexible polymer system
is (in a zeroth approximation) just the sum of a structural (entropic)
contribution, which allows for the effect of connectivity ("linear memory" in
Lifshitz terms) and an energetic contribution, which allows for the
van-der-Waals attraction and the excluded volume repulsion effects:
\begin{equation}
\label{freeL}F=F_{\mathtt{str}}\left(  \left\{  \rho_{i}(\mathbf{r})\right\}
\right)  +F^{\ast}\left(  \left\{  \rho_{i}(\mathbf{r})\right\}  \right)
\end{equation}

The precise form of the structural free energy $F_{\mathtt{str}}\left(
\left\{  \rho_{i}(\mathbf{r})\right\}  \right)  $ as a functional of the
spatially non-unform distributions $\left\{  \rho_{i}(\mathbf{r})\right\}  $
of the local number densities of the repeat units of the $i$-th sort
($i=A,B$) is determined by the actual microscopic structure of the
bottlebrush. We analyze it somewhat later when studying the bottlebrush
stability with respect to its longitudinal, angular or helical ordering. But
now, when we focus on the basically uniform cylindrical globule, the
structural term could be simply disregarded as compared to the energetic
contribution to the free energy:
\begin{equation}
\label{freeV}F=F^{\ast}\left(  \left\{  \rho_{i}(\mathbf{r})\right\}  \right)
=T\,V\,f^{\ast}\left(  \left\{  \overline{\rho}_{i}\right\}  \right)  .
\end{equation}
Here $V=\pi R^{2}H=2M\pi R^{2}a$ is the volume of the cylindrical globule
(where $2M$ and $a$ are the total number of both $A$ and $B$ side chains and
the distance between the neighboring $A$ and $B$ chains), $\overline{\rho}_{i}$
is the value of the local number density of the repeat units of the $i$-th
sort, averaged over the whole volume $V$ of the bottlebrush, and the specific
(per unit volume) free energy $f^{\ast}\left(  \left\{  \overline{\rho}%
_{i}\right\}  \right)  $ depends on the type of the interactions between the
solvent molecules and repeat units (monomers) forming the bottlebrush. For
simplicity, we assume in this paper that these interactions correspond to the
conventional incompressible Flory-Huggins lattice model, in which case
\begin{align}
\label{f_sp}v_{0}f_{\ast}\left(  \phi\right)   &  =\left(  1-\phi\right)
\ln\left(  1-\phi\right)  +\phi\nonumber\\
&  +\chi_{AB} \phi_{A} \phi_{B} + \chi_{AS} \phi_{A} \phi_{S} + \chi_{BS}
\phi_{B} \phi_{S},
\end{align}
where $v_{0}$ is the excluded volume assumed, for simplicity, to be the same
for both sorts of the repeat units and solvent molecules, $\phi_{i}\left(
\mathbf{r}\right)  =v_{0}\rho_{i}\left(  \mathbf{r}\right)  $ is the local
volume fraction of the particles of the $i$-th sort within the globular
bottlebrush, $\phi=\phi_{A}+\phi_{B}$ is the total polymer volume fraction
(volume fraction of all monomers), $\phi_{S}=1-\phi$ is the volume fraction of
the solvent molecules and $\chi_{ij}$ are the conventional Flory-Huggins
energetic parameters.

For the homogeneous bottle brush we are considering here, $\phi_{i}\left(
\mathbf{r}\right) = \phi_i = f_i \phi = v_{0}\overline{\rho}_{i}$, where
$f_{i}$ is the fraction of the monomers of the $i$-th sort, and, thus, the
function $f^{\ast}\left(  \left\{  \overline{\rho}_{i}\right\}  \right)  $ can
be rewritten in the form
\begin{align}
v_{0}f^{\ast}\left(  \left\{  \overline{\rho}_{i}\right\}  \right)   &
=v_{0}f^{\ast}\left(  \phi\right) \nonumber\\
&  =\left(  1-\phi\right)  \ln\left(  1-\phi\right)  +\phi-\widetilde{\chi
}\,\phi^{2}+\overline{\chi}\,\phi
\end{align}
where $\widetilde{\chi}=\chi_{AS}f_{A}+\chi_{BS}f_{B} -\chi_{AB}f_{A} f_{B}$
is the effective $\chi$-parameter describing the overall bottlebrush-solvent
incompatibility and the parameter $\overline{\chi}=\chi_{AS}f_{A}+\chi
_{BS}f_{B}$ does not affect the equilibrium value of $\phi$.

Indeed, taking into account that the polymer volume fraction within the
uniform bottlebrush is
\begin{equation}
\label{phi}\phi=v_{0}\left.  N\right/  \left(  2\pi R^{2}a\right)  ,
\end{equation}
where $N=N_{A}+N_{B}$ is the total average number of all monomers $A$ and $B$
per the distance $2a$ of the backbone, one can reduce the total free energy
(\ref{freeV}) to the form
\begin{equation}
F=T\,MN\,\,\frac{1}{\phi}\left[ \left(  1-\phi\right)  \ln\left(
  1-\phi\right)  +\phi-\widetilde{\chi}\,\phi^{2}+\overline{\chi}\,\phi
  \right]. \label{free0}%
\end{equation}

Minimization of the free energy (\ref{free0}) with respect to $\phi$ results,
finally, in the desired equation (it corresponds to the so-called volume
approximation of the polymer globule theory \cite{LGK}) for the equilibrium
polymer volume fraction of the two-component bottlebrush:
\begin{equation}
v_{0}p^{\ast}/T=-\ln(1-\phi)-\phi-\widetilde{\chi}\phi^{2}=0. \label{phi_bar}%
\end{equation}

Obviously, the finite equilibrium polymer volume fraction of the two-component
bottlebrush $\phi_{0}$, which is the solution of eq (\ref{phi_bar}), depends on
the value of $\widetilde{\chi}$ only and has a physically meaningful positive
value only for
\begin{equation}
\label{tau}\tau=\widetilde{\chi}-1/2>0.
\end{equation}

The corresponding asymptotics read
\begin{equation}
\left\{
\begin{array}
[c]{c}%
\phi=3\tau-(27/4)\tau^{2}+...,\quad\;\tau\ll1\\
\phi=1-\exp\left(  1+\widetilde{\chi}\right)  ,\quad\qquad\chi\gg1
\end{array}
\right.  \label{asym1}%
\end{equation}

It is worth to notice that in so-called Hildebrand approximation \cite{HS50},
in which the value of the $\chi$-parameter for the $i$-th and $j$-th species is
proportional to the squared difference of their solubility parameters:
\[
\chi_{ij}=\left.  v_{0}\left(  \delta_{i}-\delta_{j}\right)  ^{2}\right/
\left(  2T\right)  ,
\]
the effective $\chi$-parameter is always positive:%
\[
\widetilde{\chi}=\frac{\left(  2f-1+x\right)  ^{2}}{4}\chi_{AB},
\]
where $x=\left(  2\delta_{S}-\delta_{A}-\delta_{B}\right)  /\left(  \delta
_{B}-\delta_{A}\right)  $ is the selectivity parameter \cite{EAS97}. If the $A
$ and $B$ monomers are compatible ($\delta_{A}=\delta_{B}$) then
$\widetilde{\chi}=\chi_{AS}=\chi_{AB}$.

To conclude this subsection we discuss the conditions of validity of our
consideration. First, it is worth to notice that the bottlebrushes under
consideration are quasi-1D systems. Therefore, their equilibrium volume
fraction stays finite even in the $\Theta$-solvent, where $R^{2}\sim Nl^{2}$
(here $l$ is the statistical segment supposed, for simplicity, to be the same
for both sorts of the side chains). Indeed, it follows from eq (\ref{phi})
that in this case
\begin{equation}
\label{Theta}\phi_{\Theta}\sim \text{Li}/(\pi \widetilde{a})\,,
\end{equation}
where we introduced the reduced distance $\widetilde{a}=A/l$ between the neighboring side chains and
the Lifshitz number $\text{Li}=v_0/l^3$, which characterizes the chains' flexibility.
Comparing eq (\ref{Theta}) and the first of eqs (\ref{asym1}) and taking into
account that the condensed bottlebrush should be much denser than that in
$\Theta$-solvent we conclude that the first of the validity conditions reads
\begin{equation}
\label{val1}\kappa=\phi_{\Theta}/\phi=R^{2}/(Nl^{2}) \approx \text{Li}/(3\pi \widetilde{a}\tau) \ll1\,.
\end{equation}

On the other hand, the volume approximation (\ref{phi_bar}) is known
\cite{LGK} to be valid when
\begin{equation}
\label{volume}R\gg r_{c},
\end{equation}
where $r_{c}$ is the correlation radius within the globule. For dense globules
($1-\phi\ll1$ or $\tau\gg1$) the condition holds always but for semidilute
globule ($\beta/3 \ll\tau\ll1$) the correlation radius is known \cite{LGK,PGG}
to increase with decrease of $\tau$:
\begin{equation}
\label{rc}r_{c} \sim l/\tau\,.
\end{equation}

It follows from eqs (\ref{phi}), (\ref{asym1}) and (\ref{rc}) that close to
the $\Theta$-temperature the
condition (\ref{val1}) is satisfied only when the side chains are long
enough:
\begin{equation}
\label{val2}N\gg1/(\phi_{\Theta}\,\tau).
\end{equation}

Condition (\ref{val2}) is rather severe since both quantities $\phi_{\Theta}$
and $\tau$ are small. Thus, the fact that the monomer density profile usually
observed in computer simulation of bottlebrushes in poor solvents close to the
$\Theta$-temperature is not
step-like, which is expected for a condensed (globular) state, but rather
smooth \cite{51,55} is explained by insufficiently high degrees of polymerization of the side
chains, which do not satisfy condition (\ref{val2}).

\section{2.2. The Random Phase Approximation for confined polymer systems.}
\subsection{General theory.}

To consider the fluctuations within the two-component condensed bottlebrush we
are to expand the total free energy (\ref{freeL}) in powers of the density
fluctuations counted from the uniform globular state ($\Psi_{i}(\mathbf{r})$ =
$\rho_{i}(\mathbf{r})-\overline{\rho}_{i}$):
\begin{equation}
F=F_{0}+\Delta F_{2}+\Delta F_{3}+\Delta F_{4}+..., \label{F2}%
\end{equation}
where $F_{0}$ is defined by eqs (\ref{freeV}) and (\ref{free0}), the quadratic
term of the expansion reads
\begin{equation}
\Delta F_{2}=\frac{T}{2} \int\Gamma_{ij}(\mathbf{r}_{1},\mathbf{r}_{2}%
)\,\Psi_{i} (\mathbf{r}_{1})\,\Psi_{j}(\mathbf{r}_{2})\,dV_{1}\,dV_{2}
\label{F2a}%
\end{equation}
(in eq (\ref{F2a}) and thereafter we employ the rule of summation over the
repeating indices) and the next terms of the expansion (\ref{F2}) are defined
similarly \cite{73}.

Our purpose in this subsection is to find the explicit form of the kernel
$\Gamma(\mathbf{r}_{1},\mathbf{r}_{2})$ and conditions ensuring that the
quadratic form $\Delta F_{2}$ is positive definite, which implies that the
fluctuations are small. It follows from eqs (\ref{freeL}) - (\ref{f_sp}) and
(\ref{F2a}) that
\begin{align}
\Gamma_{\alpha\beta}(\mathbf{r}_{1},\mathbf{r}_{2})  &  =\frac{T^{-1}%
\,\delta^{2}F\left(  \left\{  \rho_{i}(\mathbf{r})\right\}  \right)  }%
{\delta\rho_{\alpha}(\mathbf{r_{1}})\;\delta\rho_{\beta}(\mathbf{r_{2}}%
)}\nonumber\label{Gam_0}\\
&  =\gamma_{\alpha\beta}(\mathbf{r}_{1},\mathbf{r}_{2})-c_{\alpha\beta
}(\mathbf{r}_{1},\mathbf{r}_{2}).
\end{align}
Here we introduced new designations
\begin{equation}
\gamma_{\alpha\beta}(\mathbf{r}_{1},\mathbf{r}_{2})=\frac{T^{-1}\,\delta
^{2}F_{\mathtt{str}}\left(  \left\{  \rho_{i}(\mathbf{r})\right\}  \right)
}{\delta\rho_{\alpha}(\mathbf{r_{1}})\;\delta\rho_{\beta}(\mathbf{r_{2}})},
\label{gam}%
\end{equation}%
\begin{equation}
c_{\alpha\beta}(\mathbf{r}_{1},\mathbf{r}_{2})=-\frac{T^{-1}\,\delta
^{2}F^{\ast}\left(  \left\{  \rho_{i}(\mathbf{r})\right\}  \right)  }%
{\delta\rho_{\alpha}(\mathbf{r_{1}})\;\delta\rho_{\beta}(\mathbf{r_{2}}%
)}=\delta(\mathbf{r}_{1}-\mathbf{r}_{2})\;C_{\alpha\beta}, \label{dir}%
\end{equation}
where the number matrix $\mathbf{C}=\Vert C_{\alpha\beta}\Vert$ reads
\begin{equation}
\mathbf{C}=\left|  \frac{T^{-1}\,\partial^{2} f^{*}}{\partial\rho_{\alpha
}\,\partial\rho_{\beta}}\right|  = v_{0}\;\left(
\begin{array}
[c]{cc}%
2\chi_{AS}-\phi_{S}^{-1} & k-\phi_{S}^{-1}\\
k-\phi_{S}^{-1} & 2\chi_{BS}-\phi_{S}^{-1}%
\end{array}
\right)  \label{C}%
\end{equation}
and $k=\chi_{AS}+\chi_{BS}-\chi_{AB}$.

Thus, our problem is reduced to calculating the second derivative (\ref{gam})
of the structural free energy functional. For this purpose we consider an
auxiliary thermodynamic potential
\begin{equation}
\Phi_{\mathtt{str}}\left(  \left\{  \varphi_{i}\left(  \mathbf{r}\right)
,T\right\}  \right)  =-T\,\ln Z\left(  \left\{  \varphi_{i}\left(
\mathbf{r}\right), T \right\}  \right)  \label{Phi}%
\end{equation}
which has meaning of the free energy of an ideal polymer system with a
specified architecture (in our case it is the two-component bottlebrush)
affected by a set ${\varphi_{i}(\mathbf{r})}$ of external fields applied
to the particles of the $i$-th sort. The thermodynamic potential
$\Phi_{\mathtt{str}}\left(  \left\{  \varphi_{i}\left(  \mathbf{r}\right)
,T\right\}  \right)  $ could be readily calculated (see below). On the other
hand, it is directly related to the desired thermodynamic potential
$F_{\mathtt{str}}({\rho_{i}(\mathbf{r})})$. Indeed, the partition function $Z$
appearing in eq (\ref{Phi}) is the integral over all possible non-uniform
density distributions $\rho_{i}(\mathbf{r})$ of the particles of the $i$-th
sort in the volume of the system:
\begin{equation}
Z=-\int{\prod\limits_{i}}\delta\rho_{i}(\mathbf{r})\exp\left(  -\left.
\widetilde{F}_{\mathtt{str}}\left(  \left\{  \rho_{i}(\mathbf{r})\right\}
,\left\{  \varphi_{i}\left(  \mathbf{r}\right)  \right\}  \right)  \right/
T\right)  \label{Z}%
\end{equation}
with%
\[
\widetilde{F}_{\mathtt{str}}\left(  \left\{  \rho_{i}(\mathbf{r})\right\}
,\left\{  \varphi_{i}\left(  \mathbf{r}\right)  \right\}  \right)
=F_{\mathtt{str}}\left\{  \rho_{i}(\mathbf{r})\right\}  +\int\rho
_{i}(\mathbf{r})\,\varphi_{i}(\mathbf{r})\,dV.
\]
Definition and calculation of the density integral $Z$ is, generally, rather
cumbersome but it becomes trivial in the saddle-point approximation:%
\begin{equation}
\Phi_{\mathtt{str}}\left(  \left\{  \varphi_{i}(\mathbf{r})\right\}  \right)
=\min_{\left\{  \rho_{i}(\mathbf{r})\right\}  }\widetilde{F}_{\mathtt{str}%
}\left(  \left\{  \rho_{i}(\mathbf{r})\right\}  ,\left\{  \varphi_{i}\left(
\mathbf{r}\right)  \right\}  \right)  \label{field_pot}%
\end{equation}

It follows from (\ref{field_pot}) that if the thermodynamic potential
$F_{\mathtt{str}}\left(  \left\{  \rho_{i}(\mathbf{r})\right\}  \right)  $ is
known then the thermodynamic potential $\Phi_{\mathtt{str}}\left(  \left\{
\varphi_{i}(\mathbf{r})\right\}  \right)  $ is parametrically defined as
follows:
\begin{equation}
\label{L1}\varphi_{\alpha}(\mathbf{r})=-\left.  \delta F_{\mathtt{str}}\left(
\left\{  \rho_{i}(\mathbf{r})\right\}  \right)  \right/  \delta\rho_{\alpha
}(\mathbf{r})
\end{equation}
\begin{align}
\Phi_{\mathtt{str}}\left(  \left\{  \varphi_{i}(\mathbf{r})\right\}  \right)
&  =F_{\mathtt{str}}\left(  \left\{  \rho_{i}(\mathbf{r})\right\}  \right)
\nonumber\\
&  -\int\rho_{i}(\mathbf{r})\,\dfrac{\delta F_{\mathtt{str}}\left(  \left\{
\rho_{i}(\mathbf{r})\right\}  \right)  }{\delta\rho_{\alpha}(\mathbf{r})}\;dV
\label{L2}%
\end{align}
Eqs (\ref{L1}) and (\ref{L2}) imply that the thermodynamic potential $\Phi$,
which is a functional of all external fields $\varphi_{i}(\mathbf{r})$ applied
to the particles of the system, is the Legendre transform of the
thermodynamic potential $F$, which is a functional of all number densities
$\rho_{i}(\mathbf{r})$ of these particles. Therefore, the reciprocal
relationships hold:
\begin{equation}
\rho_{\alpha}(\mathbf{r})=\left.  \delta\Phi_{\mathtt{str}} \left(  \left\{
\varphi_{i}(\mathbf{r})\right\}  \right)  \right/  \delta\varphi_{\alpha
}(\mathbf{r}) \label{rL1}%
\end{equation}
\begin{align}
F_{\mathtt{str}} \left(  \left\{  \rho_{i}(\mathbf{r})\right\}  \right)   &
=\Phi_{\mathtt{str}} \left(  \left\{  \varphi_{i}(\mathbf{r})\right\}  \right)
\nonumber\\
&  -\int\varphi_{i}(\mathbf{r})\,\dfrac{\delta\Phi\left(  \left\{  \varphi
_{i}(\mathbf{r})\right\}  \right)  }{\delta\varphi_{\alpha}(\mathbf{r})}\;dV
\label{rL2}%
\end{align}

It follows from eqs (\ref{F2a}) and (\ref{L1})-(\ref{rL2}) that
\begin{align}
\gamma_{\alpha\beta}(\mathbf{r}_{1},\mathbf{r}_{2})  &  =\left.  \frac
{\delta^{2}F_{\mathtt{str}}\left(  \left\{  \rho_{i}(\mathbf{r})\right\}
\right)  }{\delta\rho_{\alpha}(\mathbf{r_{1}})\;\delta\rho_{\beta
}(\mathbf{r_{2}})}\right\vert _{\rho_{i}\left(  \mathbf{r}\right)
=const}\label{Gam}\\
&  =\left.  -\frac{\delta\varphi_{\alpha}(\mathbf{r_{1}})}{\delta\rho_{\beta
}(\mathbf{r_{2}})}\right\vert _{\rho_{i}\left(  \mathbf{r}\right)
=const}\; .\nonumber
\end{align}%
Furthermore, if we define the structural matrix
\begin{align}
g_{\alpha\beta}(\mathbf{r}_{1},\mathbf{r}_{2})  &  =-\left.  \frac{\delta
^{2}\Phi_{\mathtt{str}}\left(  \left\{  \varphi_{i}(\mathbf{r})\right\}
\right)  }{\delta\varphi_{\alpha}(\mathbf{r}_{1})\,\delta\varphi_{\beta
}(\mathbf{r}_{2})}\right\vert _{\varphi\left(  \mathbf{r}\right)
=0}\label{Gam_1}\\
&  =-\left.  \frac{\delta\rho_{\alpha}(\mathbf{r}_{1})}{\delta\varphi_{\beta
}((\mathbf{r}_{2}))}\right\vert _{\varphi\left(  \mathbf{r}\right)
=0}\nonumber
\end{align}
we obtain
\begin{equation}
\int\gamma_{\alpha\beta}(\mathbf{r}_{1},\mathbf{r})\,g_{\beta\gamma
}(\mathbf{r},\mathbf{r}_{2})d\mathbf{r}=\delta_{\alpha\gamma}\delta
(\mathbf{r}_{1}-\mathbf{r}_{2})\; . \label{inv}%
\end{equation}

\subsection{Calculation of the structural matrix for copolymers in bulk.}

The derivation of the explicit expression for the structural matrix $\mathbf{g}$
of copolymer systems under confinement
is a straightforward extension of that in bulk. So, we remind the reader how
the structural matrix is derived in bulk.

Instead of calculating the density integral (\ref{Z}) we consider an
equivalent expression for the partition function $Z(\left\{  \varphi
_{i}\left(  \mathbf{r}\right)  \right\}  )$ of an ideal polydisperse
$n$-component system of linear macromolecules affected by the external fields
$\left\{  \varphi_{i}\left(  \mathbf{r}\right)  \right\}  $ applied to the
monomers of the $i$-th sort. The partition function $Z(\left\{  \varphi
_{i}\left(  \mathbf{r}\right)  \right\}  )$ can be written directly in terms
of the corresponding discrete microscopic model:
\begin{equation}
Z(\left\{  \varphi_{i}\left(  \mathbf{r}\right)  \right\}  )={\prod
\limits_{S}}Z_{S}^{M_{S}}\left(  \left\{  \varphi_{i}\left(  \mathbf{r}%
\right)  \right\}  \right)  .\; \label{Zm}%
\end{equation}
Here $S$ is the macromolecular structure, which for linear chains is just a
sequence of integers $\alpha\left(  i\right)  $ equal to the number
$\alpha$ of the sort of the $i$-th monomer, $M_{S}$ is the number of
chains with structure $S$ in the whole volume $V$ of the system and
$Z_{S}\left(  \left\{  \varphi_{i}\left(  \mathbf{r}\right)  \right\}
\right)  $ is the partition function of the chain $S$:
\begin{equation}
Z_{S}\left(  \left\{  \varphi\left(  \mathbf{r}\right)  \right\}  \right)
=\int d\text{\/}\Gamma_{S}\,F\left(  \Gamma_{S}\right)  \,\exp\left(
-{\sum\limits_{i=0}^{N}} \varphi_{\alpha\left(  i\right)  }\left(
\mathbf{r}_{i}\right)  /T\right)  \label{Z1}%
\end{equation}
where $\Gamma_{S}=\left(  \mathbf{r}_{0},..,\mathbf{r}_{N}\right)  $ is a
point of the configuration space of the chain $S$ and
\[
F\left(  \Gamma_{S}\right)  ={\prod\limits_{i=0}^{N-1}} g\left(
\mathbf{r}_{i+1}-\mathbf{r}_{i}\right)
\]
is the distribution function of the ideal chain $S$, the effect of
connectivity being taken into account by the bond functions $g\left(
r\right)  $, which describe the correlations between the neighboring monomers
of the chains. For simplicity, we assume that the bond function does not
depend on the sorts of the neighboring monomers and has the conventional
normalized form
\begin{equation}
g\left(  r\right)  =\left.  \exp\left(  -\left.  3r^{2}\right/  \left(
2l^{2}\right)  \right)  \right/  \left(  2\pi l^{2}/3\right)  ^{3/2},
\label{bond}%
\end{equation}
where $l$ is the statistical segment length. Thus, in the thermodynamic limit
\[
\lim_{V\rightarrow\infty}\left.  \left.  Z_{S}\left(  \left\{  \varphi
_{i}\left(  \mathbf{r}\right)  \right\}  \right)  \right\vert _{\varphi_{i}%
=0}\right/  V=1.
\]

The thermodynamic potential $\Phi_{\mathtt{str}}\left(  \left\{  \varphi
_{i}(\mathbf{r})\right\}  \right)  $ reads%
\begin{equation}
\Phi_{\mathtt{str}}\left(  \left\{  \varphi_{i}(\mathbf{r})\right\}  \right)
={\sum\limits_{S}} {M}_{S}\,\Phi_{S}\left(  \left\{  \varphi_{i}%
(\mathbf{r})\right\}  \right)
\end{equation}
where
\begin{equation}
\Phi_{S}\left(  \left\{  \varphi_{i}\left(  \mathbf{r}\right)  \right\}
\right)  =-T\ln Z_{S}\left(  \left\{  \varphi_{i}\left(  \mathbf{r}\right)
\right\}  \right)  . \label{def1}%
\end{equation}

Additivity of $\Phi_{\mathtt{str}}\left(  \left\{  \varphi_{i}(\mathbf{r}%
)\right\}  \right)  $ with respect to the terms characterizing macromolecules
with different chemical structure implies that the matrix (\ref{Gam_1}) is a
properly weighted sum of molecular response functions:%
\[
g_{\alpha\beta}(\mathbf{r}_{1},\mathbf{r}_{2})={\sum\limits_{S}}\,\rho
_{S}\,g_{\alpha\beta}^{(S)}(\mathbf{r}_{1},\mathbf{r}_{2}),
\]
Here $\rho_{S}=N_{S}/V$ is the average number density of the macromolecules
$S$ and
\[
\,g_{\alpha\beta}^{(S)}(\mathbf{r}_{1},\mathbf{r}_{2})=\left.  \frac
{\delta^{2}Z_{S}\left(  \left\{  \varphi_{i}(\mathbf{r})\right\}  \right)
}{\delta\varphi_{\alpha}(\mathbf{r}_{1})\,\delta\varphi_{\beta}(\mathbf{r}%
_{2})}\right\vert _{\varphi_{i}\left(  \mathbf{r}\right)  =0}%
\]
where we employed the definitions (\ref{Z1}) and (\ref{def1})$.$ Summing the
second field derivatives of $Z_{S}\left(  \left\{  \varphi(\mathbf{r}%
)\right\}  \right)  $ and collecting similar terms one gets%
\begin{equation}
g_{\alpha\beta}^{(S)}(\mathbf{r}_{1},\mathbf{r}_{2})=m_{\alpha}^{\left(
S\right)  }\,\delta_{\alpha\beta}\,\delta\left(  \mathbf{r}_{1}-\mathbf{r}%
_{2}\right)  +{\sum\limits_{n\geq1}}\nu_{\alpha\beta}^{\left(  S\right)
}\left(  n\right)  \,g^{\left(  n\right)  }(\mathbf{r}_{1},\mathbf{r}_{2})
\label{gab}%
\end{equation}
where
\begin{equation}
g^{\left(  n+1\right)  }\left(  \mathbf{r}_{1},\mathbf{r}_{2}\right)  =\int
g\left(  \mathbf{r}_{1}-\mathbf{r}\right)  g^{(n)}\left(  \mathbf{r}%
,\mathbf{r}_{2}\right)  d\mathbf{r},
\label{def_n}%
\end{equation}
$m_{\alpha}^{(S)}$ is the number of the monomers of the $i$-th sort, which belong to the
macromolecules $S$, and $\nu_{\alpha\beta}^{(S)}(n)$ are the numbers of
pairs of monomers of the $\alpha$-th and $\beta$-th sorts, which are
separated by $n$ bonds. The structural matrix, which was first introduced in
the form (\ref{gab}) by one of us \cite{IE79a,IE79b,74} (see also \cite{HB96}
and references therein), is a discrete analog of the expressions for the
correlators obtained in the continuous limit (like the Debye function for
homopolymer and the Leibler expressions \cite{73} for diblock copolymers).

To calculate the structural matrix explicitly we rewrite it as the inverse
Fourier transform:%
\begin{equation}
\label{RB}g_{\alpha\beta}^{(S)}(\mathbf{r}_{1},\mathbf{r}_{2})= \int
\exp\left(  i\mathbf{q}\left(  \mathbf{r}_{1}-\mathbf{r}_{2}\right)  \right)
\widetilde{g}_{\alpha\beta}^{\left(  S\right)  } \left(  q\right)  \frac
{d^{3}q}{\left(  2\pi\right)  ^{3}},\
\end{equation}
where the direct Fourier transform
%$\phia$%
\begin{align}
\label{RBF}\widetilde{g}_{\alpha\beta}^{(S)}(q)  &  =V^{-1}\int g_{\alpha
\beta}^{(S)}(\mathbf{r}_{1},\mathbf{r}_{2})\exp\left(  i\mathbf{q}\left(
\mathbf{r}_{2}-\mathbf{r}_{1}\right)  \right)  \,d\text{\/}\mathbf{r}%
_{1}\,d\text{\/}\mathbf{r}_{2}\nonumber\\
&  =m_{\alpha}^{\left(  S\right)  }\,\delta_{\alpha\beta}+{\sum\limits_{n\geq
1}}\nu_{\alpha\beta}^{\left(  S\right)  }\left(  n\right)  \,g^{n}(q)
\end{align}
is an algebraic function of the Fourier transform of the bond function:%
\begin{equation}
\label{bondF}g\left(  q\right)  =\int d\text{\/}V\,g\left(  r\right)
\exp\left(  i\mathbf{qr}\right)  =\exp\left(  -l^{2}q^{2}/6\right)  .
\end{equation}

The explicit finite expressions for homo- and some block copolymers can be
found in refs \cite{IE79a,IE79b,74} and \cite{DE93}.

\subsection{Calculation of the structural matrix for confined copolymers.}

Now, we make use of the fact that $q^{2}$ and $\exp (i\mathbf{qr})$ are the
eigen values and eigen functions of the Laplace operator $-\Delta$ acting in
the infinite space. The natural idea \cite{67,71} to generalize expression
(\ref{RB}) for the structural matrix to confined copolymers is to rewrite it
in the form of a somewhat more general series%
\begin{equation}
g_{\alpha\beta}(\mathbf{r}_{1},\mathbf{r}_{2})={\sum\limits_{s}}\,\left(
\widetilde{\mathbf{g}}(\lambda_{s})\right)  _{\alpha\beta}\,Y^{\left(
s\right)  }\left(  \mathbf{r}_{1}\right)  \,\overline{\,Y^{\left(  s\right)
}\left(  \mathbf{r}_{2}\right)  } \label{series}%
\end{equation}
where now $\left\{  \lambda_{s}\right\}  $ and $\left\{  Y_{\alpha}^{\left(
s\right)  }\left(  \mathbf{r}\right)  \right\}  $ are the full sets of the
eigen values and the orthonormalized eigen functions of the Laplace operator
$-\Delta$ acting in a finite volume under specified boundary conditions and
$\widetilde{\mathbf{g}}\left(  \lambda\right)  $ is a matrix-function of the
parameter $\lambda$.

Then the kernels $\gamma_{\alpha\beta}(\mathbf{r}_{1},\mathbf{r}_{2})$ and
$c_{\alpha\beta}(\mathbf{r}_{1},\mathbf{r}_{2})$ read%
\begin{align}
\gamma_{\alpha\beta}(\mathbf{r}_{1},\mathbf{r}_{2})  &  ={\sum\limits_{s}%
}\,\left(  \widetilde{\mathbf{g}}^{-1}\left(  \lambda\right)  \right)
_{\alpha\beta}\,Y^{\left(  s\right)  }\left(  \mathbf{r}_{1}\right)
\,Y^{\left(  s\right)  }\left(  \mathbf{r}_{2}\right)  ,\\
c_{\alpha\beta}(\mathbf{r}_{1},\mathbf{r}_{2})  &  =C_{\alpha\beta}%
{\sum\limits_{s}}\,\,Y^{\left(  s\right)  }\left(  \mathbf{r}_{1}\right)
Y^{\left(  s\right)  }\left(  \mathbf{r}_{2}\right)  ,
\end{align}
which results in the following final expression for the desired kernel
$\Gamma_{\alpha\beta}(\mathbf{r}_{1},\mathbf{r}_{2})$:
\begin{equation}
\Gamma_{\alpha\beta}(\mathbf{r}_{1},\mathbf{r}_{2})={\sum\limits_{s}%
}\,\,Y^{\left(  s\right)  }\left(  \mathbf{r}_{1}\right)  \left(
\mathbf{G}^{-1}\left(  \lambda_{s}\right)  \right)  _{\alpha\beta}Y^{\left(
s\right)  }\left(  \mathbf{r}_{2}\right)  \label{G1}%
\end{equation}
with
\begin{equation}
\mathbf{G}^{-1}\left(  \lambda\right)  =\widetilde{\mathbf{g}}^{-1}\left(
\lambda\right)  -\mathbf{C}\,. \label{G2}%
\end{equation}
which is a generalization of the well known RPA equation for polymers in
bulk:
\begin{equation}
\mathbf{G}^{-1}\left(  q^{2}\right)  =\widetilde{\mathbf{g}}^{-1}\left(
q^{2}\right)  -\mathbf{C}. \label{G2b}%
\end{equation}

We conclude that the condition ensuring that the quadratic form $\Delta F_{2}$
is positive definite is that ensuring that the matrix $\mathbf{G}$ has
positive eigen values for any allowable value of $\lambda$.

In particular, for confinement in a cylindrical capillary of the radius $R$
the eigen values and the eigen functions of the Laplace operator read:
\begin{equation}
-\lambda=q^{2}+(s_{n}^{m}/R)^{2},\quad Y_{n}^{m}\left(  \mathbf{r}\right)
=\mathcal{N}\,J_{n}\left(  s_{n}^{m}\,\frac{r}{R}\right)  e^{i\left(
qz-n\alpha\right)  }, \label{tube}%
\end{equation}
where $\mathbf{r=}(\alpha,r,z)$ is the radius-vector in cylindrical
coordinates, $\mathcal{N}^{2}=\left(  \int_{0}^{R}J_{n}^{2}\left(  s_{n}%
^{m}\frac{r}{R}\right)  \,r\,dr\right)  ^{-1}$ is a normalization factor
and $s_{n}^{m}$ is location of the $m$-th extremum of the Bessel function
$J_{n}(x)$ of the order $n$.

Accordingly, the quadratic form (\ref{F2a}) takes the form
\begin{align}
\Delta F_{2}  &  =\frac{T}{2}{\sum\limits_{n=0}^{\infty}\sum\limits_{m=1}%
^{\infty}}\,\,\int\frac{dq}{2\pi}A_{\alpha}(q,n,m)\,\overline{A_{\beta
}(q,n,m)}\nonumber\\
&  \times\left(  \mathbf{G}^{-1}\left(  q,n,m\right)  \right)  _{\alpha\beta},
\label{F2t}%
\end{align}
where the coefficient $A_{\alpha}(q,n,m)$ reads
\[
A_{\alpha}(q,n,m)=\mathcal{N}\,\int r^{2}dr\,d\alpha\,dz\,\Psi_{\alpha}\left(
\mathbf{r}\right)  J_{n}\left(  s_{n}^{m}\frac{r}{R}\right)  e^{i\left(
qz-n\alpha\right)  }.
\]

For future comparison with ordering in two-component bottlebrushes under
investigation we present here the explicit form of the structural matrix for
diblock copolymers $A_{n}B_{m}$:%
\begin{align}
g_{\alpha\alpha}^{\text{free}}\left(  s_{n}^{m},q\right)   &  =\phi\,N\,f_{D}\left(
\widetilde{Q}^{2}\,x_{\alpha}\right)  ,\label{g_dib}\\
g_{AB}^{\text{free}}\left(  s_{n}^{m},q\right)   &  =\phi\,N\,\psi\left(  \widetilde{Q}^{2}%
\,x_{A}\right)  \psi\left(  Q^{2}\,x_{B}\right)  .\nonumber
\end{align}
Here $N=n+m,\ x_{A}=1-x_{B}=n/N$, $\widetilde{Q}^{2}=\kappa_{n}^{m}+Q^2$, $Q^2=q^{2}l^{2}N/6$,
$\kappa_{n}^{m}=\left(  s_{n}^{m}l/R\right)^{2}N/6=\left(  s_{n}^{m}\right)
^{2}/(6\kappa)$, $s_{n}^{m}\ $is location of the $m$-th extremum of the Bessel
function $J_{n}(x)$ of the order $n$, parameter $\kappa$ is defined by eq
(\ref{val1}) and we introduced functions
\begin{equation}
f_{D}(y)=2\frac{\exp(-y)-1+y}{y^{2}},\quad\psi(y)=\frac{(1-\exp(-y))^{2}%
}{y^{2}}\;. \label{def_f}%
\end{equation}

\section{2.3. The Random Phase Approximation for two-component bottlebrushes}

The derivation presented above for the confined block copolymers can be now
extended to the tethered copolymers. For this purpose it is sufficient to
replace the expression (\ref{Zm}) for the partition function of the
corresponding confined system by that of the tethered system under
consideration. In this paper we consider for definiteness the two-component
bottlebrush consisting of $2M$ pairs of $A$ and $B$ side chains regularly
distributed along the backbone. The partition function of such an ideal
bottlebrush affected by the external fields $\left\{  \varphi_{i}\left(
\mathbf{r}\right)  \right\}  ,$ reads:
\begin{align}
Z  &  ={\prod\limits_{n=0}^{M-1}}Z_{N_{A}}\left(  \left\{  \varphi_{A}\left(
\mathbf{r}\right)  \right\}  ,2na\right)  \times\nonumber\label{ZB}\\
&  {\prod\limits_{n=0}^{M-1}}Z_{N_{B}}\left(  \left\{  \varphi_{B}\left(
\mathbf{r}\right)  \right\}  ,\left(  2n+1\right)  a\right)
\end{align}
where $M$ is the number of side chains of each sort, $Z_{N}\left(
\left\{  \varphi\left(  \mathbf{r}\right)  ,z_{0}\right\}  \right)  $ is the
partition function of an $N$-mer, which is attached by one of its ends to the
rigid backbone aligned along the $z$-axes at the point $z_{0}$ and affected by
an external field $\varphi\left(  \mathbf{r}\right)  $:
\begin{equation}
Z_{N}\left(  \left\{  \varphi\left(  \mathbf{r}\right)  \right\}
,z_{0}\right)  ={\prod\limits_{i=1}^{N}}\left(  d\mathbf{r}_{i}\exp\left(
-\frac{\varphi\left(  \mathbf{r}_{i}\right)  }{T}\right)  g\left(
\mathbf{r}_{i-1}-\mathbf{r}_{i}\right)  \right)  \label{Z_s}%
\end{equation}
Here $\mathbf{r}_{0}=\left(  0,0,z_{0}\right)  $ and $g\left(  r\right)  $ is
the bond function defined above by eq (\ref{bond}), the statistical segment
length $l$ being the same for $A$ and $B$ chains.

Then the thermodynamic potential $\Phi_{\mathtt{str}}\left(  \left\{
\varphi_{i}(\mathbf{r})\right\}  \right)  $ reads%
\begin{align}
\Phi_{\mathtt{str}}\left(  \left\{  \varphi_{i}(\mathbf{r})\right\}  \right)
&  ={\sum\limits_{n=0}^{M-1}}\Phi_{N_{A}}\left(  \left\{  \varphi
_{A}(\mathbf{r})\right\}  ,\mathbf{r}_{n}^{A}\right) \nonumber\\
&  +{\sum\limits_{n=0}^{M-1}}\Phi_{N_{B}}\left(  \left\{  \varphi_{B}\left(
\mathbf{r}\right)  \right\}  ,\mathbf{r}_{n}^{B}\right)  \label{Phi_m}%
\end{align}
where $\mathbf{r}_{n}^{A}=(0,0,2na),\ \mathbf{r}_{n}^{B}=\left(  0,0,\left(
2n+1\right)  a\right)  $ and
\begin{equation}
\Phi_{N}\left(  \left\{  \varphi\left(  \mathbf{r}\right)  \right\}
,z_{0}\right)  =-T\ln Z_{N}\left(  \left\{  \varphi\left(  \mathbf{r}\right)
\right\}  ,z_{0}\right)  . \label{def_s}%
\end{equation}

Additivity of $\Phi_{\mathtt{str}}\left(  \left\{  \varphi_{i}(\mathbf{r}%
)\right\}  \right)  $ with respect to the terms depending on different
external fields implies that the matrix (\ref{Gam_1}) of the molecular
response functions contains diagonal elements only:%
\begin{align}
g_{AA}(\mathbf{r}_{1},\mathbf{r}_{2})  &  ={\sum\limits_{n=0}^{M-1}}\left.
\frac{\delta^{2}\Phi_{N_{A}}\left(  \left\{  \varphi(\mathbf{r})\right\}
,\mathbf{r}_{n}^{A}\right)  }{\delta\varphi(\mathbf{r}_{1})\,\delta
\varphi(\mathbf{r}_{2})}\right\vert _{\varphi\left(  \mathbf{r}\right)
=0},\nonumber\\
g_{BB}(\mathbf{r}_{1},\mathbf{r}_{2})  &  ={\sum\limits_{n=0}^{M-1}}\left.
\frac{\delta^{2}\Phi_{N_{B}}\left(  \left\{  \varphi(\mathbf{r})\right\}
,\mathbf{r}_{n}^{B}\right)  }{\delta\varphi(\mathbf{r}_{1})\,\delta
\varphi(\mathbf{r}_{2})}\right\vert _{\varphi\left(  \mathbf{r}\right)
=0},\nonumber\\
g_{AB}(\mathbf{r}_{1},\mathbf{r}_{2})  &  =g_{BA}(\mathbf{r}_{1}%
,\mathbf{r}_{2})=0. \label{g3}%
\end{align}

On the other hand,
\begin{equation}
\left.  \frac{\delta^{2}\Phi_{N}\left(  \left\{  \varphi(\mathbf{r})\right\}
,z_{0}\right)  }{\delta\varphi(\mathbf{r}_{1})\,\delta\varphi(\mathbf{r}_{2}%
)}\right\vert _{\varphi\left(  \mathbf{r}\right)  =0}=\left.  \frac{\delta
^{2}Z_{N}\left(  \left\{  \varphi(\mathbf{r})\right\}  ,z_{0}\right)  }%
{\delta\varphi(\mathbf{r}_{1})\,\delta\varphi(\mathbf{r}_{2})}\right\vert
_{\varphi\left(  \mathbf{r}\right)  =0}\nonumber
\end{equation}%
\begin{equation}
\left.  -\frac{\delta Z_{N}\left(  \left\{  \varphi(\mathbf{r})\right\}
,z_{0}\right)  }{\delta\varphi(\mathbf{r}_{1})}\frac{\delta Z_{N}\left(
\left\{  \varphi(\mathbf{r})\right\}  ,z_{0}\right)  }{\delta\varphi
(\mathbf{r}_{2})}\right\vert _{\varphi\left(  \mathbf{r}\right)  =0},
\label{der1}%
\end{equation}
where we employed the definitions (\ref{Z_s}) and (\ref{def_s}). Summing the
second field derivatives of $Z_{N}\left(  \left\{  \varphi(\mathbf{r}%
)\right\}  \right)  $ and collecting the similar terms one gets%
\begin{align}
g_{ii}(\mathbf{r}_{1},\mathbf{r}_{2})  &  =A_{i}\left(  \mathbf{r}_{1}\right)
\delta\left(  \mathbf{r}_{1}-\mathbf{r}_{2}\right)  +B_{i}\left(
\mathbf{r}_{1},\mathbf{r}_{2}\right) \nonumber\\
&  -C_{i}\left(  \mathbf{r}_{1},\mathbf{r}_{2}\right)  , \label{gii}%
\end{align}
where%
\begin{equation}
A_{i}\left(  \mathbf{r}_{1}\right)  ={\sum\limits_{n=0}^{M-1}}\;{\sum
\limits_{l=1}^{N}}g^{\left(  l\right)  }\left(  \mathbf{r}_{n}^{i}%
,\mathbf{r}_{1}\right)  , \label{def_A}%
\end{equation}%
\begin{align}
\quad B_{i}\left(  \mathbf{r}_{1},\mathbf{r}_{2}\right)   &  ={\sum
\limits_{n=0}^{M-1}\,}{\sum\limits_{l=1}^{N}}g^{\left(  l\right)  }\left(
\mathbf{r}_{n}^{i},\mathbf{r}_{1}\right)  \,g^{\left(  m\right)  }\left(
\mathbf{r}_{1},\mathbf{r}_{2}\right) \nonumber\label{def_B}\\
&  +{\sum\limits_{n=0}^{M-1}\,}{\sum\limits_{l=1}^{N}}g^{\left(  l\right)
}\left(  \mathbf{r}_{n}^{i},\mathbf{r}_{2}\right)  \,g^{\left(  m\right)
}\left(  \mathbf{r}_{2},\mathbf{r}_{1}\right)
\end{align}%
\begin{equation}
C_{i}\left(  \mathbf{r}_{1},\mathbf{r}_{2}\right)  ={\sum\limits_{n=0}^{M-1}%
}{\sum\limits_{l=1}^{N}}g^{\left(  l\right)  }\left(  \mathbf{r}_{n}%
^{i},\mathbf{r}_{1}\right)  {\sum\limits_{m=1}^{N}}g^{\left(  m\right)
}\left(  \mathbf{r}_{n}^{i},\mathbf{r}_{2}\right)  \label{def_C}%
\end{equation}
and the kernel $g^{\left(  k\right)  }\left(  \mathbf{r}_{1},\mathbf{r}%
_{2}\right)  \ $is defined by eq (\ref{def_n}).

Based on the previous consideration, we should use for $g^{\left(  k\right)
}\left(  \mathbf{r}_{1},\mathbf{r}_{2}\right)  $ the representation
\begin{align}
g^{\left(  k\right)  }\left(  \mathbf{r}_{1},\mathbf{r}_{2}\right)   &
={\sum\limits_{n=0}^{\infty}}\,{\sum\limits_{m=1}^{\infty}} \int\frac{dq}%
{2\pi}\exp\left(  -\frac{k\lambda\left(  q,n,m\right)  }{6}\right) \nonumber\\
&  \times Y_{n}^{m}\left(  \mathbf{r}_{1}\right)  \overline{Y_{n}^{m}\left(
\mathbf{r}_{2}\right)  }%
\end{align}
with the eigen values $\lambda$ and eigenfunctions $Y_{n}^{m}\left(
\mathbf{r}_{1}\right)  $ defined by eq (\ref{tube}).

However, unlike the situation with free (untethered) copolymers under
confinement, the correlators (\ref{gii}) for bottlebrushes are, in general,
not reducible to the form (\ref{series}). Accordingly, the quadratic form
(\ref{F2a}) does not take the desired diagonal form (\ref{F2t}).

Fortunately, it could be shown that for the condensed state the correlators can
be expanded in powers of the small parameter $\kappa$ defined by eq (\ref{val1}).
Therewith, keeping only the dominant (zeroth approximation in $\kappa$) term of
this expansion one arrives precisely at the diagonal form (\ref{F2t}):
\begin{align*}
\Delta F_{2}  &  =\frac{T}{2}{\sum\limits_{n=0}^{\infty}\sum\limits_{m=1}%
^{\infty}}\,\,\int\frac{dq}{2\pi}A_{\alpha}(q,n,m)\,\overline{A_{\beta
}(q,n,m)}\\
&  \times\left(  \mathbf{G}^{-1}\left(  q,n,m\right)  \right)  _{\alpha\beta},
\end{align*}
with%
\begin{equation}
\mathbf{G}^{-1}\left(  q,n,m\right)  =\mathbf{g}^{-1}\left(  q,n,m\right)
-\mathbf{C}\,, \label{G2bb}%
\end{equation}
where the matrix $\mathbf{C}$ is defined in eq (\ref{C}) and
\begin{align}
g_{\alpha\alpha}^{\text{bb}}\left(n,s_{n}^{m},q\right)   &  =\phi\,N%
\,f_{D}\left(  \widetilde{Q}^{2}x_{\alpha}\right) \label{g_bb0}\\
&  -\phi\,N\,\psi^{2}\left(  \widetilde{Q}^{2}x_{\alpha}\right)  ,\quad
n=0\nonumber
\end{align}%
\begin{align}
g_{\alpha\alpha}^{\text{bb}}\left(n,s_{n}^{m},q\right)   &  =\phi\,N%
\,f_{D}\left(  \widetilde{Q}^{2}x_{\alpha}\right)  ,\quad n\geq1\label{g_bb1}\\
g_{AB}^{\text{bb}}  &  =0,\qquad\qquad\qquad n\geq0, \label{g_bbab}%
\end{align}
with all parameters and functions defined after eqs (\ref{g_dib}).

\section{2.4. Correlations and onset of ordering in two-component bottlebrushes:
theoretical predictions.}

The general condition of the multi-component block copolymer disordered state
stability with respect to microphase separation is well known \cite{74}. For
block copolymers confined in a cylindrical tube and the condensed
two-component bottlebrushes it should be reformulated as follows:
\begin{equation}
\Lambda=\min\Lambda_{i}\left(  n,m,q\right)  \geq0. \label{stab}%
\end{equation}
where $\Lambda_{i}\left(  n,m,q\right)  $ are the eigen values of the
correlation matrix $\mathbf{G}^{-1}$ defined by eq (\ref{G2bb}) given the
values of the "quantum numbers" $n$ and $m$ characterizing the types of the
angular and radial concentration modulation and wave number $q$ characterizing
the wave length of the longitudinal concentration oscillation \cite{71}, and
the minimum is to be sought on the set of all possible values of $n$ and $m$
and on the half-axes $q \geq 0$.

Accordingly, the condition, which determines the surface (in the space of the
structure and energetic parameters of the system under study) where the
uniform state of the latter loses its stability (spinodal), reads%
\begin{equation}
\Lambda\left(  n^{\ast},m^{\ast},q^{\ast}\right)  =0, \label{spin}%
\end{equation}
where we label with asterisks those values of $n,m,q$ that actually satisfy eq
(\ref{spin}).

Based on the WST experience \cite{rev1,rev2,rev3,rev4,57} one can arrive
readily at the following conclusions. \textit{i}) The value of $n^{\ast}$
defines the type of the angular modulation of the morphology arising after the
uniform state looses its stability. In particular, if $n^{\ast}=0$ then a
radially symmetric morphology arises, if $n^{\ast}=1$ then the
Janus-cylinder-type morphology would occur, if $n^{\ast}=2$ then the
cross-like $ABAB$ morphology would arise etc. \textit{ii}) If $q^{\ast}=0$
then the arising morphology is expected to be homogeneous along the backbone
direction, otherwise a 1D morphology periodic along this direction would occur
with the period $L\approx2\pi/q^{\ast}$. 3) the value of $m$ (more precisely,
$s_{n}^{m}$) defines the type of the radial modulation of the arising
morphology. (It is worth to notice that such a straightforward extension of
the results of the bulk stability analysis to the quasi-one-dimensional
ordering in tubes and bottlebrushes is only approximate \cite{46,71} as
mentioned above. We address this issue in more detail elsewhere.)

To demonstrate the general results let us compare ordering in bottlebrushes
and free diblock copolymers confined in a cylindrical tube. It follows from
eqs (\ref{C}) and (\ref{G2bb}) - (\ref{g_bbab}) that for bottlebrushes
\begin{equation}
(v_{0}\mathbf{G})^{-1}=\left(
\begin{array}
[c]{cc}%
\left(  g_{AA}^{\text{bb}}\right)  ^{-1}-2\widetilde{\chi}_{AS} & \phi_{S}^{-1}-k\\
\phi_{S}^{-1}-k & \left(  g_{BB}^{\text{bb}}\right)  ^{-1}-2\widetilde{\chi}_{BS}%
\end{array}
\right)  \label{GGt}%
\end{equation}
whereas for free diblocks in a tube
\begin{equation}
(v_{0}\mathbf{G})^{-1}=\left(
\begin{array}
[c]{cc}%
\left(  \mathbf{g}^{\text{free}}\right)  _{AA}^{-1}-2\widetilde{\chi}_{AS} & \left(
\mathbf{g}^{\text{free}}\right)  _{AB}^{-1}+\phi_{S}^{-1}-k\\
& \\
\left(  \mathbf{g}^{\text{free}}\right)  _{BA}^{-1}+\phi_{S}^{-1}-k & \left(
\mathbf{g}^{\text{free}}\right)  _{BB}^{-1}-2\widetilde{\chi}_{BS}%
\end{array}
\right)  \label{GGb}%
\end{equation}
with the components of the matrix $\mathbf{g}^{\text{free}}$ defined in eq (\ref{g_dib}),
$k=\chi_{AS}+\chi_{BS}-\chi_{AB}$ and $\widetilde{\chi}_{\alpha S}%
=\chi_{\alpha S}-(2\phi_{S})^{-1}$.

For symmetric blocks the eigen vectors of the matrices (\ref{GGt}) and
(\ref{GGb}) are $\mathbf{e}_{-}=(1,-1)/\sqrt{2}$ and $\mathbf{e}_{+}%
=(1,1)/\sqrt{2}$ that correspond to the order parameters
\begin{align*}
\Psi_{-}\left(  \mathbf{r}\right) & =\left( \delta \phi_{A}\left(  \mathbf{r}%
\right) - \delta \phi_{B}\left(  \mathbf{r}\right)  \right)  /\sqrt{2}\\
\Psi_{+}\left(\mathbf{r}\right)&=\left( \delta \phi_{A}\left(  \mathbf{r}%
\right) + \delta \phi_{B}\left(  \mathbf{r}\right)  \right)  /\sqrt{2}=-\delta\phi_{S}\left(  \mathbf{r}\right)  /\sqrt{2}
\end{align*}.  The corresponding eigen values are
\begin{align}
\Lambda_{-}^{\text{bb}}\left(  n,m,q\right) & =\left(  g_{AA}^{\text{bb}}\left(n,s_{n}^{m},q\right)  \right)  ^{-1}-\chi_{AB}, \label{l-} \\
\Lambda_{+}^{\text{bb}}\left(  n,m,q\right) & =\left(  g_{AA}^{\text{bb}}\left(n,s_{n}^{m},q\right)  \right)  ^{-1}-\chi_+ , \label{l+}
\end{align}
for bottlebrushes (with $\chi_+=4\chi_{AS}-\chi_{AB}-2\phi_{S}^{-1}$)  and
\begin{align}
\Lambda_{-}^{\text{free}}\left(s_{n}^{m},q\right) & =\left(g_{AA}^{\text{free}}\left(  s_{n}^{m},q\right) -
g_{AB}^{\text{free}}\left(  s_{n}^{m},q\right) \right)  ^{-1}-\chi_{AB}, \label{l--} \\
\Lambda_{+}^{\text{free}}\left(s_{n}^{m},q\right) & =\left(g_{AA}^{\text{free}}\left(  s_{n}^{m},q\right) +
g_{AB}^{\text{free}}\left(  s_{n}^{m},q\right) \right)  ^{-1}-\chi_+ , \label{l++}
\end{align}
for free diblocks in a tube.

\subsection{Polymer-polymer and polymer-solvent correlations}
Substituting eqs (\ref{g_dib}) and (\ref{g_bb0}) - (\ref{g_bbab}) into
expressions (\ref{l-}) and (\ref{l--}) we see that the functions
$\Lambda_{-}(0,0,Q^2)$ are the same for free and tethered diblocks
(bottlebrushes) in the case of a purely longitudinal ordering
($n=0,\;s_0^1=0$) (see the well known \cite{73} bold solid line \textit{1} in
Fig. 1 with a minimum at $Q^2=Q_*^2=3.785$). It is worth to emphasize that the
location of the minimum depends only on the length $N$ of the side block
(which scales the dimensionless quantity $Q^2=q^2l^2N/6$) and does depend
neither on the grafting density nor on the value of the $\chi$-parameter
(temperature). This surprising similarity is due to the very nature of our
approximation, which is valid only for strongly compressed bottlebrushes
($\kappa\ll1$). Indeed, in this case the entropic loss due to fixing one of
the chain ends on the backbone is negligible as compared to that due to chain
squeezing into a narrow tube, which is the same for free and tethered chains.
For other concentration modes having no angular but some radial modulation
($n=0,\,s_n^m>0$) one has
\begin{equation}
\label{Lam-}
\Lambda_{-}(0,s_n^m,Q^2)=\Lambda_{-}\left(0,0,Q^2+(s_n^m)^2/(6\kappa)\right)
\end{equation}
(see definitions after eq (\ref{g_dib}). In other words, for these modes the
instability curve $\Lambda_{-}(Q^2)=0$ looks like that for the lamellar-like
mode ($n=0,\,s_0^1=0$) but shifted by $(s_n^m)^2/(6\kappa)$ (thin solid lines
\textit{1} and \textit{2} in Fig. 1).

\begin{figure*}
\includegraphics{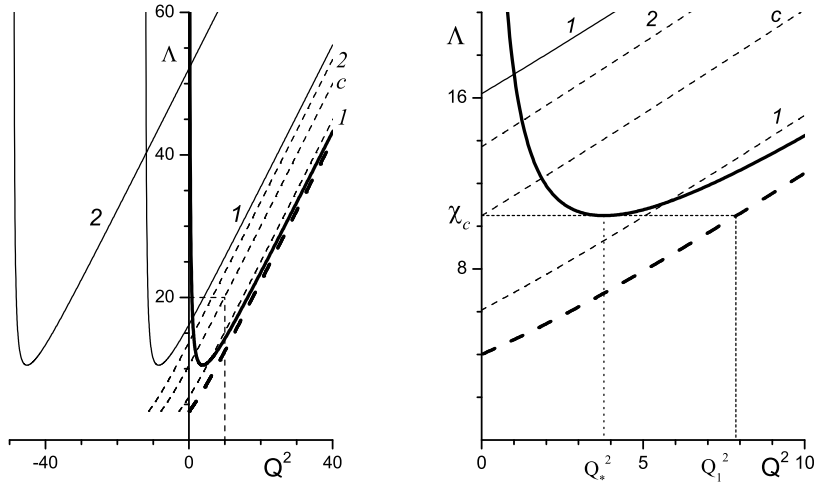}
\caption{\label{fig1}The plots $\Lambda_{-}(Q^2)$ for $n=0, m=1$ (the bold
  solid line), $n=0, m=2$ (thin solid lines) and $n=1, m=1$ (the dashed
  lines). The labels $1$, $2$ and $c$ correspond to the values of
  $\kappa=0.2$, $\kappa=0.05$ and $\kappa=\kappa_1$, respectfully. The bold
  dashed line corresponds to  $\kappa=0$. The right figure shows the enhanced
  inset from the left one. The dotted straight lines are the guides for eyes
  to explain the meaning of the quantities $\chi_c$, $Q_*^2$ and $Q_1^2$.}
\end{figure*}

Therefore, for the radially symmetric modes with $(s_0^m)^2/(6Q_*^2>\kappa$
the function $\Lambda_{-}(0,s_n^m,\widetilde{Q}^2)$ reaches its minimum for
$Q=0$. For $\kappa<\kappa_0=(s_0^m)^2/(6Q_*^2\approx 0.65$, which is surely
true within our approximation valid for $\kappa\ll1$, the lamellar-like mode
is the only one for which the function $\Lambda_{-}(0,s_n^m,Q^2)$ reaches its
minimum at a finite value of $Q$. Moreover, in this case the lamellar mode is
dominant among those with no angular modulation. Indeed, as shown in Fig. 2,
it is only the square averaged amplitude of this mode, which strongly
increases when
\begin{equation}\label{AB}
\phi N \chi_{AB}\rightarrow \chi_{c}=10.495.
\end{equation}

The situation is rather different for the modes with angular modulation ($n\geq1$). The
relationship (\ref{Lam-}) holds for these modes of free diblock copolymers in a tube also but
for bottlebrush angular concentration modes one has
\begin{align}
\label{Lamm-}
\Lambda_{-}(n,s_n^m,Q^2)&=\overline{\Lambda}_{-}(Q^2+(s_n^m)^2/\kappa), \\
\overline{\Lambda}_{-}(Q^2)&=\left(\phi N f_D(Q^2/2)\right)^{-1}-\chi_{AB}.  \nonumber
\end{align}
In other words, the instability condition $\Lambda_{-}=0$ looks like that for
the mixture of free blocks A and B but shifted by $(s_n^m)^2/\kappa$ (see the
dashed lines in Fig. 1.).

\begin{figure*}
\includegraphics{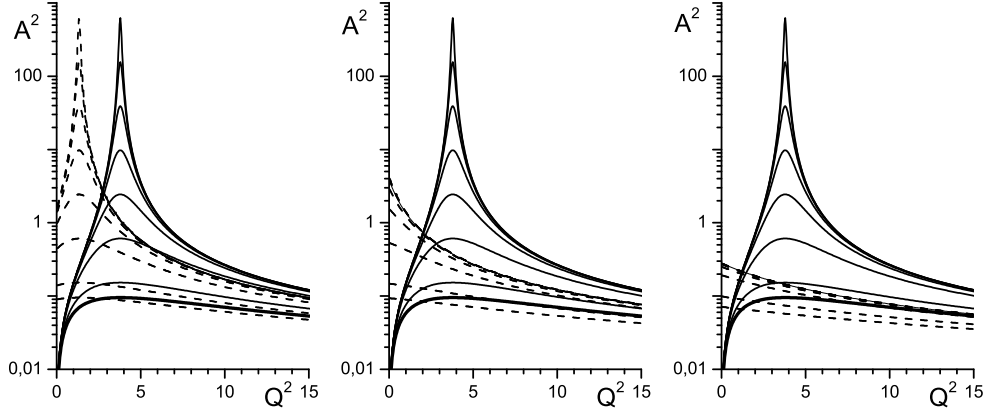}
\caption{\label{fig2}The square averaged amplitude of the lamellar mode
  (solid) and first radially (but not angular) modulated mode with $n=0, m=2,
  s_0^m\approx 3.832$ (dashed) for $\kappa=1$ (left), $\kappa=0.5$ (middle)
  and $\kappa=0.25$ (right). The first two cases are beyond validity of our
  approximation and are only suggestive. The values of $A^2(Q^2)$ monotonously
  increase with increase of $\chi$ which enables us not to label them. The
  bottom lines correspond to $\chi=0$, the other lines do to
  $\chi=\chi_c-4^{i+1}/10000$, where \textit{i} is the number of the curve
  from top to bottom.}
\end{figure*}

As described above, passing the instability threshold (\ref{spin}) results in
formation of a lamellar-like or Janus-cylinders morphology depending on which
function, (\ref{l-}) or (\ref{l+}), has a deeper minimum.

\begin{figure*}
\includegraphics{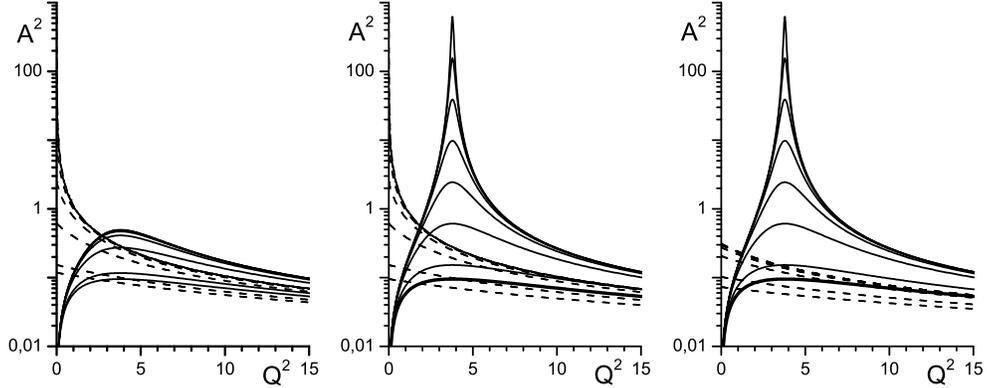}
\caption{\label{fig3}The square averaged amplitude of the lamellar mode
  (solid) and first angular modulated mode with $n=1, m=1, s_0^m\approx 1.841$
  (dashed) for $\kappa=0.11$ (left), $\kappa=\kappa_1$ (middle) and
  $\kappa=0.05$ (right). For $\kappa=0.1$ the angular "Janus" mode appears
  (within the RPA) at $\widetilde{\chi}=\chi_1=
  4/f_D\left((s_0^1)^2/12\kappa\right)\approx8.471$. The bottom lines
  correspond to $\chi=0$, the other ones do to $\chi=\chi_1-4^{i+1}/10000$ for
  $\kappa=0.1$ and $\chi=\chi_c-4^{i+1}/10000$ otherwise.}
\end{figure*}

As shown in Fig. 3, if the shift $(s_n^m)^2/\kappa$ is small then the
instability with respect to forming of Janus-cylinders appears first (with
increase of $\chi_{AB}$). With decrease of $\kappa$ the shift increases and we
arrive at the value of  $\kappa=\kappa_1=(s_1^1)^2/(6Q_1^2)\approx 0.0717$
where both the lamellar and first angular modulated mode loose their stability
simultaneously at $\chi=\chi_c$. With further decrease of $\kappa$ the
lamellar phase becomes more thermodynamically stable. It follows from eq
(\ref{val1}) that $\kappa$ decreases when the distance $a$ between the
neighboring side chains of the bottlebrush increases (and, thus, grafting
density decreases). Thus, the Janus-cylinders could be formed, in agreement
with predictions of refs \cite{8,39,41} in bottlebrushes with high grafting
density. In bottlebrushes with lower grafting density we predict, contrary to
refs \cite{8,39,41}, longitudinal lamellar-like (pearl-necklace) ordering
rather than forming Janus-cylinder.

It is worth to emphasize that our theory predicts a deep similarity between
the polymer-polymer and polymer-solvent ordering in condensed
bottlebrushes. Indeed, comparison of eqs (\ref{l-}) and (\ref{l+}) leads to
conclusion that in symmetric bottlebrush the only difference in formation of
the polymer concentration modulation described by the order parameter
$\Psi_{-}(\mathbf(r))$ and solvent concentration modulation described by the
order parameter $\Psi_{+}(\mathbf(r))$ is energetic. More precisely, depending
on whether the limit (\ref{AB}) is reachable earlier or later than
\begin{equation}\label{S}
\phi N \chi_{+}\rightarrow \chi_{c}=10.495,
\end{equation}
the polymer-polymer or polymer-solvent inhomogeneities will occur but the type
of morphology arising in both cases is expected to be the same.

\section{3. MOLECULAR DYNAMICS SIMULATIONS OF TWO-COMPONENT BOTTLE-BRUSHES}

\subsection{Model and methods to analyze the results}

The strictly rigid backbone of the simulated bottle-brush polymer is simply
taken as an immobile straight line in $z$-direction, where we also apply
periodic boundary conditions, thus disregarding any end-effects. Using units
of length such that the length parameter $\sigma_{LJ}$ of the Lennard-Jones
potential (see below) is taken to be unity, $\sigma_{LJ}=1$, we graft side
chains (alternatingly of type A and type B) at regular positions $z_{\nu}%
=(\nu-1)/\sigma, \;\nu=1,2,\ldots,2M$, with $\sigma= 1.14$ and a number 2M = 50
of grafted chains. The side chain length was chosen as N = 35. For comparison,
also some results were obtained for a higher grafting density $(\sigma= 1.51)$
and two lower ones $(\sigma=0.57$ and $\sigma= 0.76$); in the latter case,
three chain lengths were studied $(N_{A}=N_{B}=N=20$, $35$, and $50$,
respectively). We are fully aware of the fact that our side chain lengths
would not be long enough to study the asymptotic scaling behavior under good
solvent conditions \cite{47}, of course; however, for poor solvent conditions
(as studied here) very large relaxation times would prevent us from reaching
thermal equilibrium for longer side chains. Moreover, the range of N that we
explore nicely corresponds to the range that can be probed in experiments
\cite{10,11,53}.

The side chains are modeled by the standard bead-spring model \cite{24}
already used in our previous work \cite{51,55}. All beads interact with a
truncated and shifted Lennard-Jones potential%
\begin{equation}
\label{eq60}
U_{LJ}(r)=\left\{
\begin{array}
[c]{r@{\; , \;}l}%
4\epsilon_{LJ}[(\sigma_{LJ}/r)^{12}-(\sigma_{LJ}/r)^{6}]+C & r\leq r_{c}\\
0 & r>r_{c}%
\end{array}
\right.
\end{equation}
where $r_{c}=2.5\sigma_{LJ}$, and the constant C is defined such
that $U_{LJ}(r=r_{c})$ is continuous at the cutoff. Units are
chosen such that $\sigma_{LJ}=1$ irrespective of whether we deal with
AA, BB or AB pairs, and also Boltzmann's constant $k_{B}=1$, as
well as the masses of the beads
$m_{A}=m_{B}=m=1$. For the energy parameters, however, we choose%

\begin{equation}
\label{eq2}\epsilon^{AA}_{LJ}= \epsilon^{BB}_{LJ}=1,\; \epsilon^{AB}_{LJ}=1/2;
\; 3/4; \; 7/8;\; \text{and} \; 15/16,
\end{equation}
respectively. Remember that for $\epsilon_{LJ}^{AB}=1$ the problem reduces to
a single-component brush.

As usual \cite{24,51,55} the connectivity of the beads along a side chain is
maintained by the ``finitely extensible nonlinear elastic'' (FENE) potential,%

\begin{equation}
\label{eq3}U_{FENE}(r)= - \frac1 2 kr_{0}^{2} \ln[1-(r/r_{0})^{2}], \quad0<r
\leq r_{0}\;,
\end{equation}
with the standard choice of parameters, $r_{0}=1.5$, and $k=30$, respectively.
Since $U_{FENE}(r \geq r_{0})=\infty$ the bond lengths are constrained.

Simulations were carried out controlling temperature by the Langevin
thermostat \cite{24,51,55}. The equation of motion of a bead then reads%

\begin{equation}
\label{eq4}m \frac{d^{2}\vec{r}_{i}}{dt^{2}} = - \nabla U_{i} - m \gamma
\frac{d\vec{r}_{i}}{dt} + \vec{\Gamma}_{i}(t),
\end{equation}
where $t$ denotes the time, $U_{i}$ is the total potential acting
on the i-th bead (which is at the position $\vec{r}_{i}$),
$\gamma$ is the friction coefficient, and $\vec{\Gamma}_{i}(t)$
the random force. As is well known, $\gamma$ and
$\vec{\Gamma}_{i}(t)$ are related by the
fluctuation-dissipation relation%

\begin{equation}
\label{eq5}\langle\vec{\Gamma}_{i}(t) \cdot\vec{\Gamma}_{j}(t^{\prime}%
)\rangle= 6 k_{B}T \gamma\delta_{ij} \delta(t^{\prime}-t).
\end{equation}
Eqs.~\ref{eq4},\ref{eq5} are numerically integrated using the
GROMACS package, applying the leap frog algorithm \cite{87,88}.
Following previous work \cite{24,51,55}, $\gamma= 0.5$ was chosen,
and a time step $\Delta t = 0.006 \tau$ where the
Molecular Dynamics(MD) time unit was chosen as $\tau= (m \sigma_{LJ}%
^{2}/\epsilon_{LJ})^{1/2}=1$.

As was briefly mentioned above, equilibration of our model system
is rather difficult. Typically we started by equilibrating the
system first at a temperature $T=3.0$, which is known \cite{89} to correspond roughly
to the Theta temperature of homopolymers in dilute solution. 500 different
configurations at $T=3.0$ were equilibrated using runs extending
over a time range of $30\times10^{6} \tau$. All these configurations
are then cooled down from $T=3.0$ to $T=1.5$ with temperature
steps $\Delta T =0.1$. At each temperature $2\times10^{6}$ steps were
used for further equilibration. As an example, Figs.~4 and 5 show
typical snapshot pictures of the simulated bottle brushes. At
$T=3.0$ the individual side chains take
irregular conformations, stretching out in the directions
perpendicular to the backbone. Obviously, these conformations are
rather disordered, the contacts between monomers of different
chains are too scarce to induce significant microphase separation, and even on
local scales the composition of the bottle brush is rather
randomly mixed. At $T = 1.5$ the situation is different,
however: for $\sigma= 0.57$ one can clearly see the presence of a
pearl-necklace type structure, as far as the total density is
concerned; typically these pearls sometimes are almost exclusively
formed from one type of chain only, implying a periodic
concentration variation along the $z$-axis. For $\sigma= 0.76$,
however, the interpretation of many snapshot pictures rather
suggest that "Janus pearls" (containing an A-rich and a B-rich
part, separated by an interface) have formed. For $\sigma= 1.14$,
already rather elongated A-rich and B-rich parts (reminiscent of
"Janus cylinders") are visible, but there still occur strong
undulations in the local monomer density in $z$-direction. These
undulations for $\sigma= 1.51$ clearly are somewhat weaker,
although the density variation still is far from a uniformly
filled cylinder. While some snapshots suggest that a "Janus
cylinder" type ordering persists over the full lengths of the
studied backbone length, we emphasize the caveat that in Figs.~4,
5 one cannot really distinguish whether a local cross section
perpendicular to the cylinder axis resembles more a sphere (as
appropriate for a cylinder) or a dumbbell, for instance.

In order to provide a quantitative analysis of the various types of short
range order we define correlation functions adapted to the cylindrical
symmetry of our problem, by suitable generalization of the two-point
correlation function in fluid systems,%

\begin{eqnarray}
G_{\alpha\beta}(\delta z,\delta\phi)=&\frac{\overline{N}}{\rho \overline{N}_{\alpha}\overline{N}_{\beta}}%
\langle\sum\limits_{i=1}^{\overline{N}_{\alpha}}\sum\limits_{j=1}^{\overline{N}_{\beta}}
\delta(z_{i}-z_{j}-\delta z)\\ &\delta(\phi_{i}-\phi_{j}-\delta \phi)\rangle,\;
\alpha,\beta=A,B \nonumber \label{eq51}%
\end{eqnarray}

In Eq.~\ref{eq51} the first sum runs over all $\overline{N}_{\alpha}$ particles of type
$\alpha$, which have the cylindrical coordinates $z_{i},r_{i},
\phi_{i}$, while the second some runs over all $\overline{N}_{\beta}$ particles of type
$\beta$, which have the cylindrical coordinates
$z_{j},r_{j},\phi_{j}$, $\overline{N}_{\alpha}=\overline{N}_{\beta}=NM$. Fig.~6 shows this
correlation function $G_{AA}(\delta z,\delta\phi)$ for a typical
case: as expected, in the temperature region close to the
$\theta$-temperature this correlation decays to its asymptotic
value for large $\delta z$, $G_{\alpha\beta} (\delta z
\rightarrow\infty, \delta\phi)=1$ if $\delta z \geq10$, while a
nontrivial angular dependence occurs for smaller $\delta z$.
Basically, this structure is due to intra-chain-correlations. The
correlation function $G_{AB}(\delta z,\delta\phi)$ in the region
$\delta z \leq10$ is predominantly smaller than unity, the chains
avoid unfavorable contacts already at this rather high
temperature. However, the angular dependence occurring at $T=2.9$
in $G_{AB}(\delta z , \delta\phi)$ for $\delta z> 10$ clearly is
due to a lack of statistical accuracy, as a comparison of data for
individual runs shows. Similarly, the weaker deviations of $G_{AA}
(\delta z \geq15, \delta\phi)$ from unity should be disregarded
for the same reason either. While at low temperatures ($T=1.5$,
see Fig.~6 clearly a more systematic correlation over a larger
range of $\delta z$ develops, the systematic finite size effects
due to the periodic boundary condition in $z$-direction become a
concern, since

\begin{equation}
\label{eq52}G_{\alpha\beta} (\delta z, \delta\phi) = G_{\alpha\beta}%
(L_{b}-\delta z, \delta\phi)
\end{equation}
where the length $L_{b}$ of the backbone is $L_{B} = 2M/\sigma$ (in the
examples shown in Fig. 6 we have $L_{b}=87.72$, i.e., $G_{\alpha\beta}(\delta z,
\delta\phi)$ is shown for $0 \leq \delta z \leq L_{b}/2$).

The conclusion of the discussion of Fig. 6 is that a substantially larger
statistical effort would be needed to sample $G_{\alpha\beta}(\delta z ,
\delta\phi)$ with meaningful accuracy. Therefore it was decided to focus on a
more meaningful average information, defined as%

\begin{equation}
\label{eq53}\ G_{\alpha\beta} (\delta z,m) = \int\limits_{0} ^{\pi}%
G_{\alpha\beta} (\delta z, \delta\phi) \cos(m \delta\phi)d \phi, m = 0.1,2
\end{equation}

The choice $m=0$ simply means that the angular difference $\delta\phi$ between
the coordinates $\vec{r}_{i},\vec{r}_{j}$ is not at all taken into
consideration. The choice $m=1$ is used in the expectation that this
correlation then will yield some information whether or not Janus-cylinder
type ordering occurs, while the choice $m=2$ could be useful, if
``Janus-dumbbell'' and double-cylinder-like structures occur (see also the
discussion in Sec. 2.2).

We have also used these correlations to transform to correlation functions
that reflect fluctuations of total number density $(g_{nn}(\delta z, m))$ and
relative concentration $(g_{cc}(\delta z,m))$. Following Bhatia and Thornton
\cite{90} we obtain%
\begin{equation}
\label{eq54}g_{nn}(\delta z,m)=\sum_{\alpha,\beta=A,B}   x_{\alpha} x_{\beta}G_{\alpha\beta}(\delta
z,m)
\end{equation}
and
\begin{equation}
\label{eq55}g_{cc}(\delta z,m) = \sum_{\alpha,\beta=A,B} y_{\alpha} y_{\beta
}G_{\alpha\beta}(\delta z,m)
\end{equation}
where in our case $x_{A}=x_{B}=1/2$ and $y_{A}=-y_{B}=1/2$. Of course,
$g_{nn}(\delta z,m)$ can be computed directly, considering a correlation
$G_{\alpha\beta}$ where the sums over i and j (Eq.~\ref{eq51}) both run over
all particles, irrespective of whether they are of type A or B.

Since it is not clear whether measurements will become possible
that will yield information directly on $g_{\alpha\beta}(\delta
z,m)$, we also consider the structure factor $S_{\alpha\beta}(q)$
of the bottle brushes; as is well-known, for one-component
bottle-brushes ample information on the structure factor $S(q)$ is
available from scattering experiments \cite{10,11}, and for
two-component bottle-brushes relevant information could be
obtained from scattering experiments as well if one species (A or
B) is a deuterated polymer. Assuming that the wave vector of the
scattering $\vec{q}$ is also oriented along the $z$-axis of the
bottle-brush polymer, the partial structure factor
$S_{\alpha\beta}(q)$ becomes%

\begin{equation}
\label{eq56}S_{\alpha\beta} (q) = \frac{1}{\overline{N}} \sum
\limits_{k=1}^{\overline{N}_{\alpha}} \sum\limits_{\ell= 1}^{\overline{N}_{\beta}} \langle
\exp[iq(z_{k} -z_{\ell})]\rangle\;,
\end{equation}
where $\alpha\beta\in A,B, \quad \overline{N} = \overline{N}_{A}+\overline{N}_{B}$. Again it is
useful to transform to structure factors that relate to total density
fluctuations $(S_{nn}(q))$ and concentration fluctuations $(S_{cc}(q))$, respectively,%

\begin{equation}
\label{eq57}S_{nn}(q)=S_{AA}(q)+2S_{AB}(q)+S_{BB}(q),
\end{equation}

\begin{equation}
\label{eq58}S_{cc}(q)=x_{B}^{2}S_{AA}(q)+x_{A}^{2}S_{BB}(q)-2x
_{A}x_{B}S_{AB}(q).
\end{equation}

In an actual experiment, of course, it would be more natural to consider the
situation when the orientation of the scattering vector $\vec{q}$ is fixed by
the experimental setup, while the orientations of the rigid backbones of the
bottlebrush polymers in solutions are randomly oriented. Then one would need
to consider a structure factor%

\begin{equation}
\label{eq59}\tilde{S}_{\alpha\beta} (q) = \frac{1}{\overline{N}}
\sum\limits_{k=1}^{\overline{N}_{\alpha}} \sum\limits_{\ell= 1}^{\overline{N}_{\beta}} \langle
\exp[i\vec{q}\cdot(\vec{r}_{k} -\vec{r}_{\ell})]\rangle\;,
\end{equation}
where it is understood that the average $\langle\ldots\rangle$ includes an
average over the orientation of $q$. This structure factor $\tilde{S}%
_{\alpha\beta}(q)$ can be interpreted as the Fourier transform of a
correlation function $\tilde{g}_{\alpha\beta}(\Delta r)$, where $\Delta r$ is the absolute
value of the distance between two sites of monomers $\vec{r}_{i},\vec{r}_{j}$
in the bottle brush, $\Delta r =\sqrt{(x_{j}-x_{i})^{2}+(y_{j}-y_{i})^{2}+(z_{j}%
-z_{i})^{2}}$. This correlation $g_{\alpha\beta}(\Delta r)$ was presented
and discussed already in our preliminary communication \cite{55},
and hence will not be discussed further here. We only note that
for large values of $\Delta r$ we
essentially must have $\Delta r \approx|z_{j}-z_{i}|$ and then $\tilde{g}%
_{\alpha\beta}(\Delta r)\approx g_{\alpha\beta}(\delta z = \Delta r)$, since the differences
$x_{j}-x_{i}, y_{j}-y_{i}$ must be relatively much smaller (of the order of
the cross-sectional radius $R$ of the cylindrical brush). However, for small
$\Delta r$
the behavior of $\tilde{g}_{\alpha\beta}(\Delta r)$ is needed to estimate the number
of contacts $n_{con}^{\alpha\beta}$ which is defined as%

\begin{equation}
\label{eq92}n_{con}^{\alpha\beta} = 4 \pi\int\limits_{0}^{r_{con}}\tilde
{g}_{\alpha\beta} (\Delta r) (\Delta r)^{2}d(\Delta r)
\end{equation}
Eq.~\ref{eq92} means that a pair of particles $(\alpha, \beta)$ is defined to
have a pairwise contact if their distance is less than $r_{con}$. We have
followed the criterion due to Stillinger \cite{91} to take $r_{con}= 1.5 \sigma_{LJ}$.

\section{RESULTS}

We first focus on the dependence of the number of contacts
$n_{con}^{AA} =n_{con}^{BB}$ and $n_{con}^{AB}$ on temperature $T$
and interaction energy $\epsilon_{AB}$ (remember
$\epsilon_{AA}=\epsilon_{BB} = \epsilon= 1$), Fig.~7, to clarify
what is the range of interest of those parameters, where the
microphase separation occurs. We recognize that for $T
> 3$ all contact numbers start to become independent of $T$ when
$T$ is increased. This behavior is easily interpreted by the fact
that for $T > \theta(\theta \approx3)$ the repulsive interactions
among the beads dominate, the side chains essentially take on
self-avoiding walk-like configurations. Of course, each interior
monomer of a side chain has at least two monomers of the same type
(A or B) which are nearest neighbors along the same side chains,
so we expect that $n_{con}^{AA}=n_{con}^{BB}$ cannot be less than
two. Fig.~7 shows that the actual numbers
$n_{con}^{AA}=n_{con}^{BB}$ at temperatures slightly above
$T=\theta$ are about twice as large,
$n_{con}^{\alpha\alpha}\approx4$. In contrast, the numbers of
heterocontacts between chains of different type are at least an
order of magnitude smaller, from $n_{con}^{AB}(T=3.3,
\epsilon_{A}=\frac1 2) \approx0.20$ to $n_{con}^{AB}(T=3.3,
\epsilon _{AB}=15/16) \approx0.42$. It is interesting to note that
the number $n_{con}^{AA}=n_{con}^{BB}$ increases strongly with
decreasing temperature for $T<2.0$, which we identify as the
regime of temperatures where for the chosen (short!) chain length
of the side chains a rather dense and approximately cylindrical
structure of the bottle-brush can be identified. Note that
$n_{con}^{AA}=n_{con}^{BB}$ at these low temperatures decreases
systematically with increasing $\epsilon_{AB}$, while
$n_{con}^{AB}$ increases with $\epsilon_{AB}$. However, the
dependence of the total number of contacts
$(n_{con}^{AA}+n_{con}^{AB}$) on $\epsilon_{AB}$ is much weaker.

On the other hand, the temperature dependence of $n_{con}^{AB}$ depends very
strongly on $\epsilon_{AB}$ : while $n_{con}^{AB}$ for $\epsilon_{AB}\leq3/4 $
stays almost independent of $T$, even for low $T$ where the brush is in a
rather dense state, for $\epsilon_{AB}=7/8$ and $\epsilon_{AB}=15/16$ we find
a strong increase of $n_{con}^{AB}$ with decreasing $T$. Tentatively, this
increase can be attributed to the formation of AB-interfaces in the collapsed
parts of the bottle-brush.

We now turn to the behavior of the correlation functions
$g_{nn}(\delta z,m)$ and $g_{cc}(\delta z,m)$ and focus on the
grafting density $\sigma= 1.14$ and compare two choices of
$\epsilon_{AB}, \epsilon_{AB} = 3/4$ and $\epsilon _{AB}=15/16$
(Figs.~8-11). We see that for $\epsilon_{AB}=3/4$ (Fig. 8) in the
average density fluctuation at high $T$ there is a rapid decay of
$g_{nn}(\delta z, m=0)$ to unity (deviations from unity of order
0.01 are just statistical noise), while at lower temperature a
clear minimum near $\delta z=10$ followed by a maximum near
$\delta z=25-30$ develops, indicating rather long wavelength
fluctuations in the thickness of the collapsed cylinder. This
long-wavelength structure is not present in $g_{nn}(\delta z,m=1)$,
Fig. 8b, while $g_{nn}(\delta z,m=2)$, Fig. 8c, indicates that a
uniform correlation gradually develops, which does not occur for
$\epsilon_{AB}=15/16$, however (Fig. 10). While in the average
concentration correlation $g_{cc}(\delta z,m=0)$ one sees a long
wavelength periodicity for $\epsilon_{AB}=3/4$ (Fig. 9a), as well
as for $\epsilon_{AB}=15/16$, (Fig. 11a), this periodicity is
basically independent of temperature and its amplitude is very
small (of order 0.005). Thus, this variation clearly must be
discarded as being due to somewhat insufficient preparation of the
initial states. In contrast, Fig. 9b shows that $g_{cc}(\delta z,1)$
develops much stronger order of uniform sign, indicating that
this system shows a tendency to develop Janus-cylinder type short
range order. While at $\epsilon_{AB}=7/8$ (not shown) this
tendency is also present, though less developed, for
$\epsilon_{AB}=15/16$ (fig. 11b) the ordering tendency is much
weaker: this indicates that the effective $\chi $-parameter for
phase separation between A and B at $\epsilon_{AB}=15/16$ and $N =
35$ may be too weak to cause micro-phase separation. The
correlation function $g_{cc}(\delta z, m = 2)$, Figs. 9c and 11c
show a rapid decay to zero in both cases, independent of
temperature, indicative of the fact that this correlation measures
only some intra-chain correlation effects in these cases, and no
collective effects are detected by $g_{cc}(\delta z,m=2)$ at all.

When we study still larger grafting densities, such as $\sigma=
1.51$ (Figs. 12, 13), these conclusions are confirmed: there is no
significant long wave length correlation in either uniform density
or concentration, but a pronounced tendency to form Janus-cylinder
like short-range order develops, together with a long range
correlation in the density, seen in $g_{nn}(\delta z,m=2)$, which
is due to a cross-section of the cylindrical brush which is
locally non-spherical. The obvious interpretation of Fig. 8c) and
12c) is that the local Janus-type order is not strictly of Janus
cylinder type but rather of Janus dumbbell type, for
$\epsilon_{AB}=3/4$, as it was suggested already in our
preliminary communication \cite{55} where other quantities for the
case $\epsilon_{AB}=1/2$ were studied. This conclusion is
corroborated by the concentration correlations (Fig. 13), which
reveal neither a significant long range correlation in
$g_{cc}(\delta z,m=0)$ nor in $g_{cc}(\delta z,m>2)$ while
$g_{cc}(\delta z, m=1)$ exhibits the development of very
pronounced Janus-type order.

For considerably lower grafting density, such as $\sigma= 0.57$ (not shown) or
$\sigma= 0.76$ (Figs. 14,15) the behavior is very different, however: both
$g_{nn}(\delta z, m=0)$ and $g_{cc}(\delta z, m=0)$ do not exhibit much
structure at high temperature, but develop an oscillation (minimum near
$\delta z=10$, maximum near $\delta z = 20$) when $T$ is lowered towards
$T=1.5$. For longer chain lengths (N = 50, not shown) the behavior is
qualitatively similar, but the minimum now occurs near $\delta z = 12$ and the
maximum near $\delta z = 24$, and for N = 20 (not shown) $g_{cc}(\delta z,
m=0)$ has its minimum at $\delta z=6$, the maximum at $\delta z = 12$. Thus,
there is a clear dependence on this weak correlation seen in Figs. 14a, 15a
on the length of the side-chains. In $g_{nn}(\delta z, m=1)$ there occurs
only a fast decay to zero for N=20 (not shown), while for N=35 a minimum
develops for $\delta z=10$ (Fig.14b) and for N=50 this minimum occurs for
$\delta z = 12$, i.e., these structures are strongly correlated with the
structure that develops for $g_{nn}(\delta z, m=0)$. Taken together with the
observation (Fig. 5b) that the chains form a pearl-necklace structure, we
conclude that typically the ``mass'' in the pearls is not distributed
spherically symmetric around the backbone (the $z$-axis). This feature is also
suggested by visual inspection of the snapshots (Fig. 5a, 5b), but Fig. 14
implies that it is not an accidental observation of some particular snapshots,
but a statistically significant physical effect.

Turning to the concentration correlation $g_{cc}(\delta z, m=1)$
we notice that for high temperatures a rapid decay to zero occurs
but this decay becomes significantly slower as $T$ is lowered
(Fig. 15b). Again the comparison with the corresponding data for
$N=20$ (not shown) and N=50 (not shown) reveals a qualitatively
similar picture (for N = 50 the decay is so slow, however, that
our choice of 2M = 50 would not be large enough to avoid artefacts
due to the periodic boundary conditions). Turning finally
attention to $g_{nn}(\delta z,m=2)$ and $g_{cc}(\delta z,m=2)$,
Figs. 14c and 15c, we notice a monotonous decay to zero, which
becomes somewhat slower when $T$ is decreased. For N = 20 (not
shown) these correlations are almost independent of $T$, decay to
zero for $\delta z \geq5$, while for N = 50 (not shown)
$g_{nn}(\delta z,m=2)$ develops a slow decay (which is absent in
$g_{cc}(\delta z, m =2)$, however). Clearly these data are
delicate to interpret, since upon lowering the temperature for this
larger chain length one presumably crosses the boundary from an
essentially uniform cylindrical structure (as function of $z$) at
higher temperatures to rather elongated clusters \cite{56}. If one
compares the behavior to the case $\sigma= 0.57$ (here some
snapshot pictures, Fig. 5a, suggest a clear alternation of A-rich
and B-rich regions along the z-axis), one does not find any
significant structure of longer range in either $g_{nn}(\delta z,
m=1,2)$ or $g_{cc}(\delta z,m=1,2)$ [in order to save
space, the corresponding data are not shown], and only both
$g_{nn}(\delta z, m=0)$ and $g_{cc}(\delta z,m=0)$ exhibit the
development of periodic structures, similar to what is seen in
Figs. 14a, 15a.

Finally Fig.~16 shows a few typical results of structure factors
$S_{cc}(q)$. Note that due to the periodic boundary conditions $q$
is ``quantized'' in multiples of $2 \pi/L_{b} = \pi\sigma/M$, with
$2M=50$ in Fig.~16 (a,b) while $2M=100$ in Fig. 16 (c,d).

We find that the position of the main peak occurs at $q_{max}\approx0.3$,
(which corresponds to $q_{max}*R_0 = 1.92$, where $R_0$ would be the gyration radius
of a diblock macromolecule $A_NB_N$ with $N = 35$, as used here, at the Theta point,
$T = 3.0$) in all cases, it is only the peak intensity that strongly grows with increasing
$\sigma$ or changing the temperature. Such a finding is quite nontrivial,and without
guidance by the theoretical treatment of the previous section,would have been
somewhat unexpected. Moreover, it is worth to notice that the theoretical and simulation
values for $x=(q_{max}*R_{0})^2$ ($3.785$ and $1.92^{2} \approx 3.685$,
respectively) are in remarkable agreement.
Since the theory is done for temperatures slightly below the Theta
temperature but very long side chain lengths, while the simulation is done
for temperatures down to one half the Theta temperature but rather
short side chains, the agreement is perhaps better than one would
have expected.

In previous work on cylindrical bottle-brushes with a single type
of side chain under poor solvent conditions \cite{56} it was found
that for $0.4 \leq\sigma\leq0.8$ clusters elongated in
$z$-direction occur, which have a gyration radius of order $\langle R_{gz}%
^{2}\rangle^{1/2} \approx10$. These elongated clusters are also visible in the
snapshot pictures (Figs. 5a,b). For larger $\sigma$ the range over which the
concentration distribution is uniform (as measured by $g_{cc}(\delta z,m=1)$
see Fig. 9b) involves a similar length scale. However, the lack of a more
pronounced temperature dependence of $S_{cc}(q)$ is disturbing, and lacks an
explanation. We conclude that in any case it seems difficult to clarify the
nature of short range order in binary bottle-brush polymers from an analysis
of $S_{cc}(q)$.

In order to get a more precise picture of the mesophase ordering in reciprocal space
from the simulations, which could be compared to the corresponding results of the analytical
calculations of Sec.~2 (e.g., Figs.~2, 3), it is useful to carry out numerical Fourier transforms
of the data on the correlations in real space, presented in Figs.~8 - 15. For example, Fig.~17
shows the Fourier transforms $S_{nn}(q,1)$ and $S_{nn}(q,2)$ which are the Fourier transforms
of $g_{nn}(\delta z,1)$ and $g_{nn}(\delta z,2)$, for a representative selection of our data,
while Fig.~18 shows the Fourier transform of the corresponding concentration fluctuation
correlations \{$S_{cc}(q,1)$ is the Fourier transform of $g_{cc}(\delta z, 1)$,
while $S_{cc}(q,2)$ is the Fourier transform of $g_{cc}(\delta z,2)$\}.
One can see that for $\sigma = 0.76$ the density fluctuation structure factors $S_{nn}(q,1)$
and $S_{nn}(q,2)$ show a very pronounced increase at small $q$ as the temperature is lowered,
while for $\sigma = 1.14$ this increase is much weaker, and for $S_{nn}(q,1)$ it is evident
that a small peak grows at nonzero $q$ only $(q \approx 0.3)$, compatible with the $q_{\textrm{max}}$
extracted from $S_{nn}(q)$ as shown in Fig.~16. We conclude from these data that
density variations play a central role for mesophase formation for $\sigma = 0.76$,
but less so for $\sigma = 1.14$. The concentration fluctuation structure factor
$S_{cc}(q,0)$ exhibits for $\sigma =0.76$ a strongly growing peak near $q_m \approx 0.3$
(not shown since the data are qualitatively similar to Fig. 16b),
while $S_{cc}(q,0)$ for $\sigma = 1.14$ exhibits a very weak temperature dependence only (not shown),
as it must be, considering that $g_{cc}(\delta z,0)$ in this case is basically
independent of temperature (Fig.~9). However, the trend seen for $S_{cc}(q,1)$,
which exhibits a growth for small $q$ with decreasing temperature for all the cases
included in Fig.~18, indicates a tendency for local Janus cylinder type order in these cases,
while long range order clearly is not established. We have not included $S_{cc}(q,2)$ into this study,
since the data for $g_{cc}(\delta z,2)$ \{Fig.~9c, 11c, 13c\} indicated
already hardly any interesting temperature dependence to be present.

The data shown in Figs.~16 - 18 indicate that in the case $\sigma = 0.76$
density takes the leading role in the mesophase ordering of the cylindrical brush,
and the concentration ordering adjusts to it, while for $\sigma = 1.14$ concentration
variations along the backbone drive the mesophase ordering, with a strong coupling
between density variations and concentration  variations still being present.
In this case, a periodic modulation of density along the z-axis still grows
when the temperature is lowered (Fig.~17a), and correspondingly also $S_{cc}(q)$
has a pronounced peak near $q \approx q_m \approx 0.3$ (Fig.~16 c,d).
On the other hand, for $\sigma = 1.51$ a rather different behavior is found (Fig.~19):
while for $S_{nn}(q,1)$ only a weak peak at about $q \approx 0.2$ grows when $T$ is lowered (Fig.~19a),
the growth of both $S_{nn}(q,2)§$ and $S_{cc}(q,1)$ for $q \rightarrow 0$ is very strong.
We interpret these observations as a tendency towards the formation of Janus dumbbell-type structures.

\section{4. CONCLUSIONS}
In this paper the microphase separation in binary (A,B) bottlebrush polymers
with rigid backbones driven by decrease of temperature
(i.e., variation of the solvent quality from Theta solvents to poor solvents) was studied,
considering also the variation of grafting density along the backbone of these cylindrical brushes.
Two complementary theoretical methods were used,
namely,
\begin{itemize} \item[(i)] a suitable extension of the random phase approximation (RPA), and
\item[(ii)] Molecular Dynamics (MD) simulations.
\end{itemize}
The first approach has the merit that arbitrarily long chain lengths of the side chains can be considered, as well as infinitely long backbones (so that end effects or effects of boundary
conditions along the backbone do not matter). Besides, the approach results in some explicit expressions for the correlation functions under study. However, the disadvantage of the approach is that
it is basically a linear stability analysis around the homogeneous state,
strongly nonlinear effects are out of consideration.

The MD approach, on the other hand, in principle can take both nontrivial
correlations and nonlinear effects into account,
but in practice (to avoid excessive requests for computer time resources)
is limited to rather short length of the side chains, and also the length
of the (rigid) backbone is finite (end effects then are avoided
by a periodic boundary conditions, but the price to be paid is that reciprocal
space in the z-direction along the backbone is discretized). Also,
non-negligible statistical errors (and systematic errors due to insufficient
length of the MD trajectories, which affect the establishment of full thermal
equilibrium) restrict the accuracy of the results that can be gotten.

Whereas the RPA treatment assumes that the monomer density in the radial
direction perpendicular to the cylinder axis is uniform and constant
(up to the cylinder radius, while it is zero outside of the cylinder surface),
in the model studied by MD this assumption clearly does not apply,
as has been shown in our previous work \cite{51,54,56}.
Rather, it was found that there occur density oscillations very close
to the cylinder axis (reminiscent of the layering of fluid particles
adjacent to the hard wall), and for the chain lengths one can study only for a
rather small regime of radial distances the density profile then is approximately constant.
One approaches rather fast the regime where the density continuously drops to zero,
similar to the density profiles across interfaces in phase-separated polymer
solvent systems, where the density decreases from its value in the melt towards
zero over a distance of several monomer diameters. This low temperature
behavior in the simulation comes closest to the situation assumed in the RPA
calculations. At higher temperatures,
close to the $\Theta$ temperature, the interfacial profile is so broadened that
a flat region where the density has reached the melt density no longer can be identified.
Given the different temperature and chain length regimes for which theory and
simulation access the microphase separation conditions, it is very gratifying
that there is not only a good qualitative agreement between both, but
furthermore the wavelength of instability towards the microphase separation
agrees quantitatively between both.

Qualitatively, the analytical theory describes similar trends as seen in the simulation:
there is a competition between microphase separation developing periodic order in axial direction,
i.e.,  the system tends to develop long range order with a particular periodicity
(the wavelength of this periodic modulation being controlled by the
gyration radius of the side chains, while the strength of the modulation
depends on temperature and grafting density, as the analytical theory
compellingly shows, and as is confirmed by our simulations)
and microphase separation of the Janus-type,
characterized by $q=0$ (the cross section of the dumbbell, depending on the
ratio $\epsilon_{AB}/\epsilon$, in the strongly segregated regime).

The theory shows that the instability leading to the latter type of order
prevails if the grafting density is sufficiently high
(and/or the side chain length sufficiently long).
Both the snapshot pictures of the simulated bottle-brushes (Figs.~4,5)
and the quantitative analysis of the simulations in terms of the correlation
functions of density fluctuations $g_{nn}(\delta z,m)$ [with $m = 0,1,2$]
and concentration fluctuations (or their Fourier transforms) are compatible with such an interpretation.

We emphasize that both the simulation and the analytical theory have considered
a particularly symmetric case, where both types of side chains have equal length ($N_A = N_B$) and also their number was taken exactly the same (and moreover the same solvent quality was chosen for both types of chains, $\epsilon_{AA} = \epsilon_{BB}$).
Clearly, when one would like to discuss particular experimental systems,
these rather special conditions need to be relaxed. Also, we have restricted attention
to the case where the side chains are perfectly flexible while the backbone was assumed to be rigid.

For many real bottle-brush systems it is probably more realistic to assume backbones
that are semi-flexible rather than completely rigid, and also the local intrinsic stiffness
of the side chains may play a role. Thus, the treatment presented in our paper can be taken
as a first step towards a more complete description of less complex but interesting systems only.

\bigskip

\textbf{Acknowledgments}. I.E. thanks the Alexander von Humboldt Foundation
and the Ministry of education and science of the Russian federation for
support of this work. P.E.T. thanks the Max Planck Institute for Polymer
Research for supporting him with a Max Planck Fellowship. \newpage

\clearpage

\begin{figure}[ht]
\includegraphics[height=6.20cm]{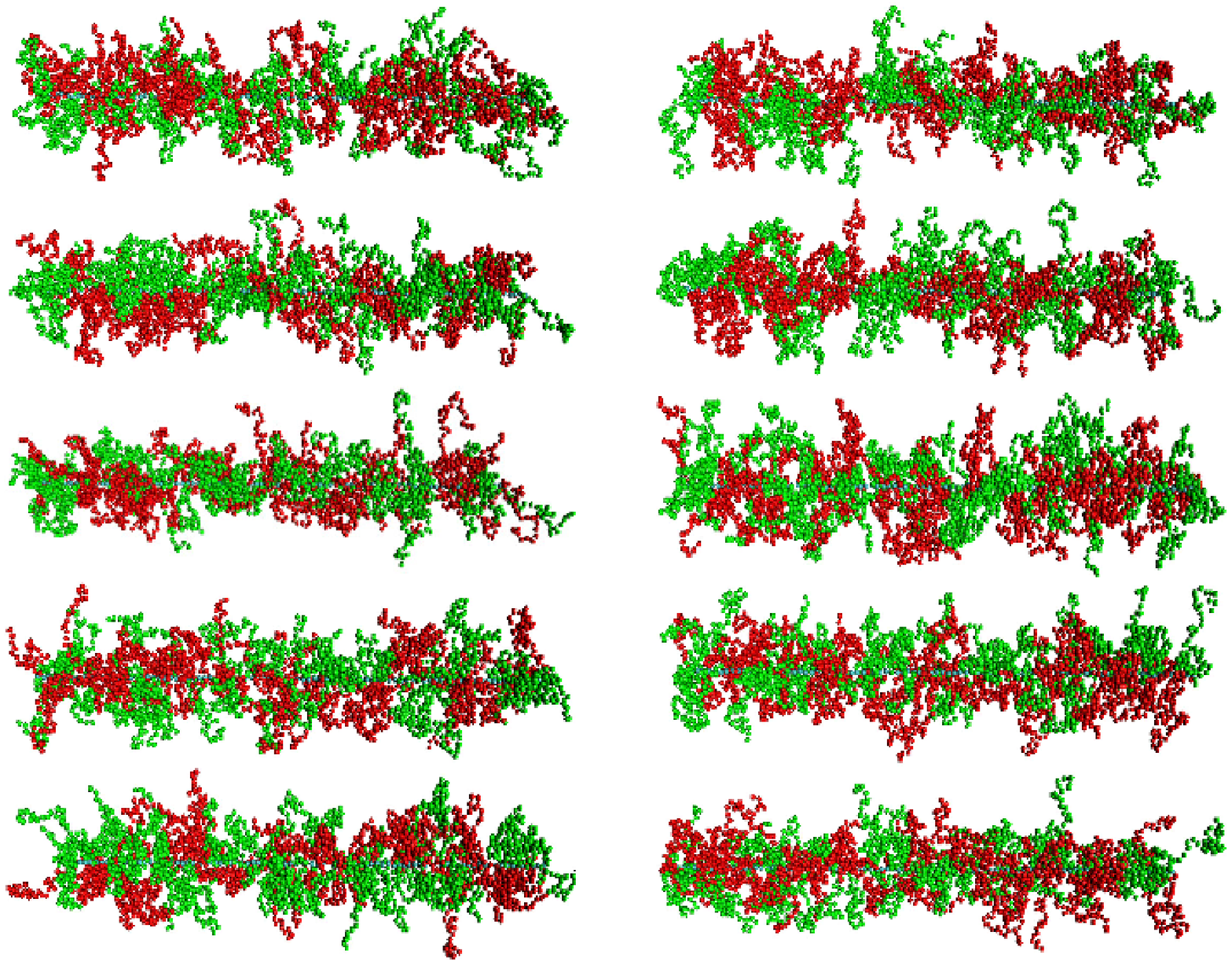}
\includegraphics[height=6.00cm]{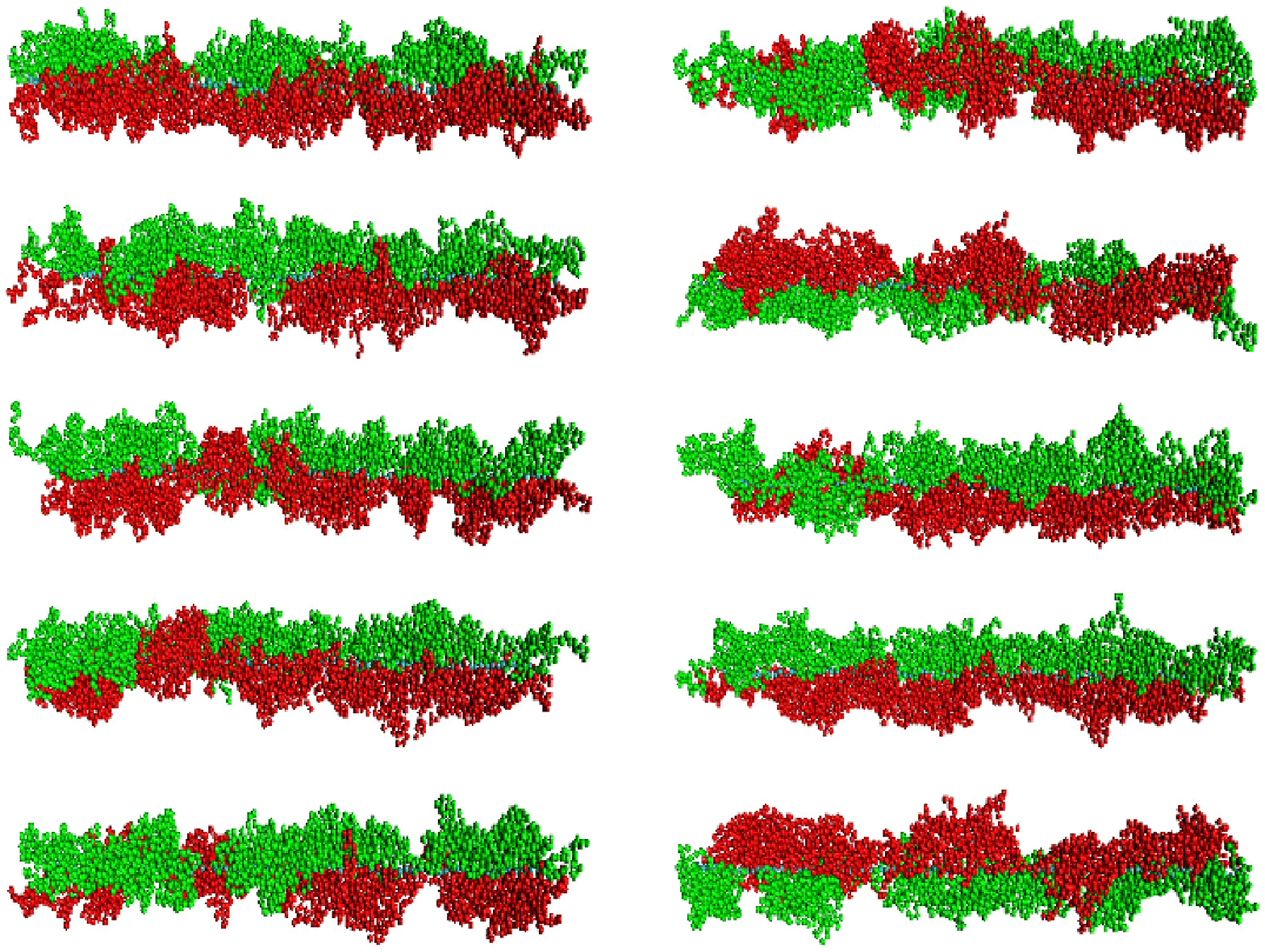}
\caption{\label{fig4} (Color online) Snapshot picture of bottle-brush polymers
at $\sigma = 1.14, N= 35, \epsilon_{AB}=3/4$ for $T=3.0$ (upper
part) and $T=1.5$ (lower part). A and B monomers are distinguished
by different color (or light grey vs. dark grey, respectively)}
\end{figure}

\begin{figure}[ht]
\mbox{
      \subfloat[]{\includegraphics[height=5.0cm]{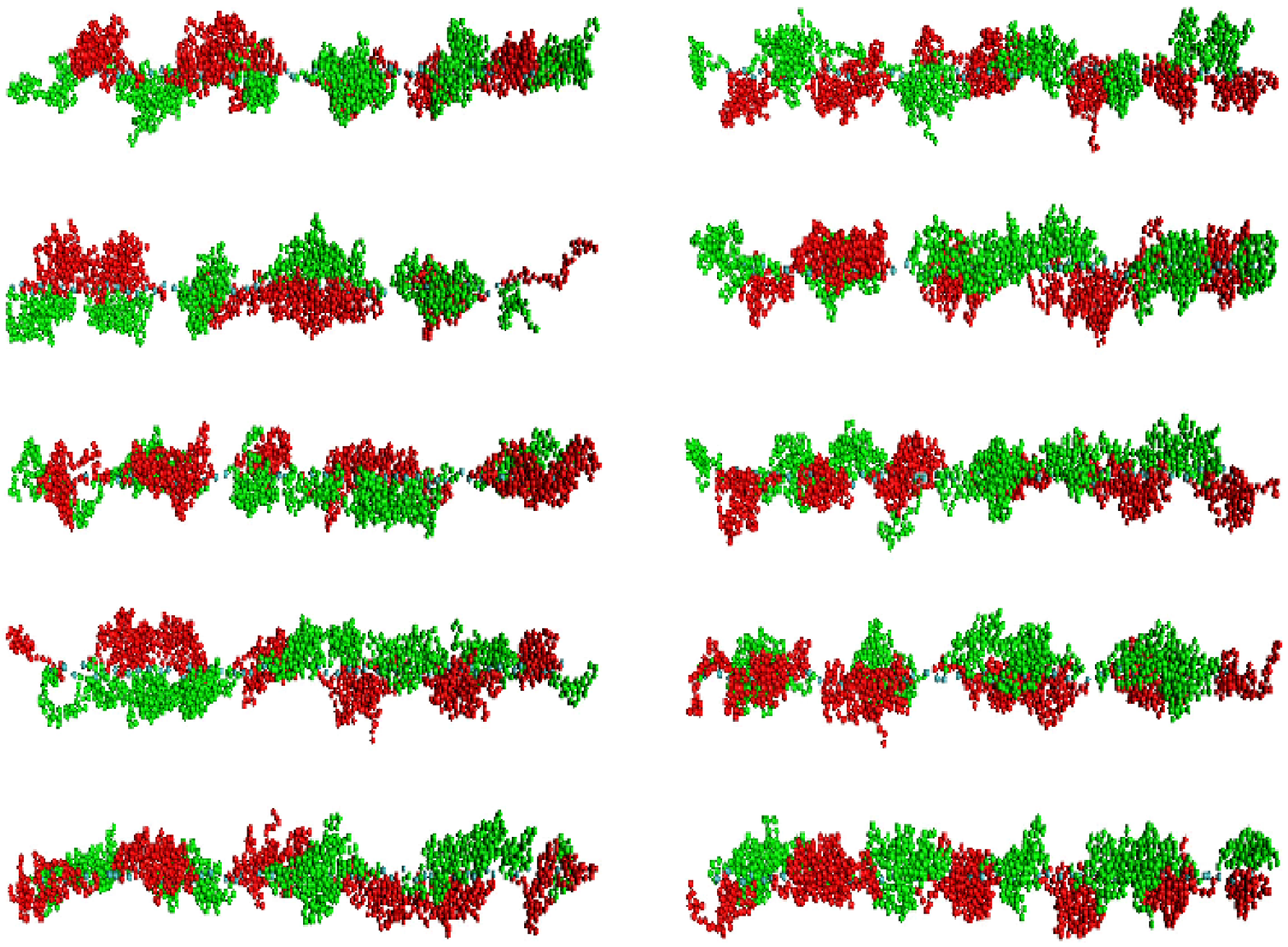}}
      }
\mbox{
      \subfloat[]{\scalebox{1.0}{\includegraphics[height=5.0cm]{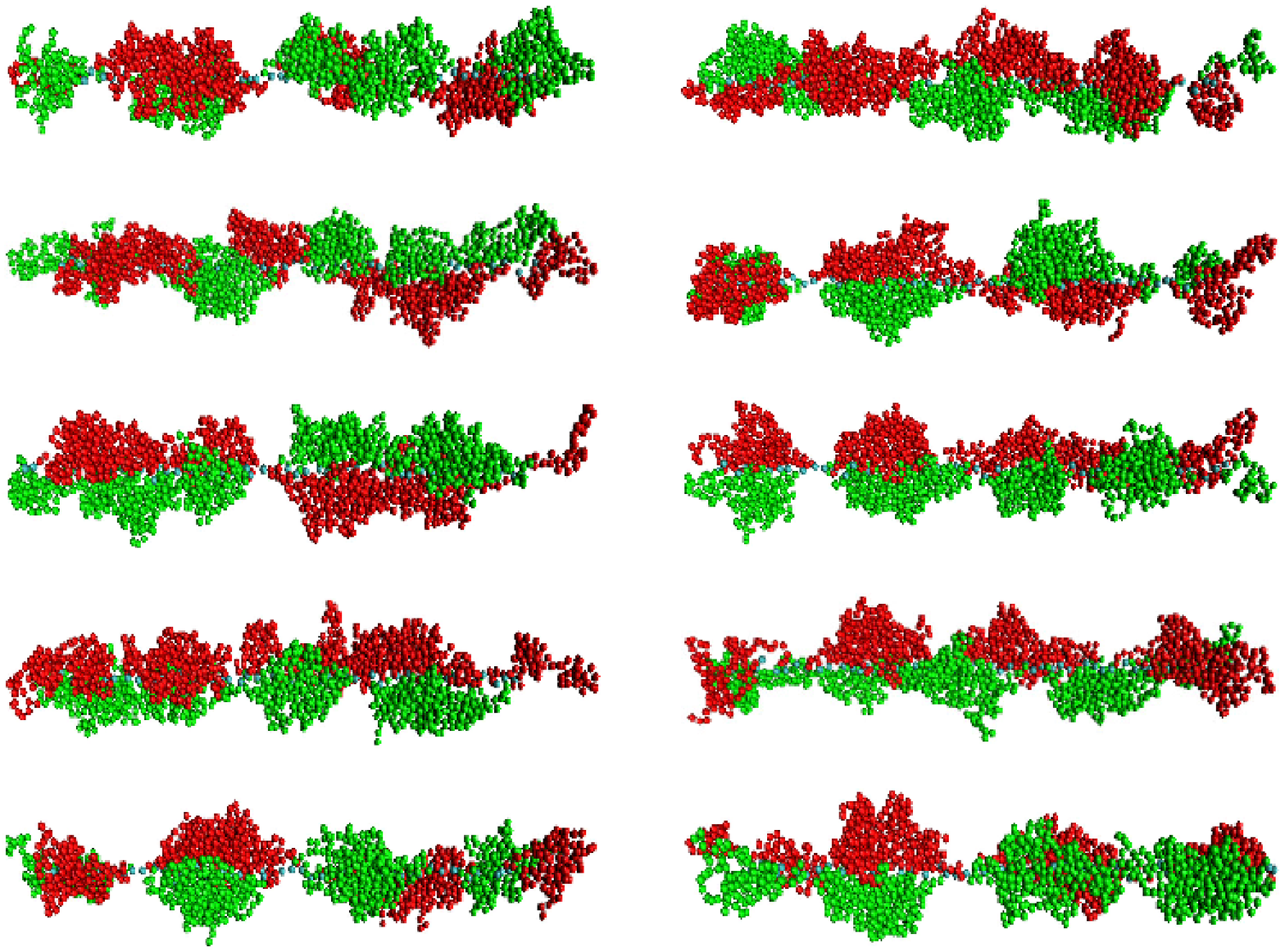}}}      }
\mbox{
      \subfloat[]{\scalebox{1.0}{\includegraphics[height=5.0cm]{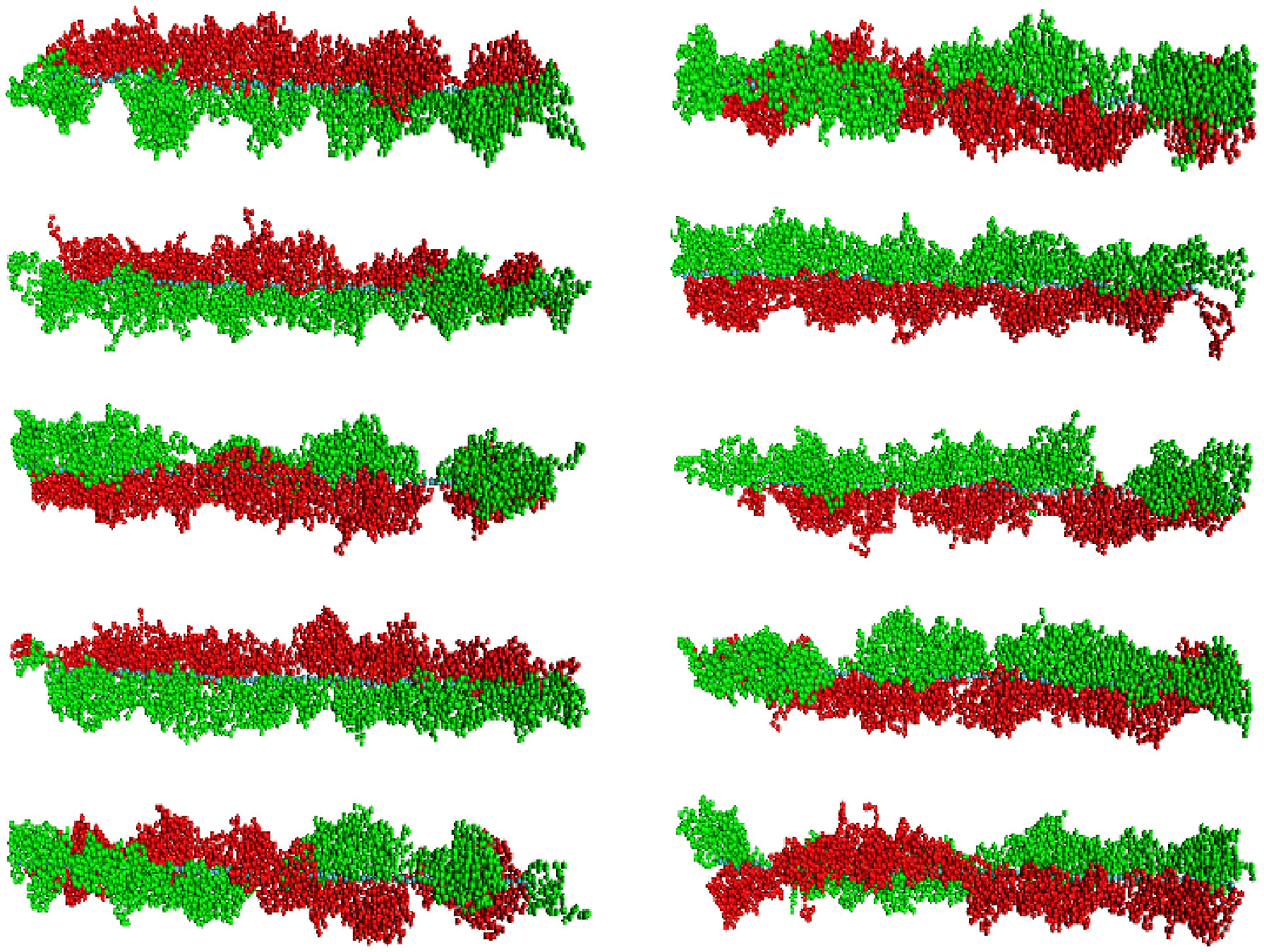}}}      }
\mbox{
      \subfloat[]{\scalebox{1.0}{\includegraphics[height=5.0cm]{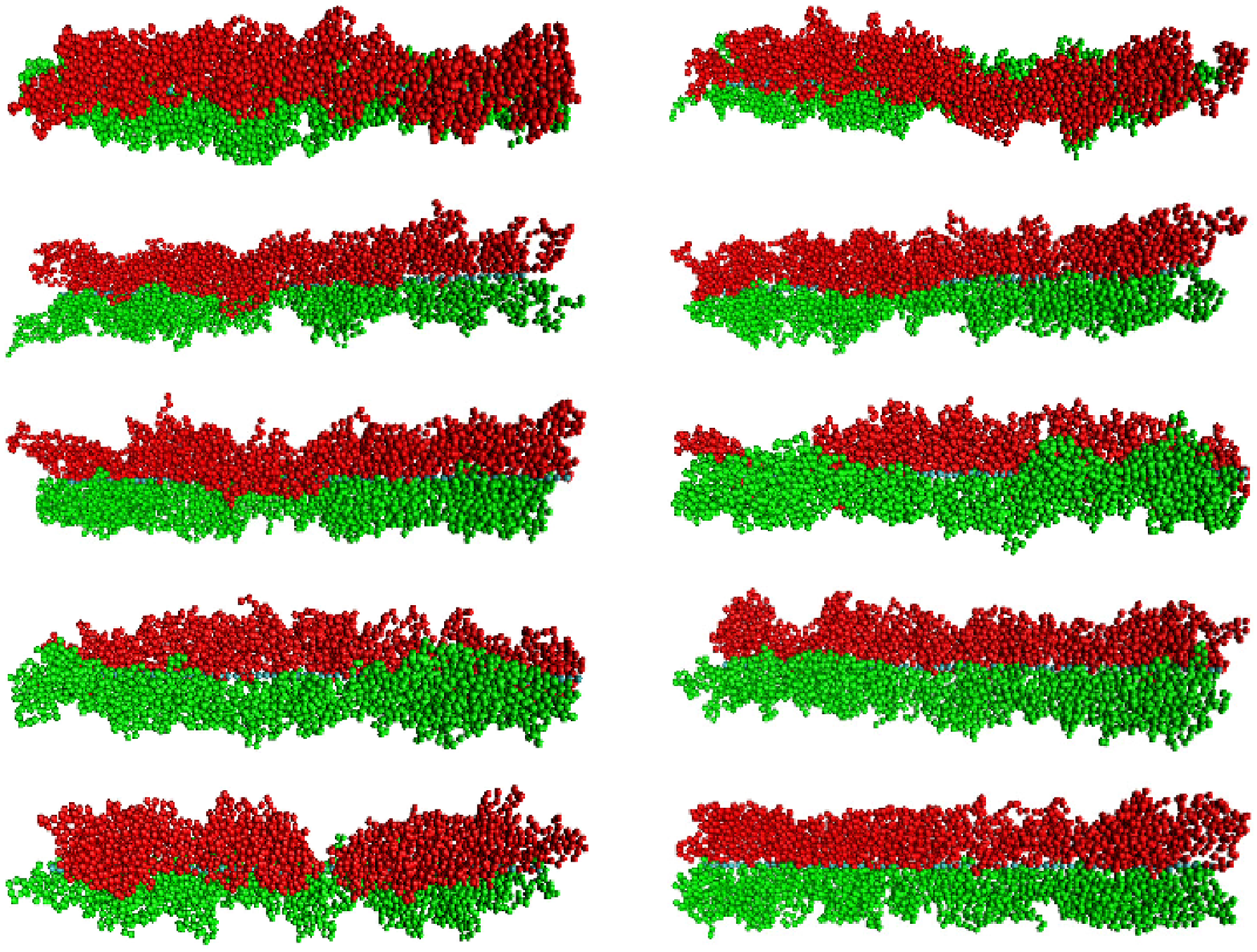}}}      }
\caption{\label{fig5} (Color online) Snapshot picture of
bottle-brush polymers at $\epsilon_{AB}=1/2$, $N=35$,$T=1.5$ and four
different values of $\sigma: \sigma =0.57$ (a), 0.76 (b), 1.14 (c)
and 1.51 (d)}
\end{figure}

\begin{figure}
\mbox{
      \subfloat[]{\scalebox{1.0}{\includegraphics[height=8.0cm,angle=270]{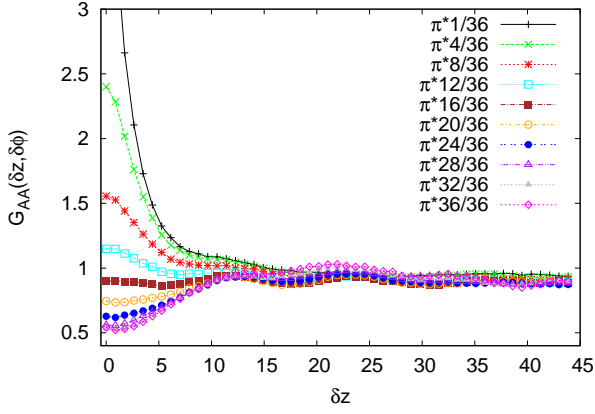}}} }
\mbox{
      \subfloat[]{\scalebox{1.0}{\includegraphics[height=8.0cm,angle=270]{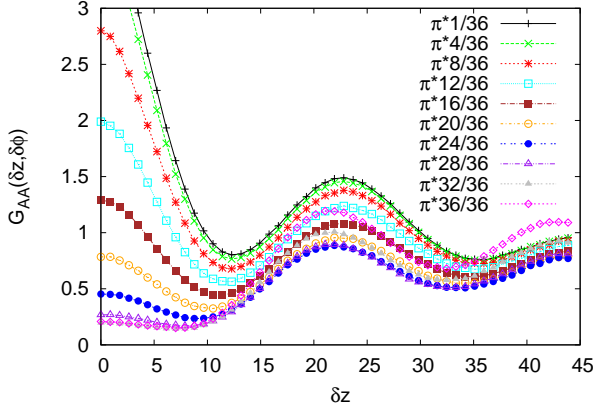}}}
      }
\mbox{
      \subfloat[]{\scalebox{1.0}{\includegraphics[height=8.0cm,angle=270]{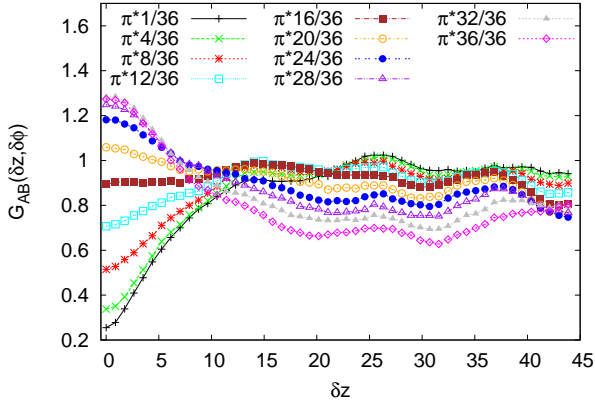}}}
      }
\caption{\label{fig6} (Color online) Correlation function
$G_{AA}(\delta z, \delta \phi)$ plotted vs. $\delta z$ for the
case $\sigma = 0.57$, N = 50, $\epsilon_{AB}=1/2$, and the
temperatures $T$ = 2.9 (a) and 1.5 (b). Several choices of the
angle $\delta \phi = k \pi/36$, $k=1,4,8,12,16,20,24,28,32$ and 36
are shown, as indicated (c). Correlation function $G_{AB}(\delta
z, \delta \phi)$ plotted vs. $\delta z$ for the case $\sigma
=0.57$, N = 50, $\epsilon_{AB}= 1/2$ and $T=2.9$ .}
\end{figure}

\begin{figure}
\mbox{
      \subfloat[]{\scalebox{1.0}{\includegraphics[height=8.0cm,angle=270]{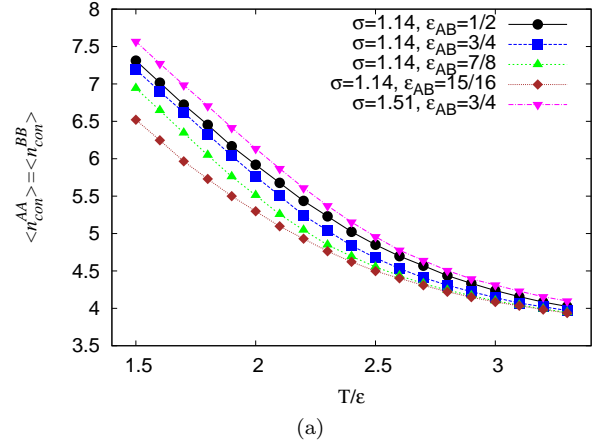}}}
      }
\mbox{
      \subfloat[]{\scalebox{1.0}{\includegraphics[height=8.0cm,angle=270]{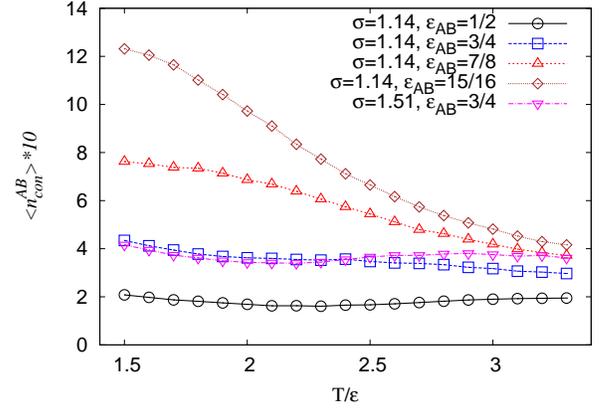}}}
      }
\caption{\label{fig7} (Color online) Number of contacts
$n_{con}^{AA}=n_{con}^{BB}$ (a) and $n_{con}^{AB}$ (b) plotted vs
temperature, for $\sigma = 1.14$ and several choices of
$\epsilon_{AB}, \epsilon_{AB} = 1/2$, $3/4$, $7/8$ and $15/16$, as
indicated. For comparison, the case $\sigma = 1.51$,
$\epsilon_{AB}=3/4$ also is included. All data refer to $N=35$ and
$2M=100$ was chosen (in order to minimize artefacts due to the
periodic boundary conditions). Note that in part (b) the numbers
of contacts are multiplied by a factor 10 throughout.}
\end{figure}

\begin{figure}
\mbox{
      \subfloat[]{\scalebox{1.0}{\includegraphics[height=8.0cm,angle=270]{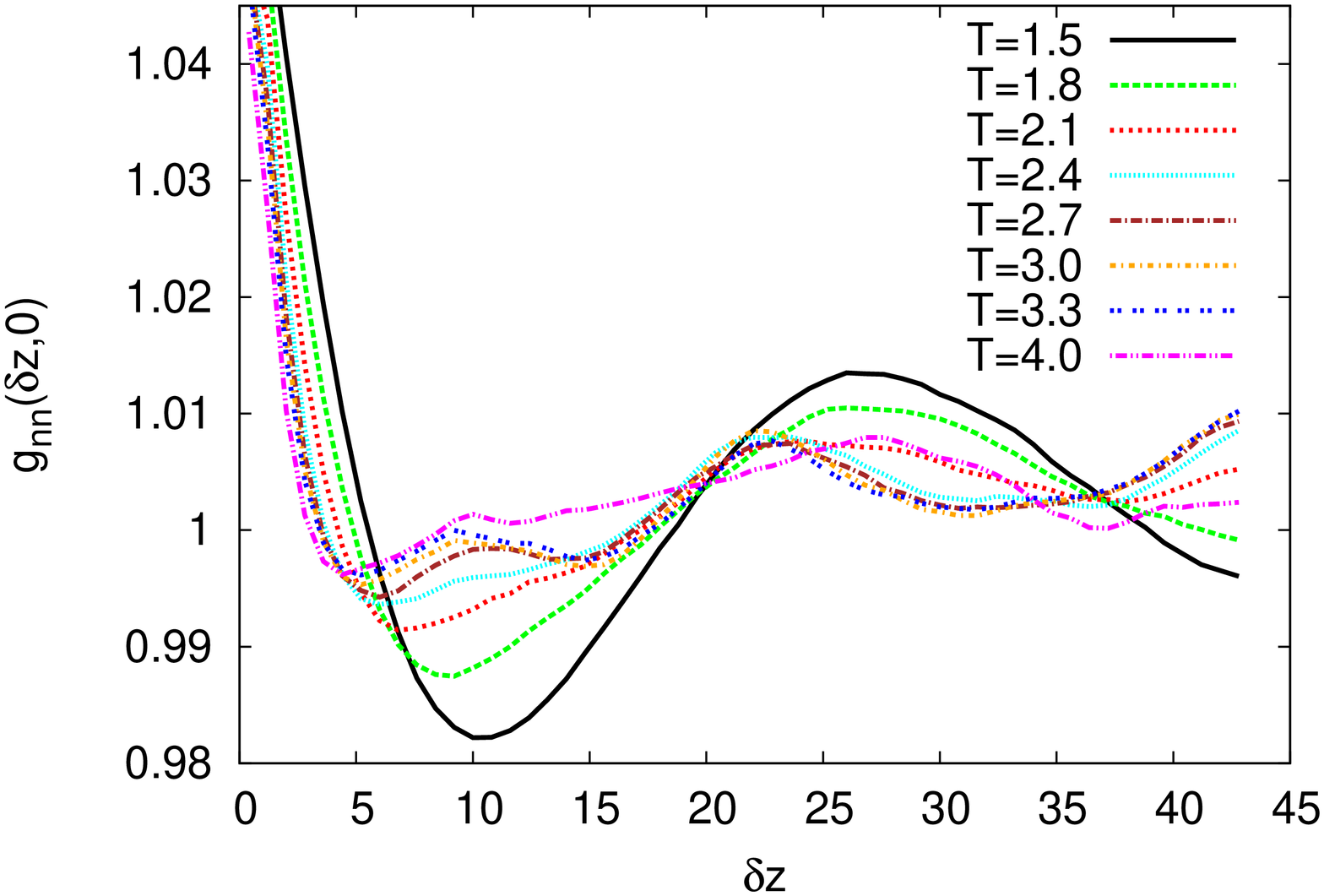}}}
      }
\mbox{
      \subfloat[]{\scalebox{1.0}{\includegraphics[height=8.0cm,angle=270]{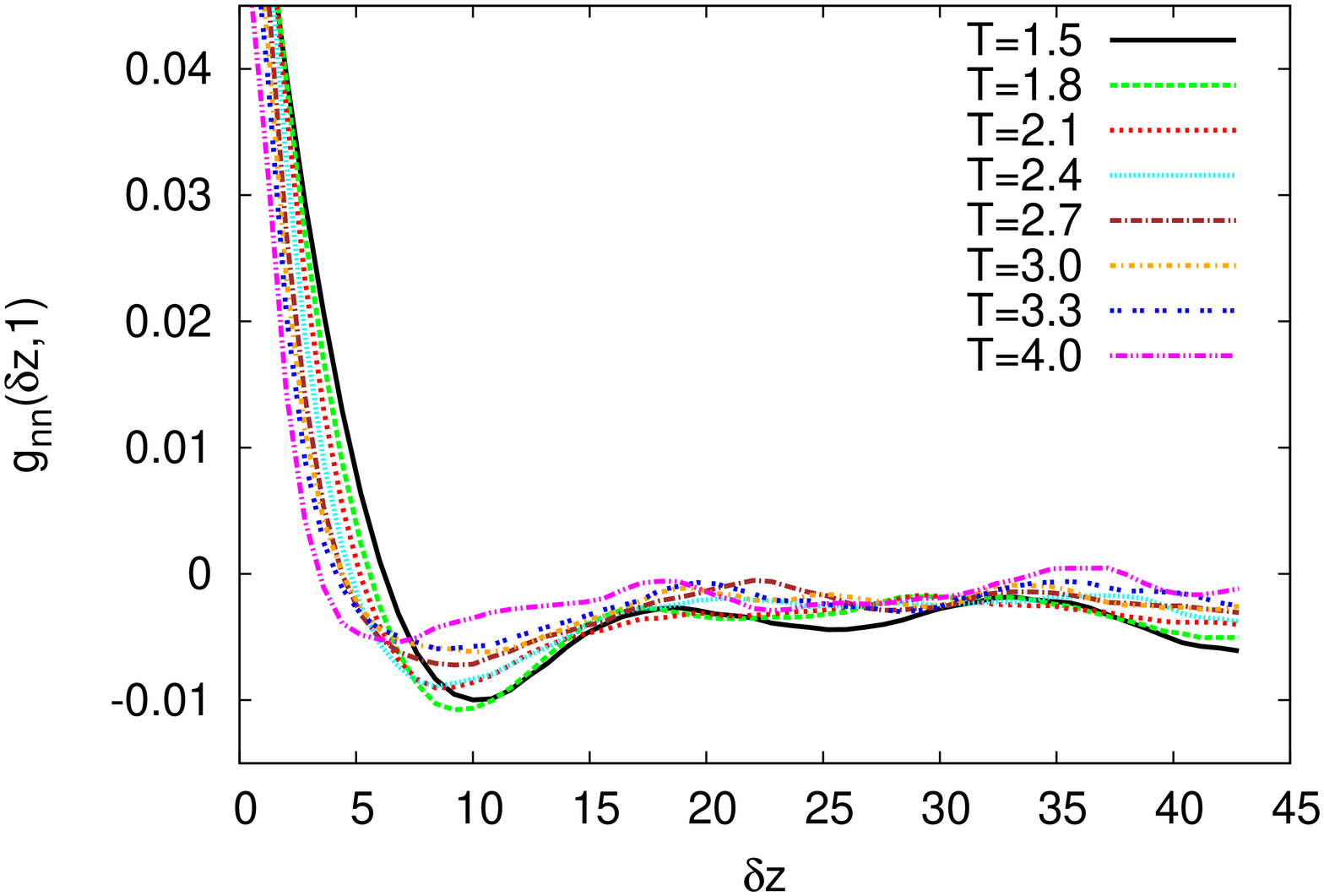}}}
      }
\mbox{
      \subfloat[]{\scalebox{1.0}{\includegraphics[height=8.0cm,angle=270]{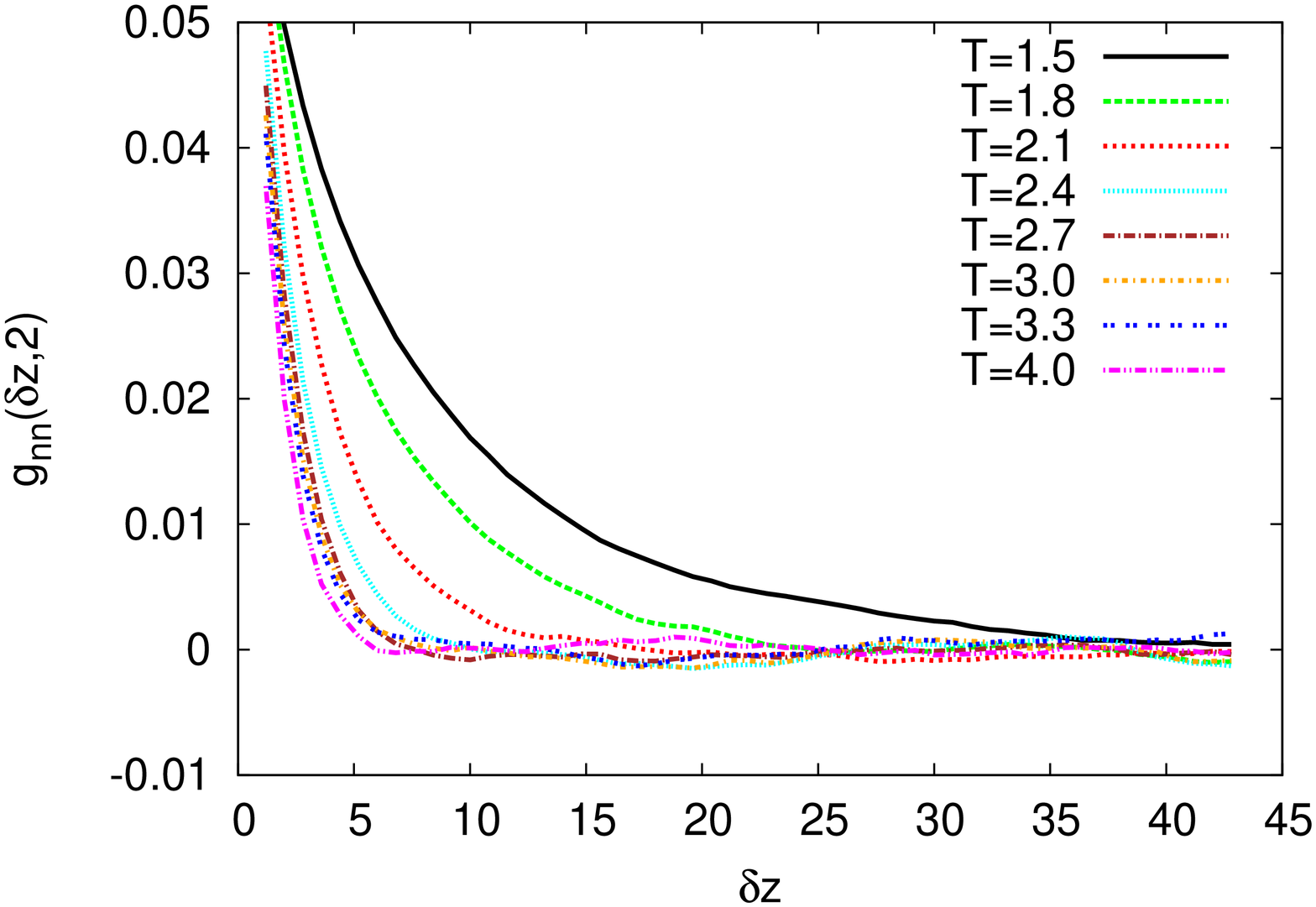}}}
      }
\caption{\label{fig8} (Color online) Correlation functions
$g_{nn}(\delta z,m)$ relating to density fluctuations plotted vs.
$\delta z$ and the case $\epsilon_{AB}=3/4$, N = 35, $\sigma =
1.14$ for $m=0$ (a), $m=1$ (b) and $m=2$ (c). Eight temperatures
$T=1.5,1.8,2.1,2.4,2.7, 3.0, 3.3$ and $4.0$ are included, as indicated.}
\end{figure}

\begin{figure}
\mbox{
      \subfloat[]{\scalebox{1.0}{\includegraphics[height=8.0cm,angle=270]{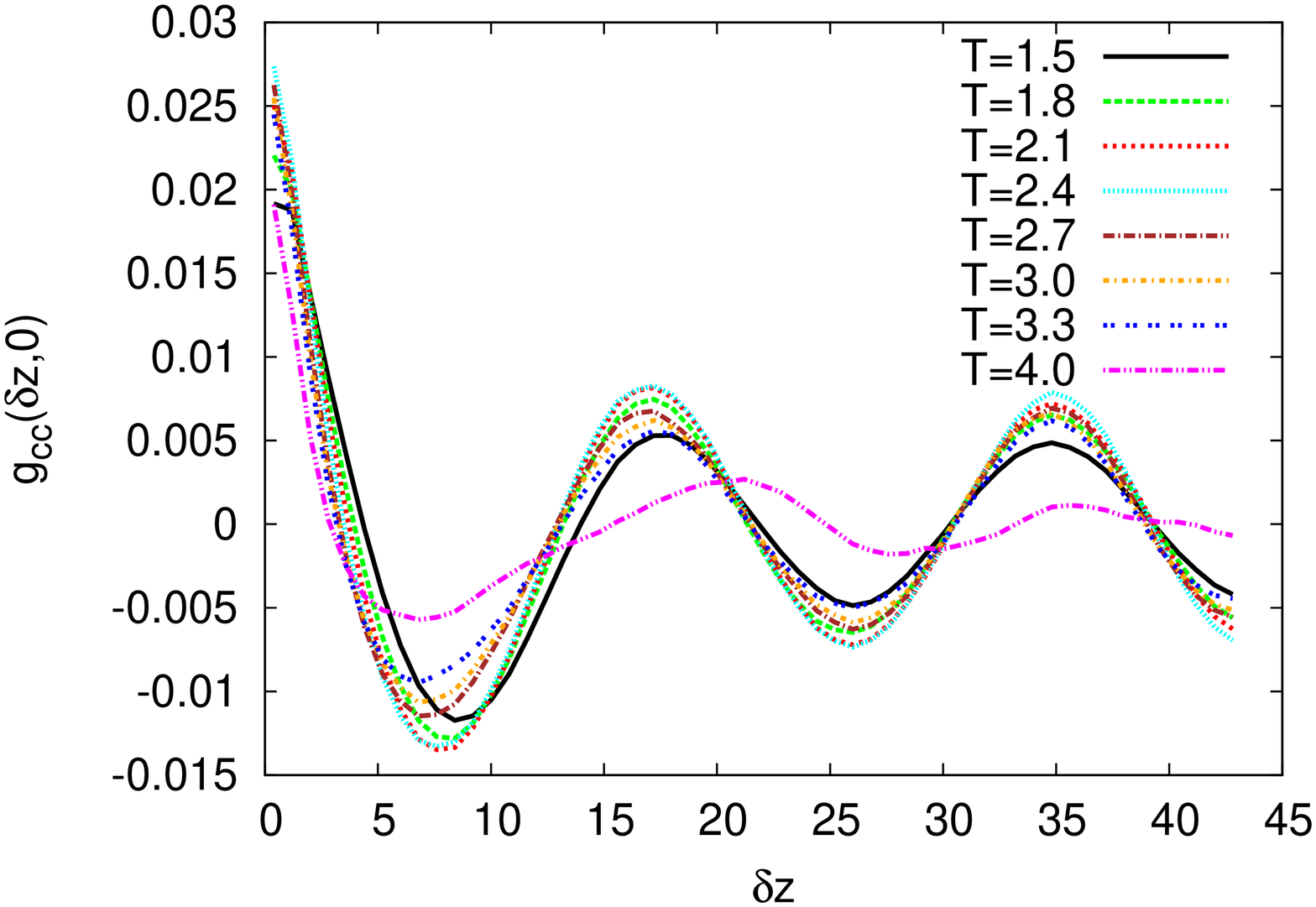}}}
      }
\mbox{
      \subfloat[]{\scalebox{1.0}{\includegraphics[height=8.0cm,angle=270]{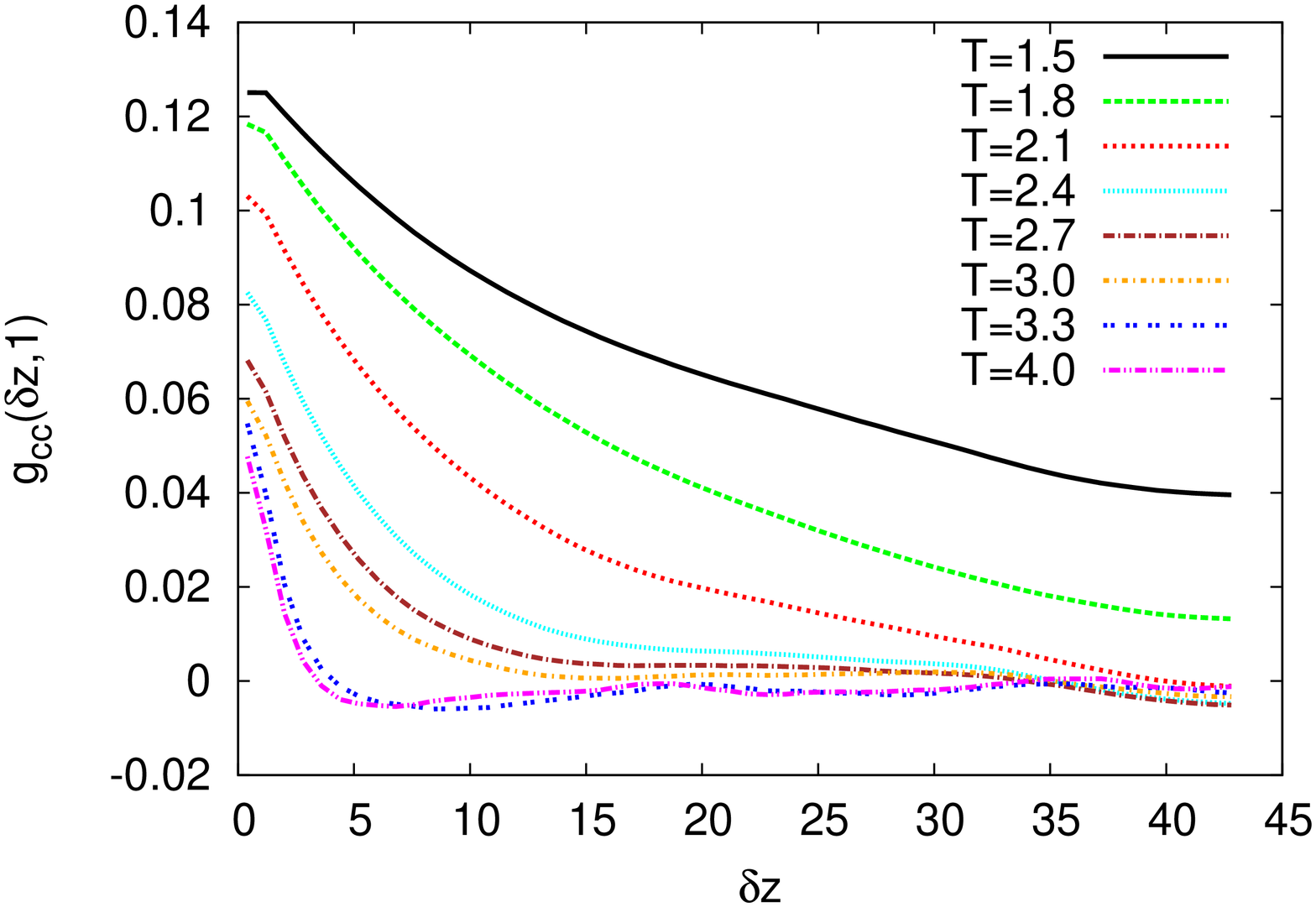}}}
      }
\mbox{
      \subfloat[]{\scalebox{1.0}{\includegraphics[height=8.0cm,angle=270]{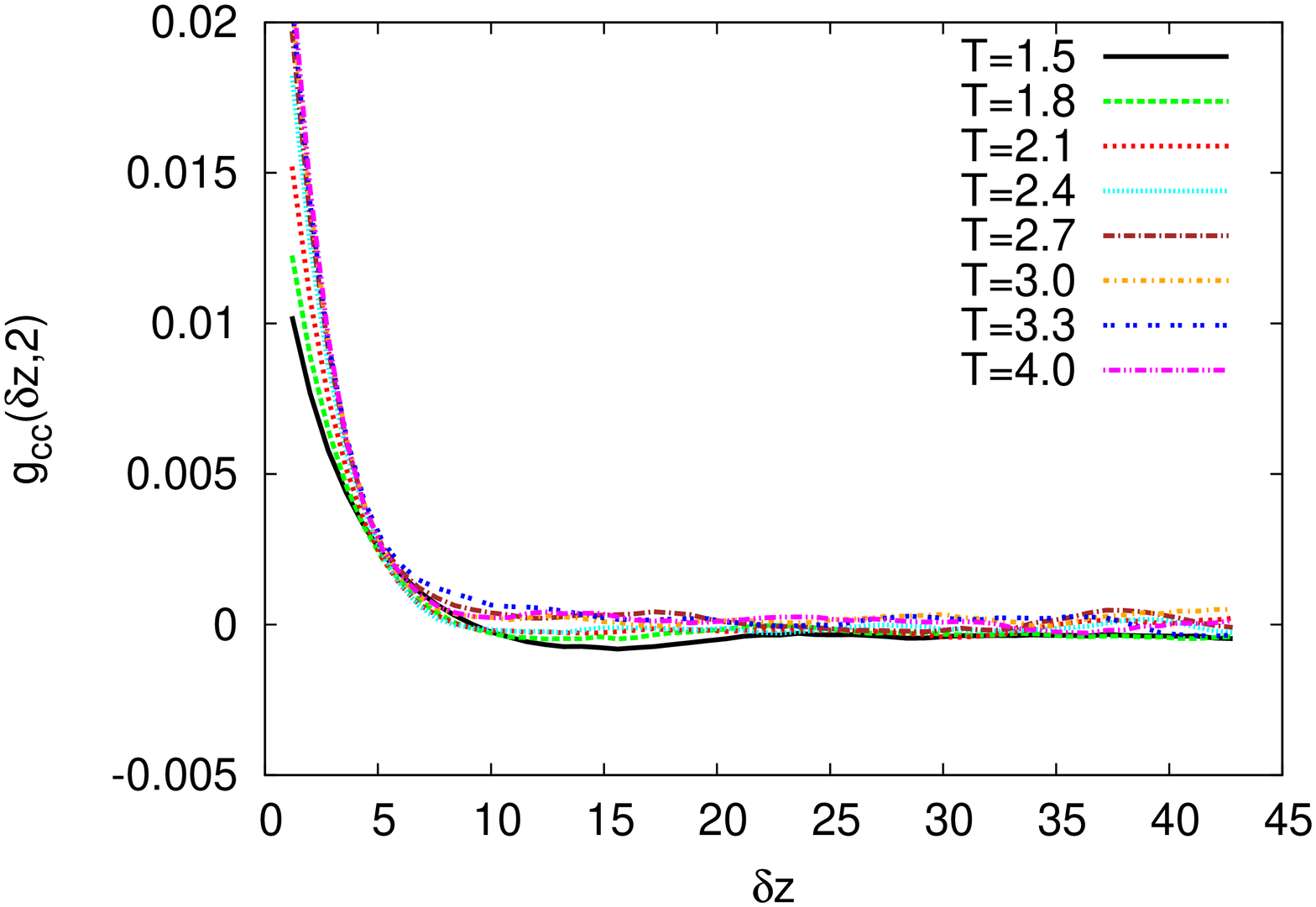}}}
      }
\caption{\label{fig9} (Color online) Same as Fig.~8, but for the correlation function $g_{cc}(\delta z,m)$ relating to concentration fluctuations.}
\end{figure}

\begin{figure}
\mbox{
      \subfloat[]{\scalebox{1.0}{\includegraphics[height=8.0cm,angle=270]{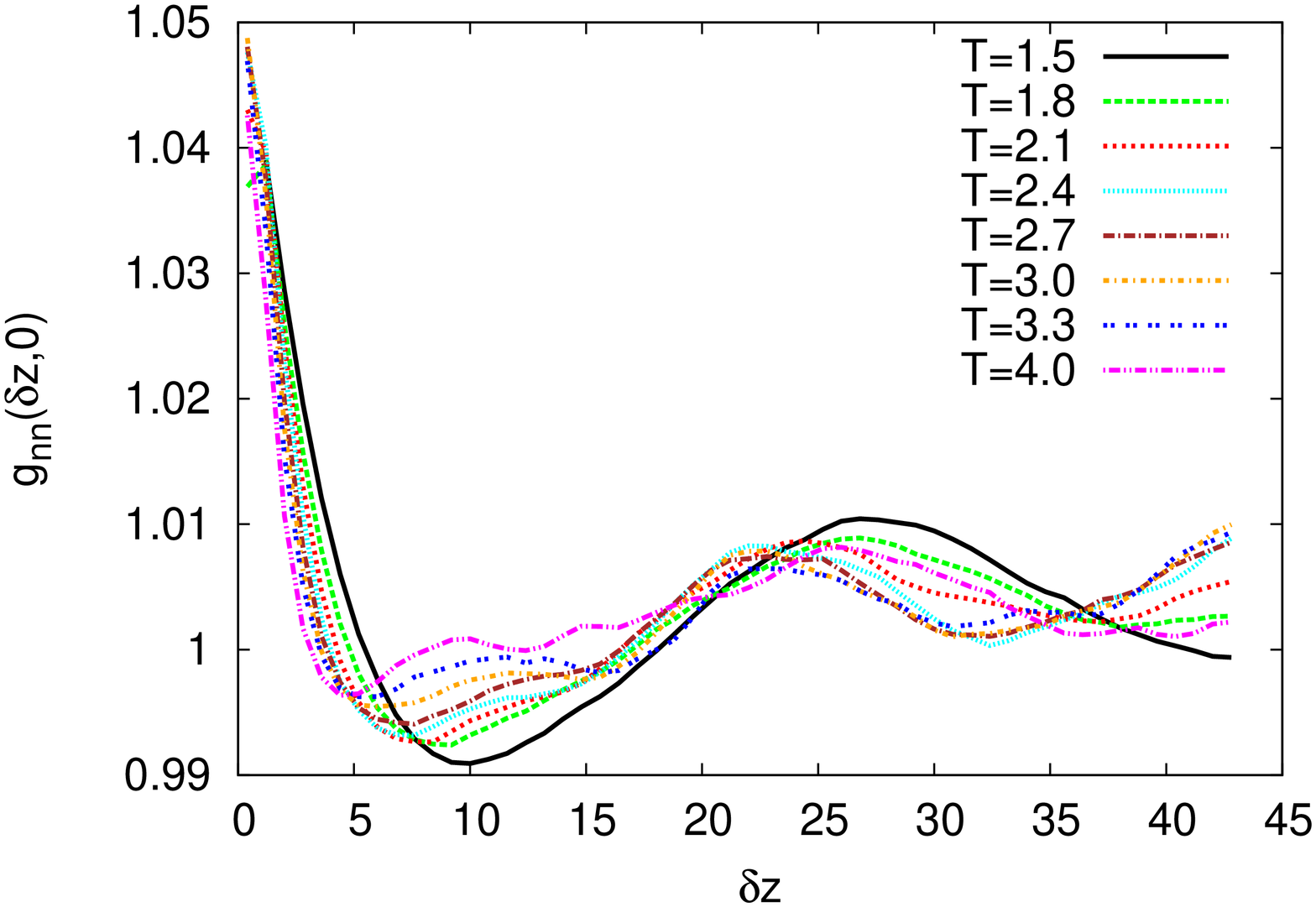}}}
      }
\mbox{
      \subfloat[]{\scalebox{1.0}{\includegraphics[height=8.0cm,angle=270]{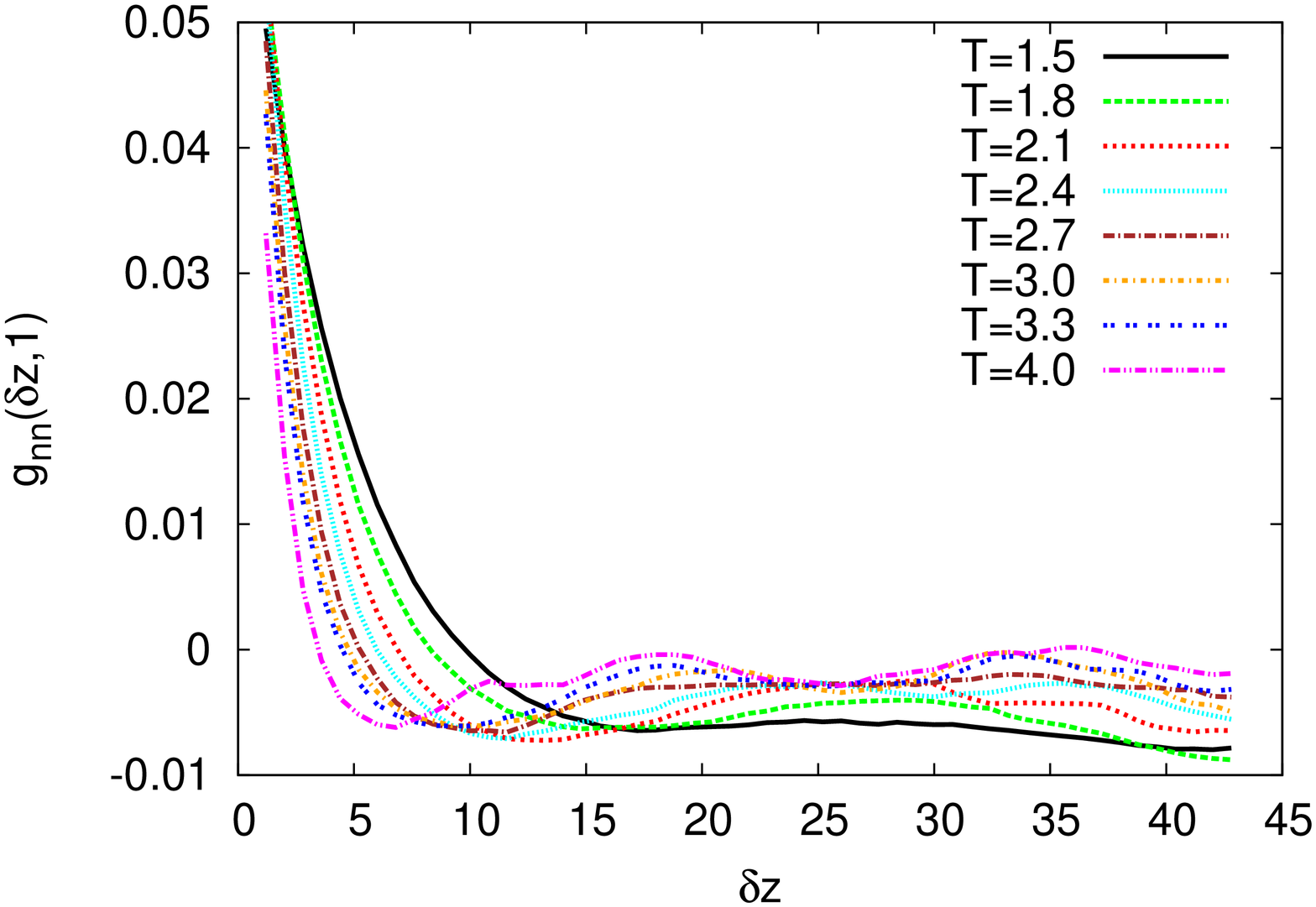}}}
      }
\mbox{
      \subfloat[]{\scalebox{1.0}{\includegraphics[height=8.0cm,angle=270]{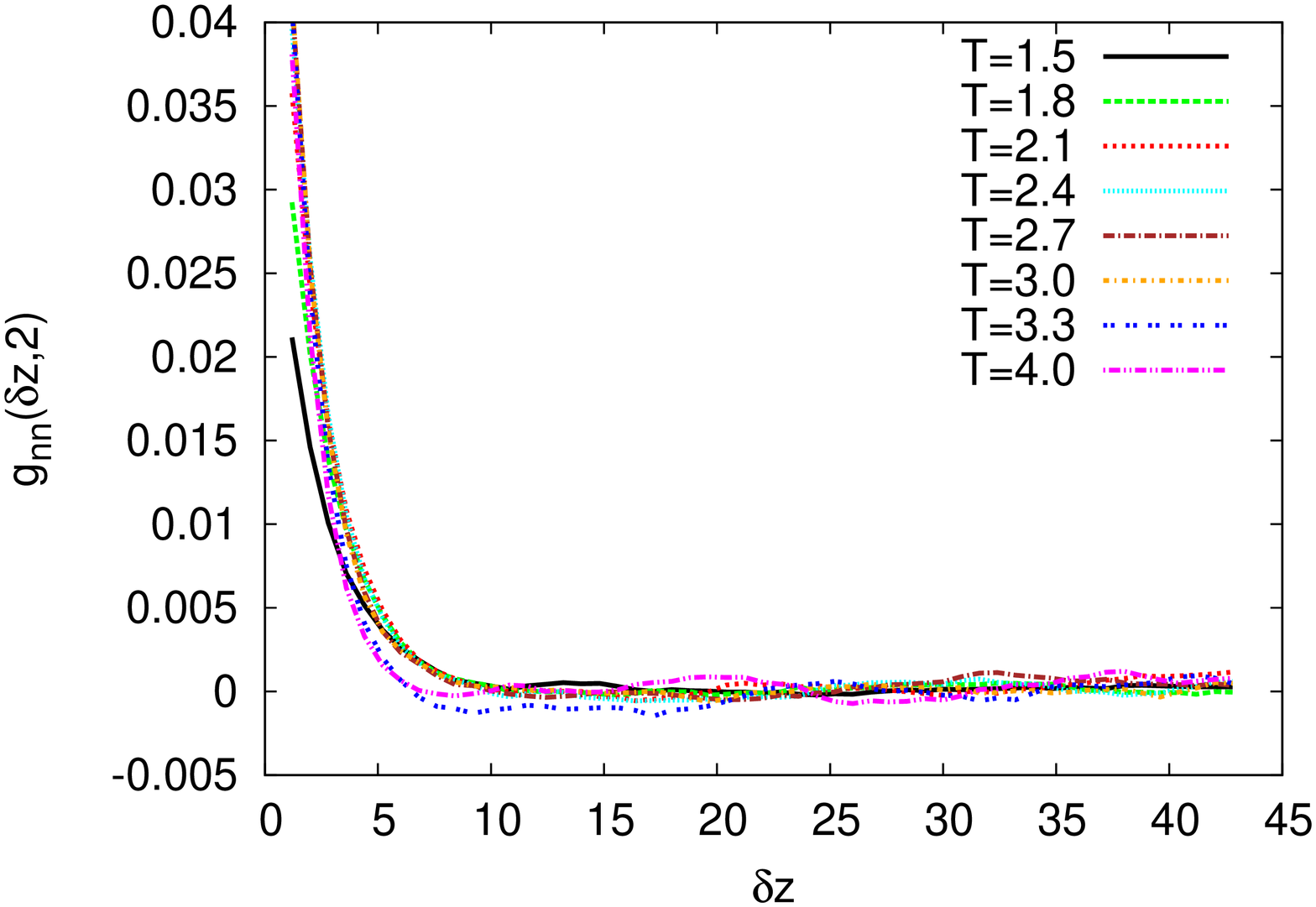}}}
      }
\caption{\label{fig10} (Color online) Same as Fig.~8, but for the case $\epsilon_{AB}=15/16$.}
\end{figure}

\begin{figure}
\mbox{
      \subfloat[]{\scalebox{1.0}{\includegraphics[height=8.0cm,angle=270]{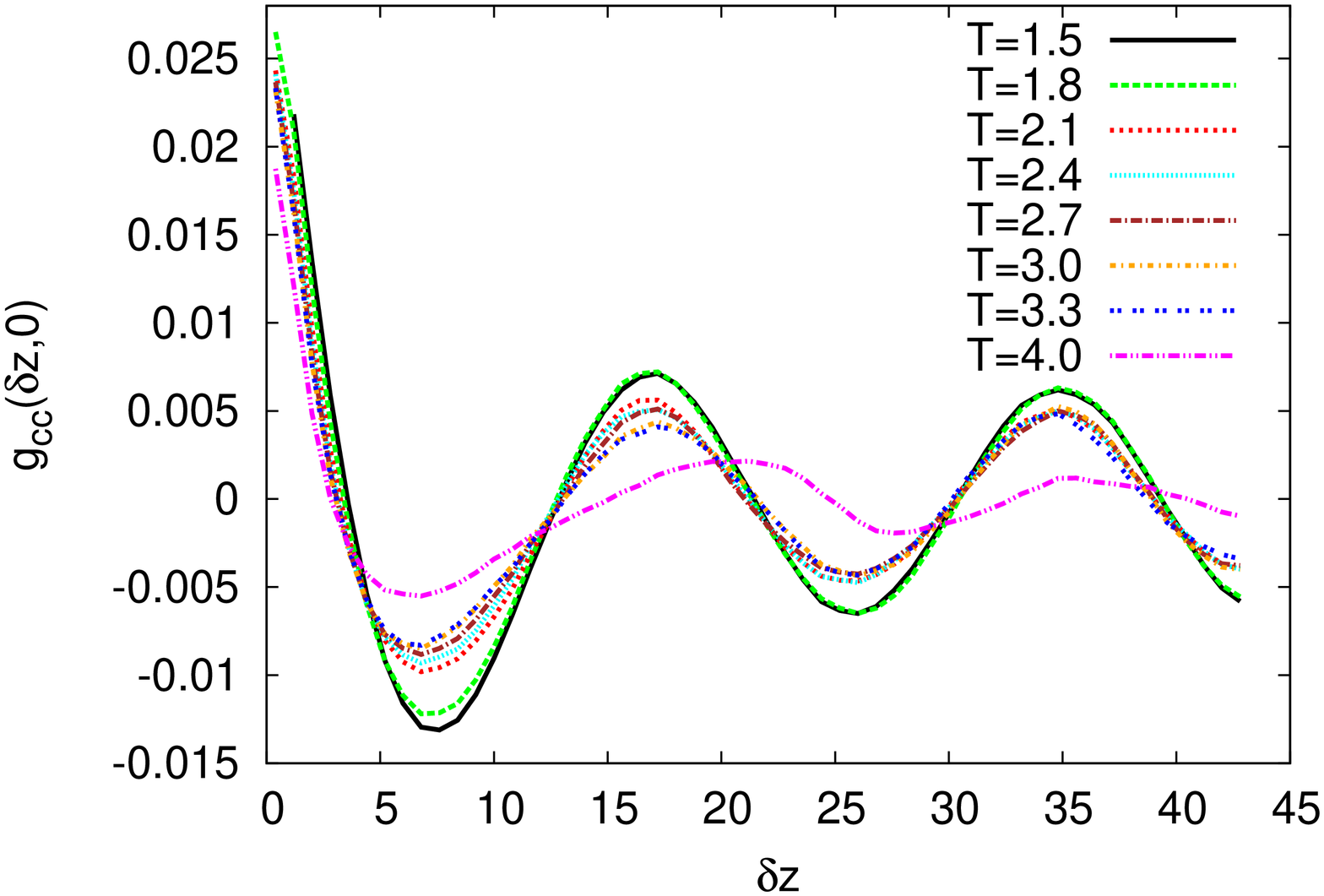}}}
      }
\mbox{
      \subfloat[]{\scalebox{1.0}{\includegraphics[height=8.0cm,angle=270]{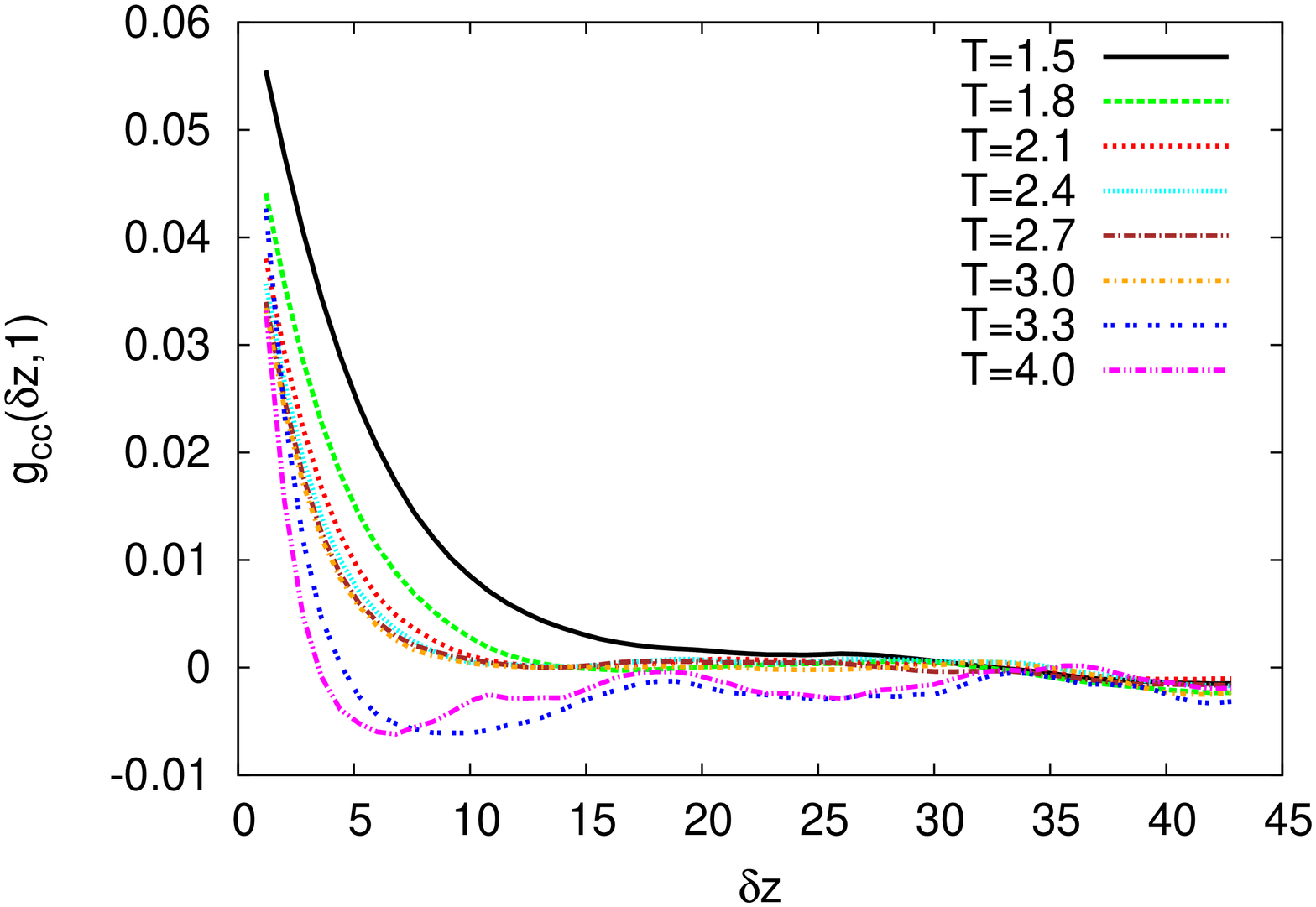}}}
      }
\mbox{
      \subfloat[]{\scalebox{1.0}{\includegraphics[height=8.0cm,angle=270]{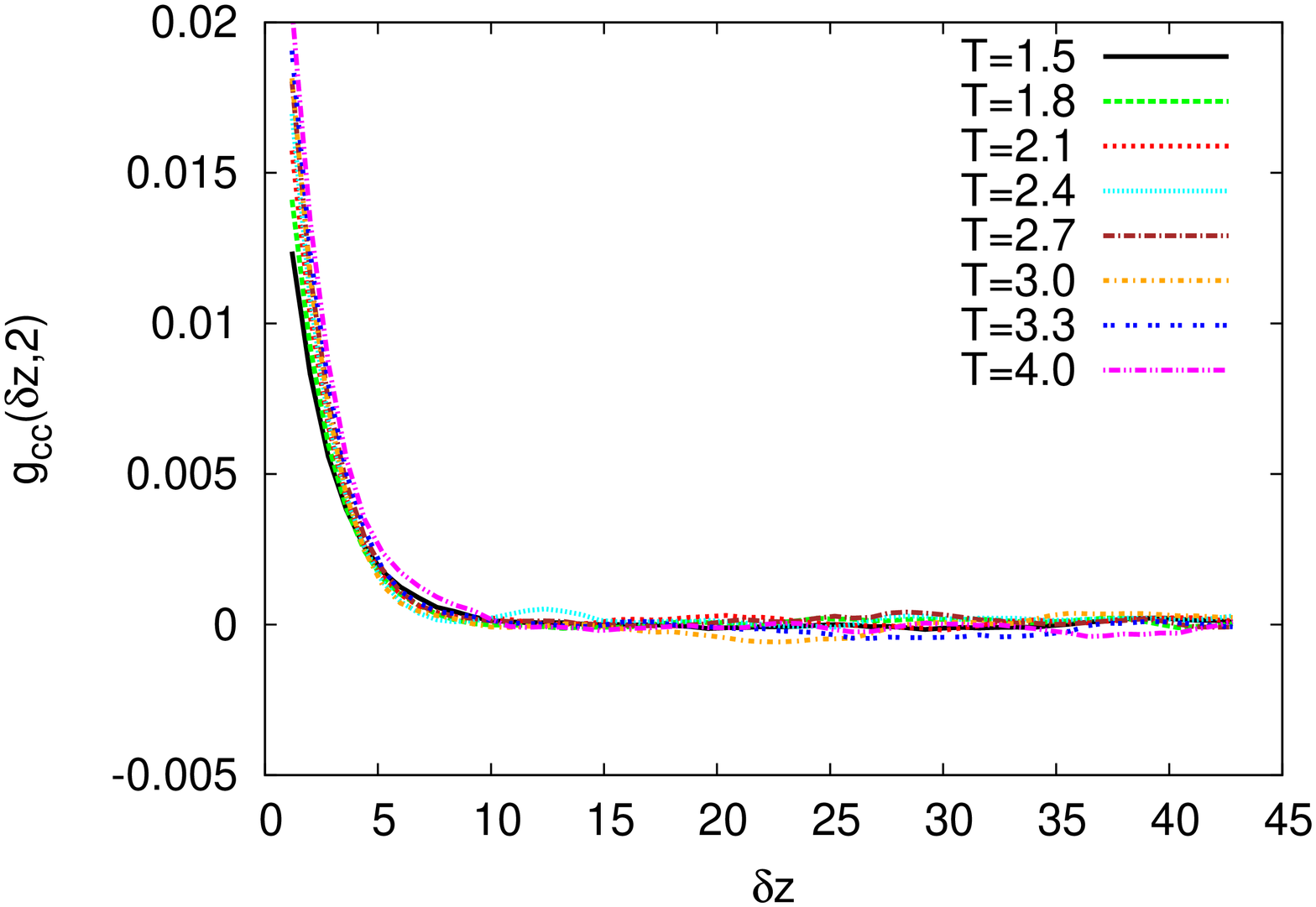}}}
      }
\caption{\label{fig11} (Color online) Same as Fig.~9, but for the case $\epsilon_{AB}=15/16$.}
\end{figure}

\begin{figure}
\mbox{
      \subfloat[]{\scalebox{1.0}{\includegraphics[height=8.0cm,angle=270]{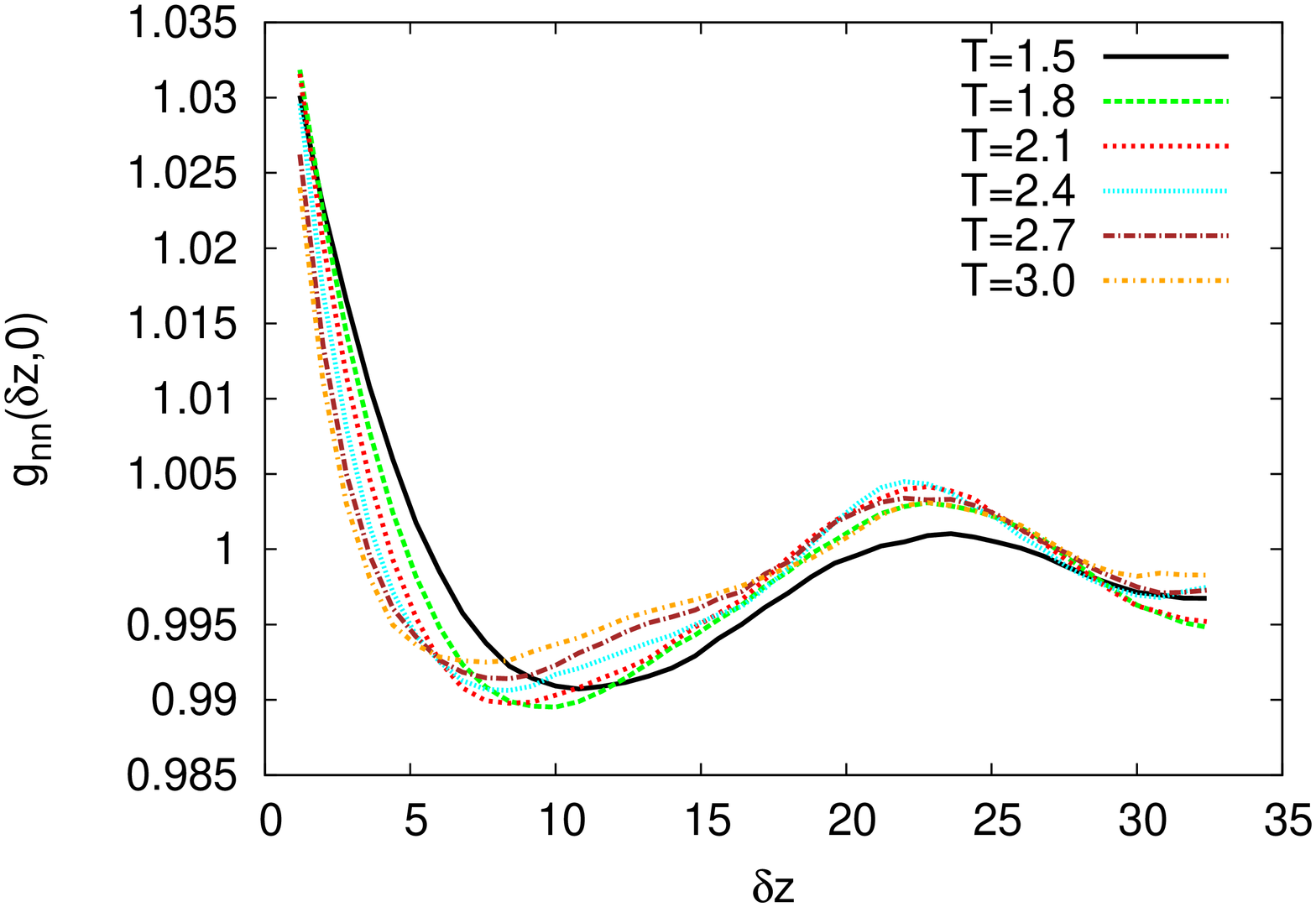}}}
      }
\mbox{
      \subfloat[]{\scalebox{1.0}{\includegraphics[height=8.0cm,angle=270]{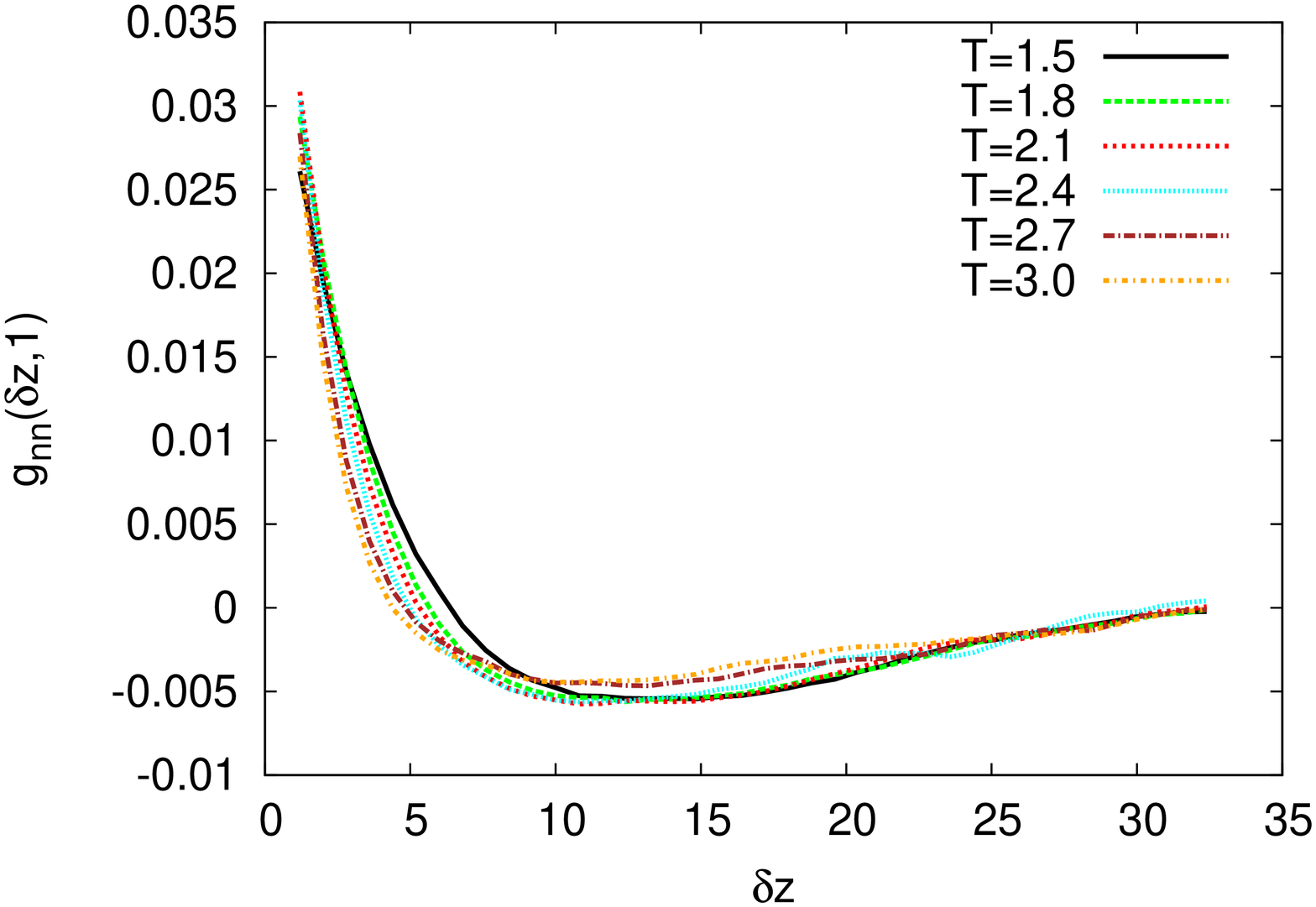}}}
      }
\mbox{
      \subfloat[]{\scalebox{1.0}{\includegraphics[height=8.0cm,angle=270]{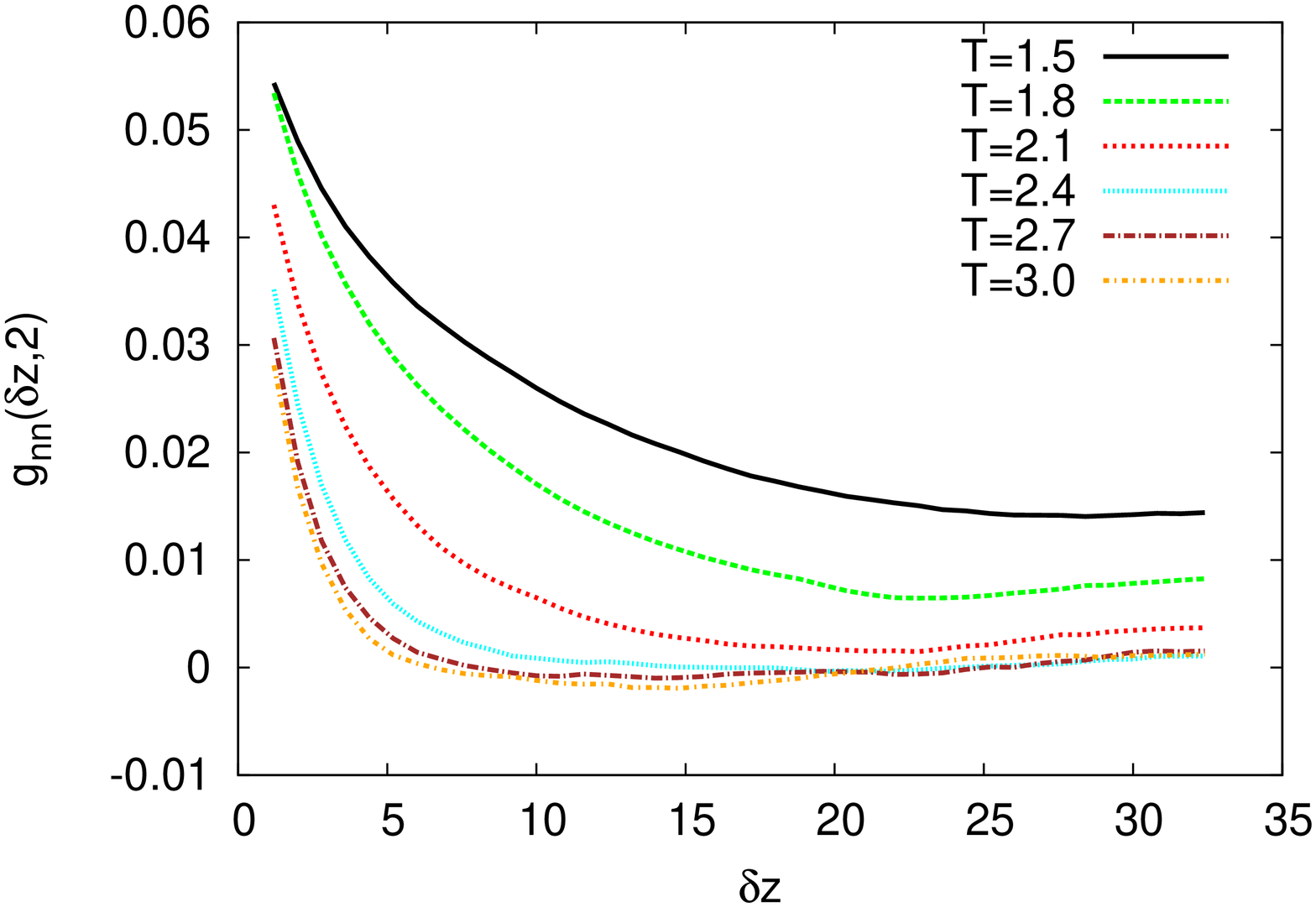}}}
      }
\caption{\label{fig12} (Color online) Same as Fig.~8, but for the case $\epsilon_{AB}=3/4$, $N=35, \sigma = 1.51$. Six temperatures
from $T=1.5$ to $T=3.0$ are shown, as indicated.}
\end{figure}

\begin{figure}
\mbox{
      \subfloat[]{\scalebox{1.0}{\includegraphics[height=8.0cm,angle=270]{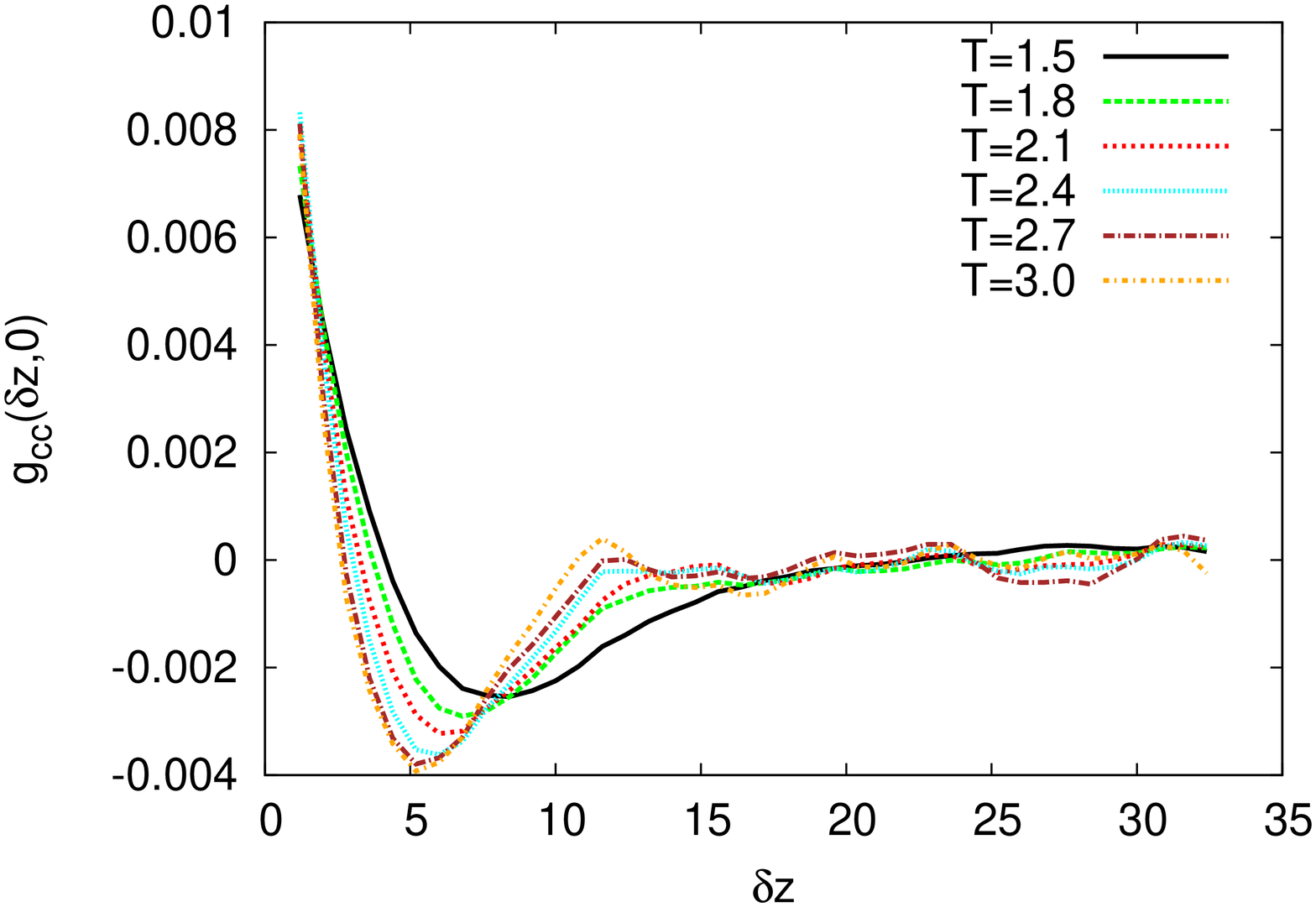}}}
      }
\mbox{
      \subfloat[]{\scalebox{1.0}{\includegraphics[height=8.0cm,angle=270]{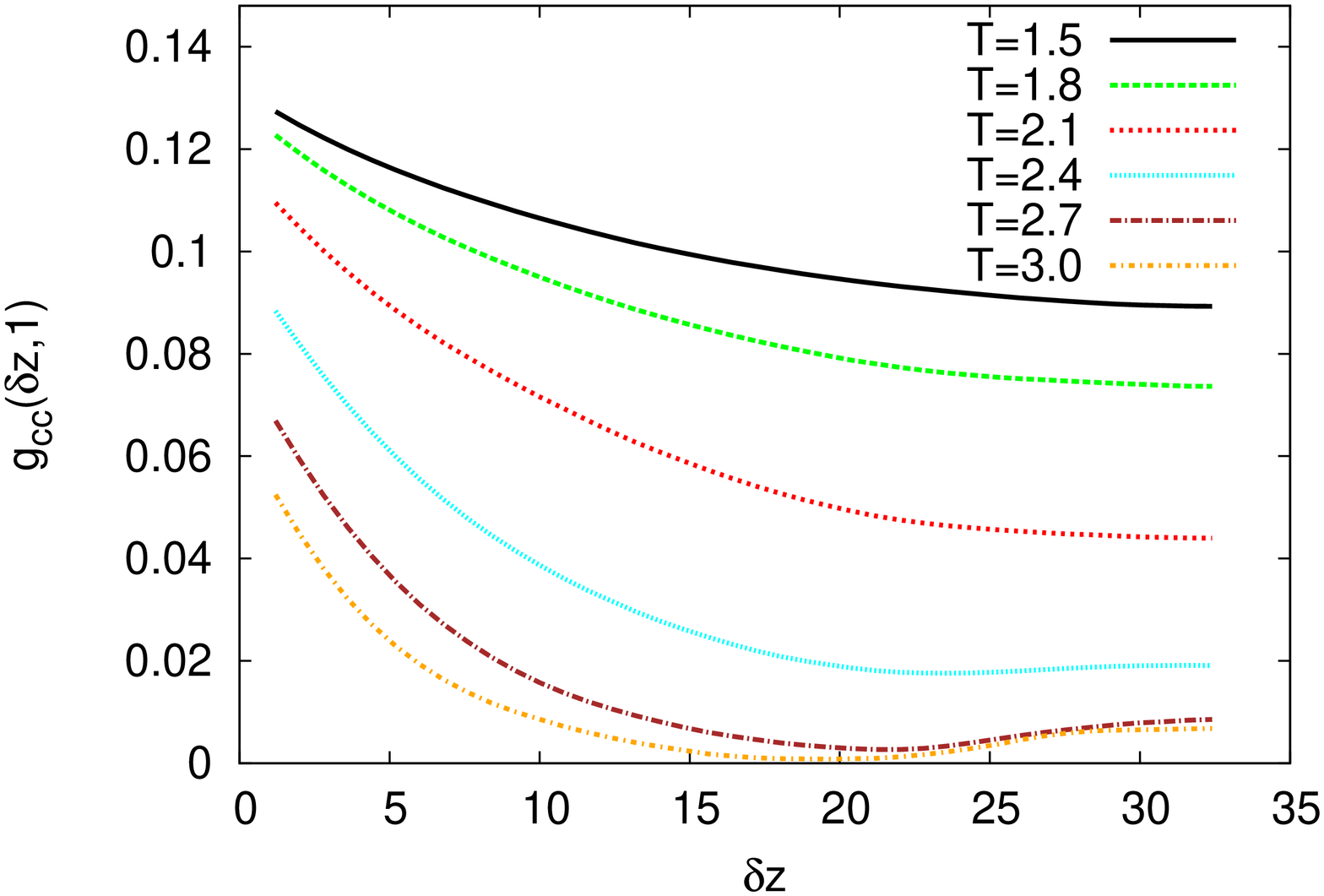}}}
      }
\mbox{
      \subfloat[]{\scalebox{1.0}{\includegraphics[height=8.0cm,angle=270]{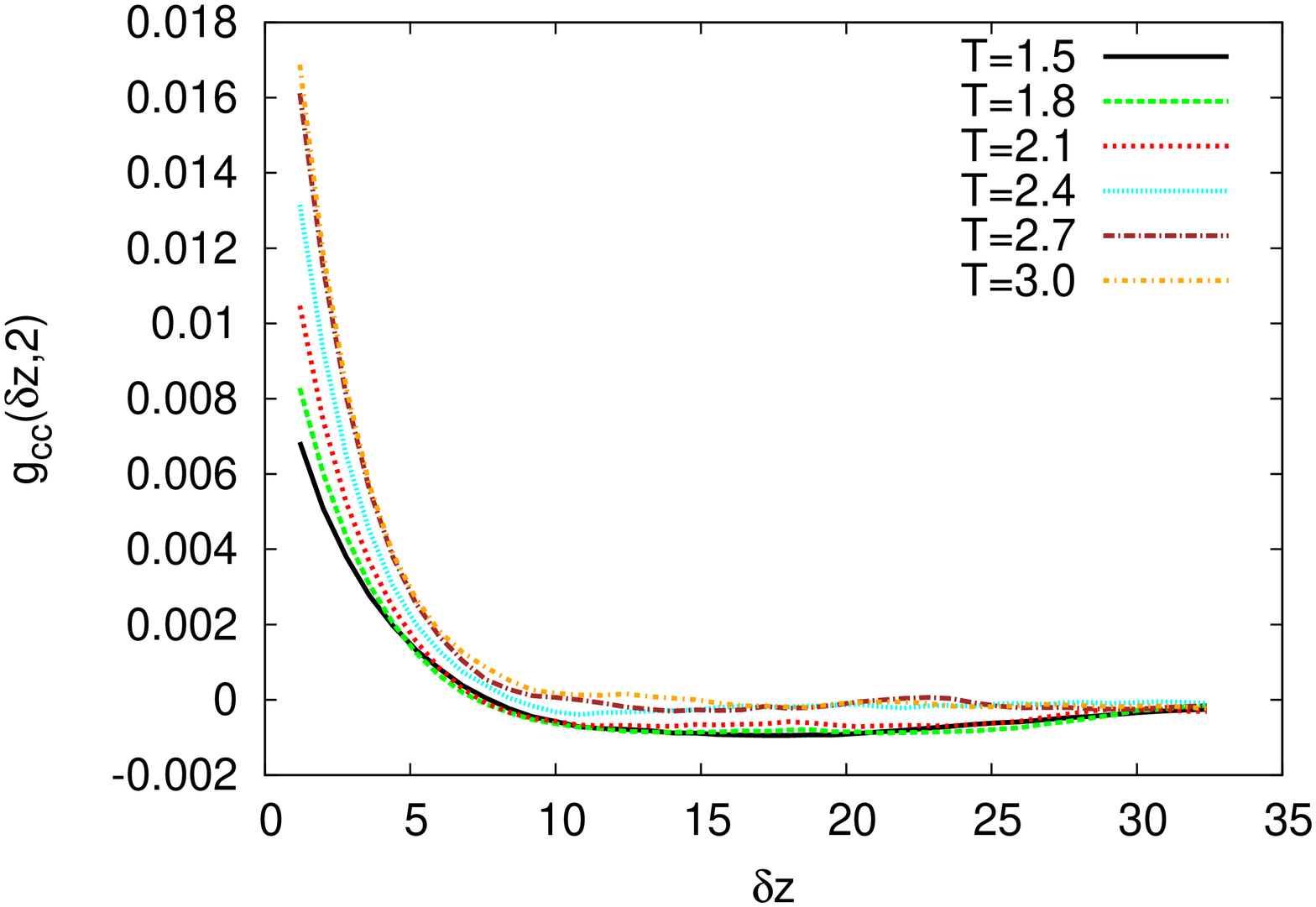}}}
      }
\caption{\label{fig13} (Color online) Same as Fig.~9, but for the case $\epsilon_{AB}=3/4,\; N = 35,\; \sigma = 1.51.$}
\end{figure}

\begin{figure}
\mbox{
      \subfloat[]{\scalebox{1.0}{\includegraphics[height=8.0cm,angle=270]{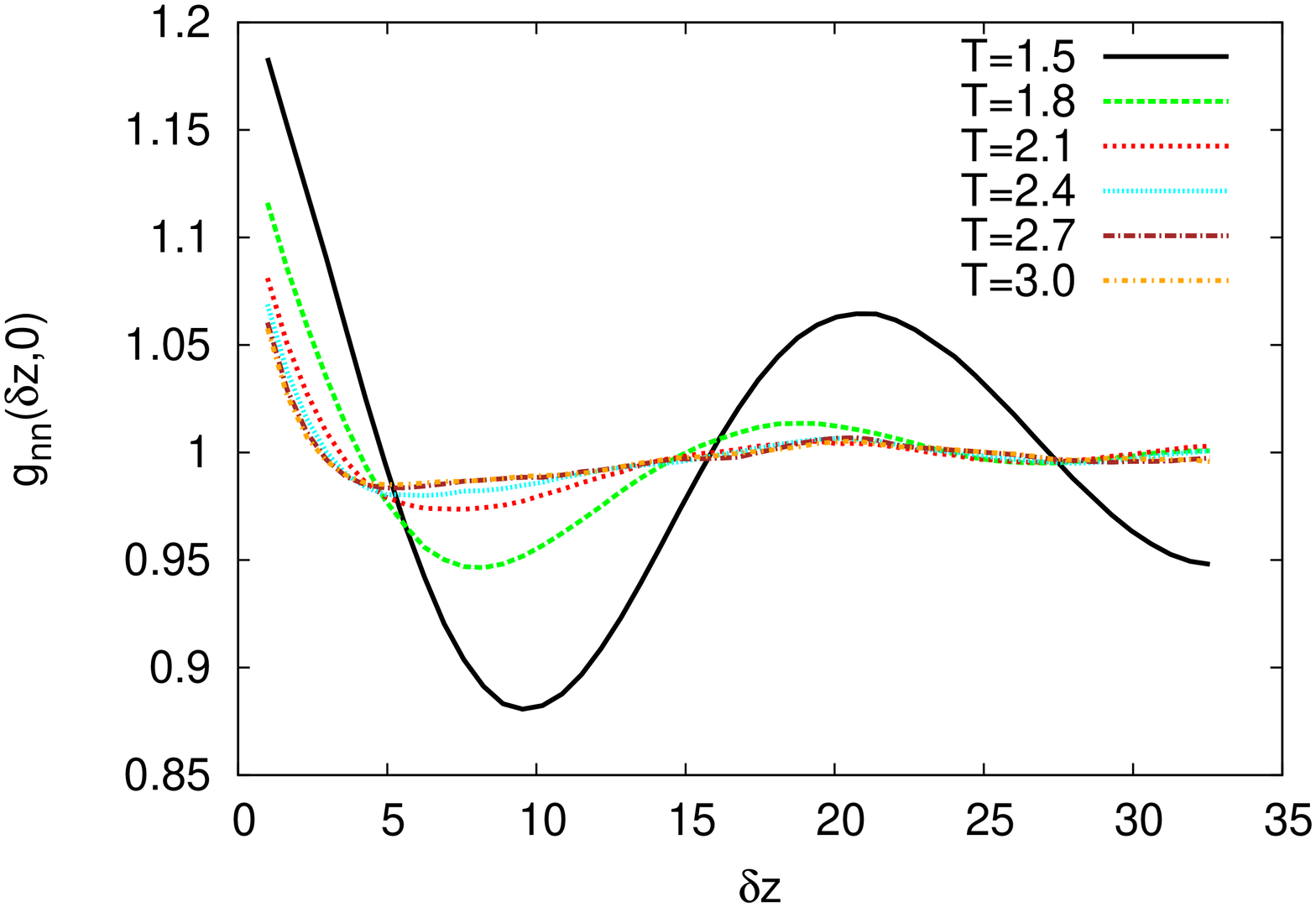}}}
      }
\mbox{
      \subfloat[]{\scalebox{1.0}{\includegraphics[height=8.0cm,angle=270]{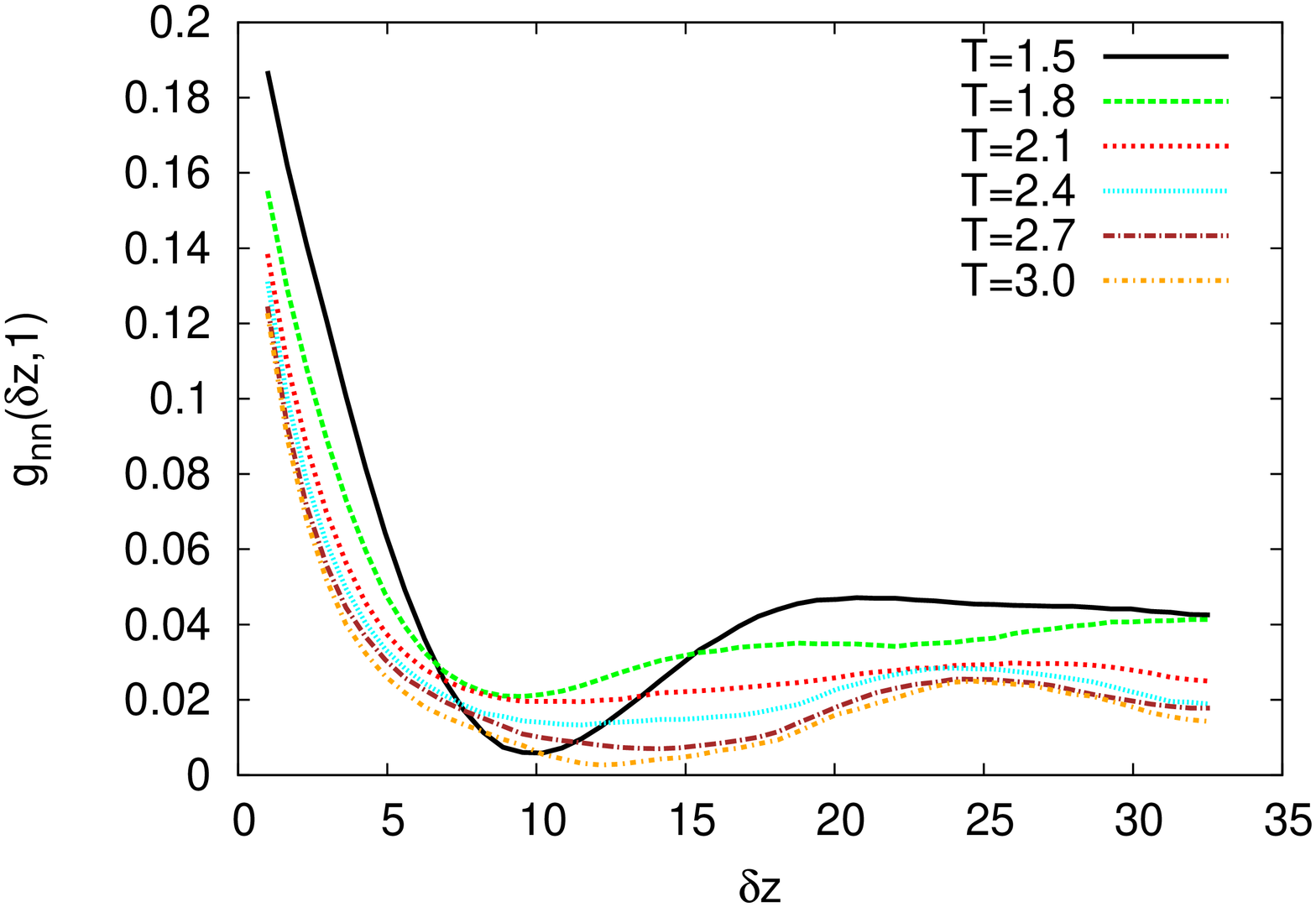}}}
      }
\mbox{
      \subfloat[]{\scalebox{1.0}{\includegraphics[height=8.0cm,angle=270]{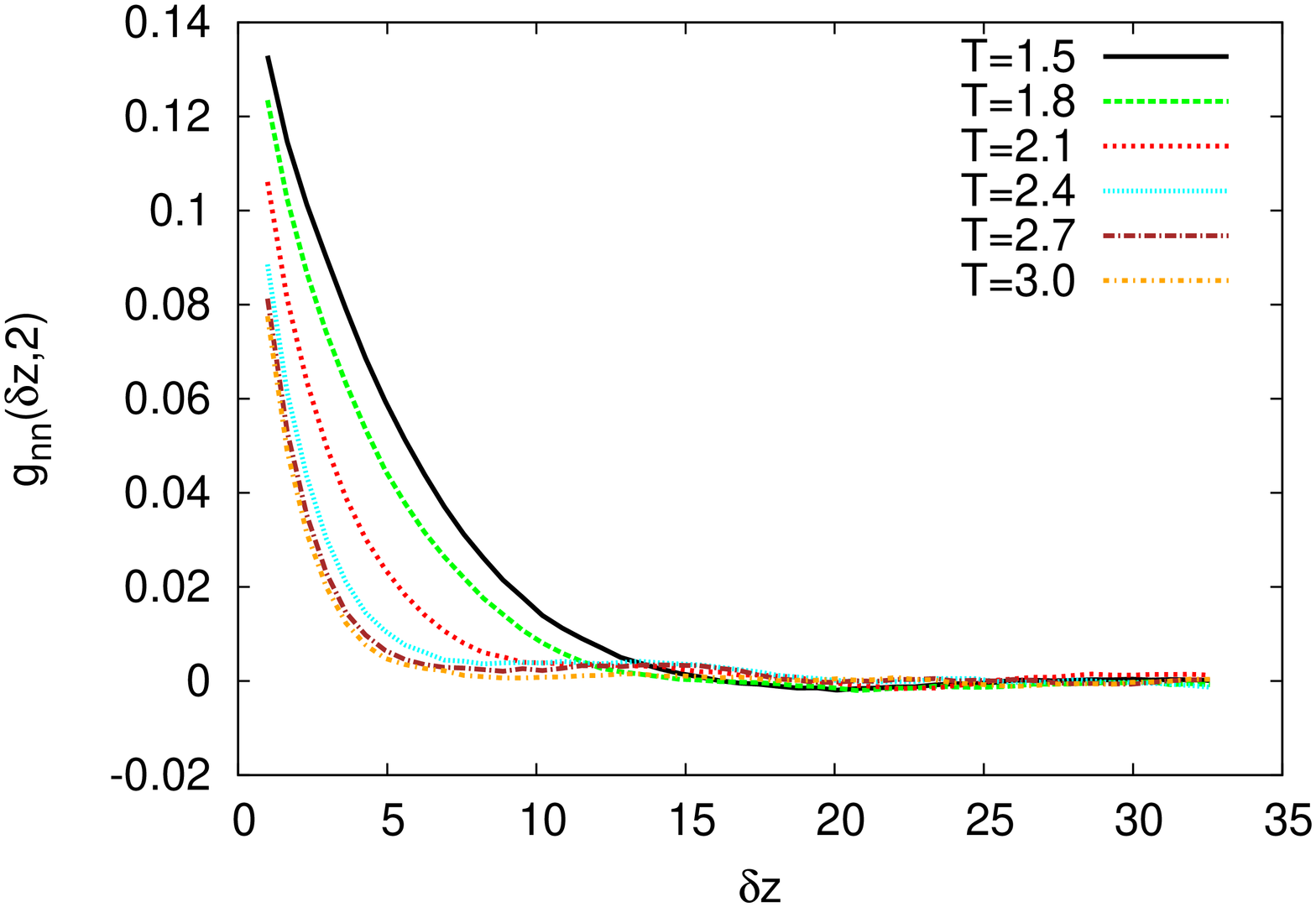}}}
      }
\caption{\label{fig14} (Color online) Same as Fig.~8, but for the case $\epsilon_{AB}=1/2,\; N = 35,\; \sigma = 0.76$.}
\end{figure}

\begin{figure}
\mbox{
      \subfloat[]{\scalebox{1.0}{\includegraphics[height=8.0cm,angle=270]{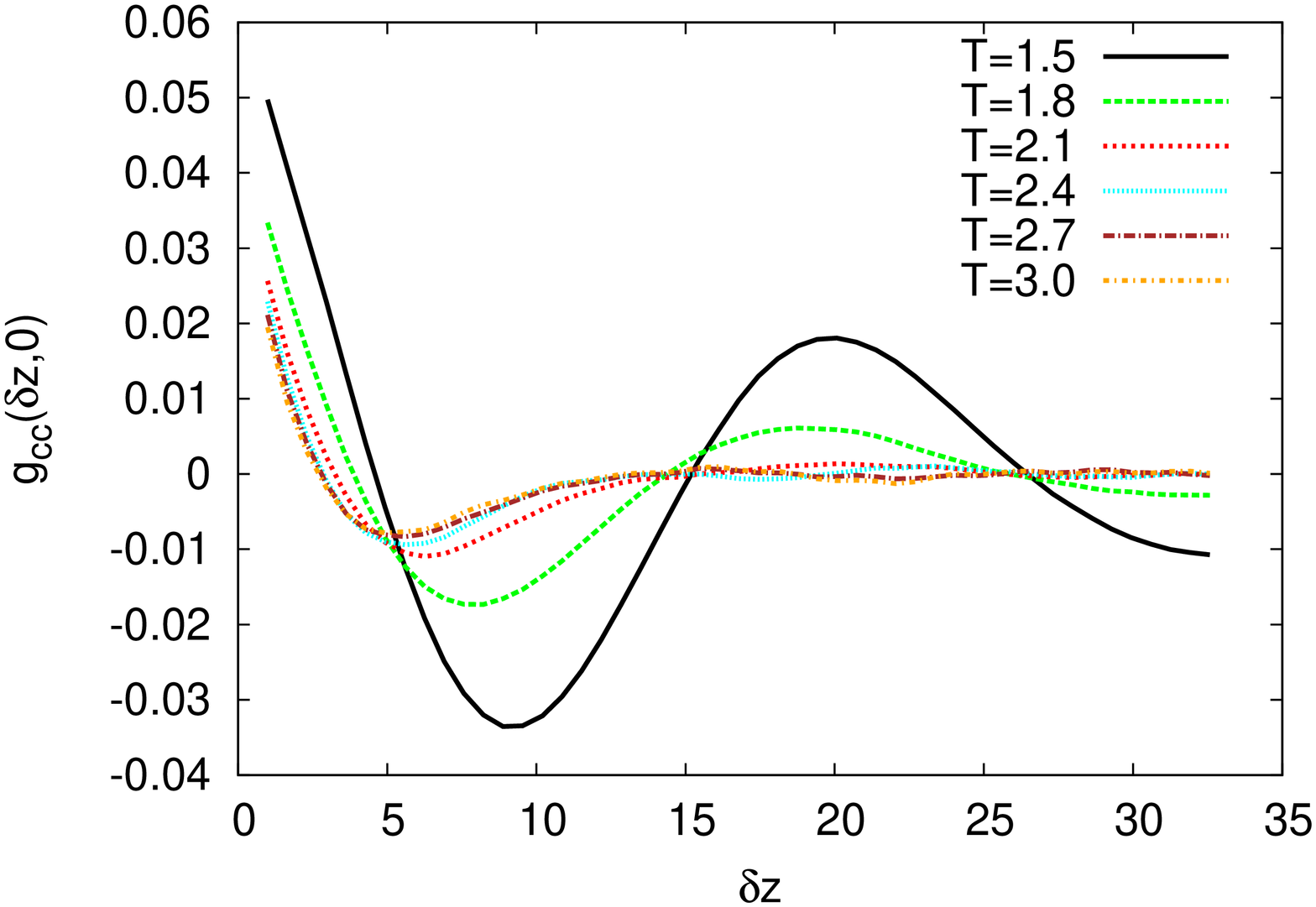}}}
      }
\mbox{
      \subfloat[]{\scalebox{1.0}{\includegraphics[height=8.0cm,angle=270]{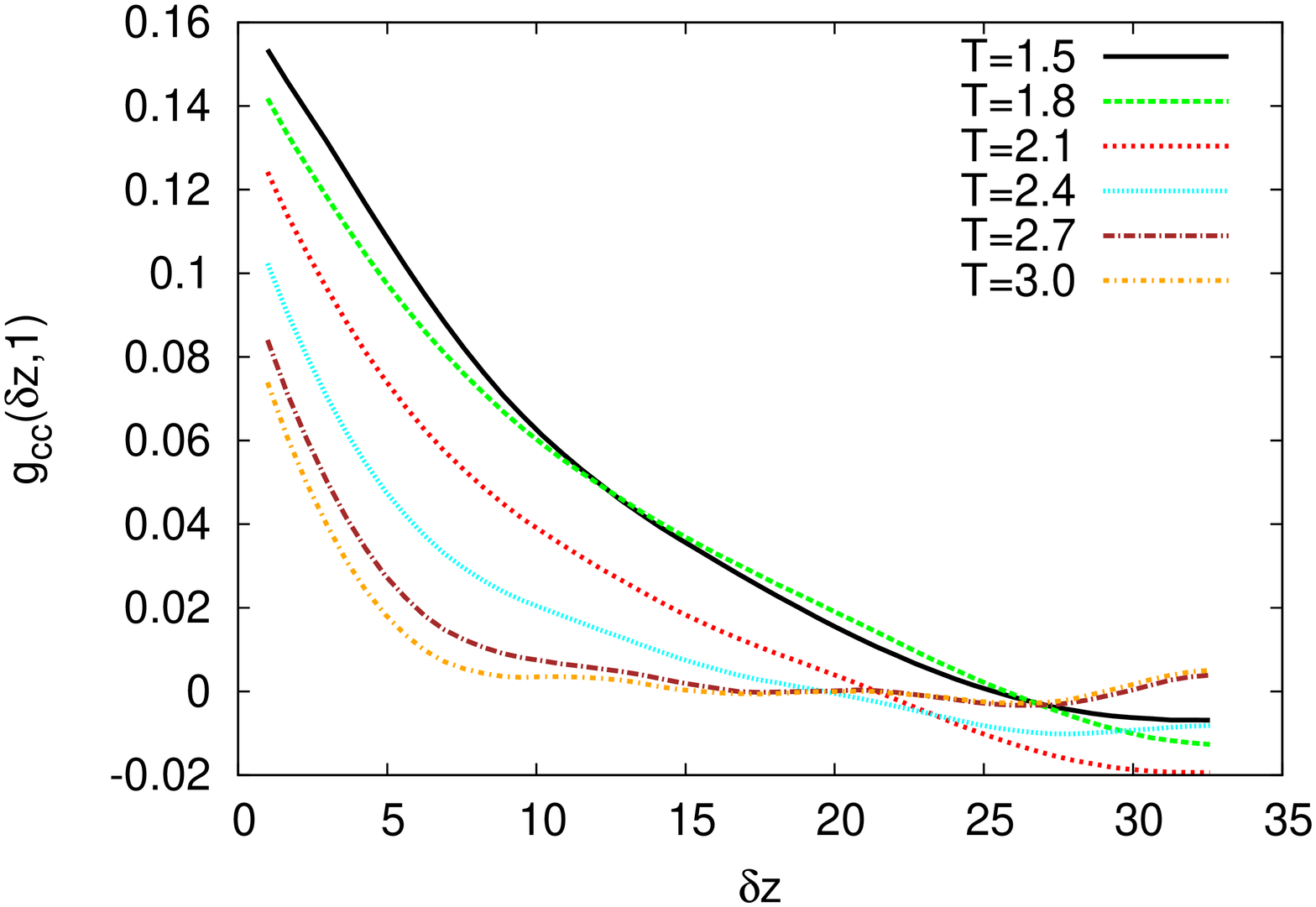}}}
      }
\mbox{
      \subfloat[]{\scalebox{1.0}{\includegraphics[height=8.0cm,angle=270]{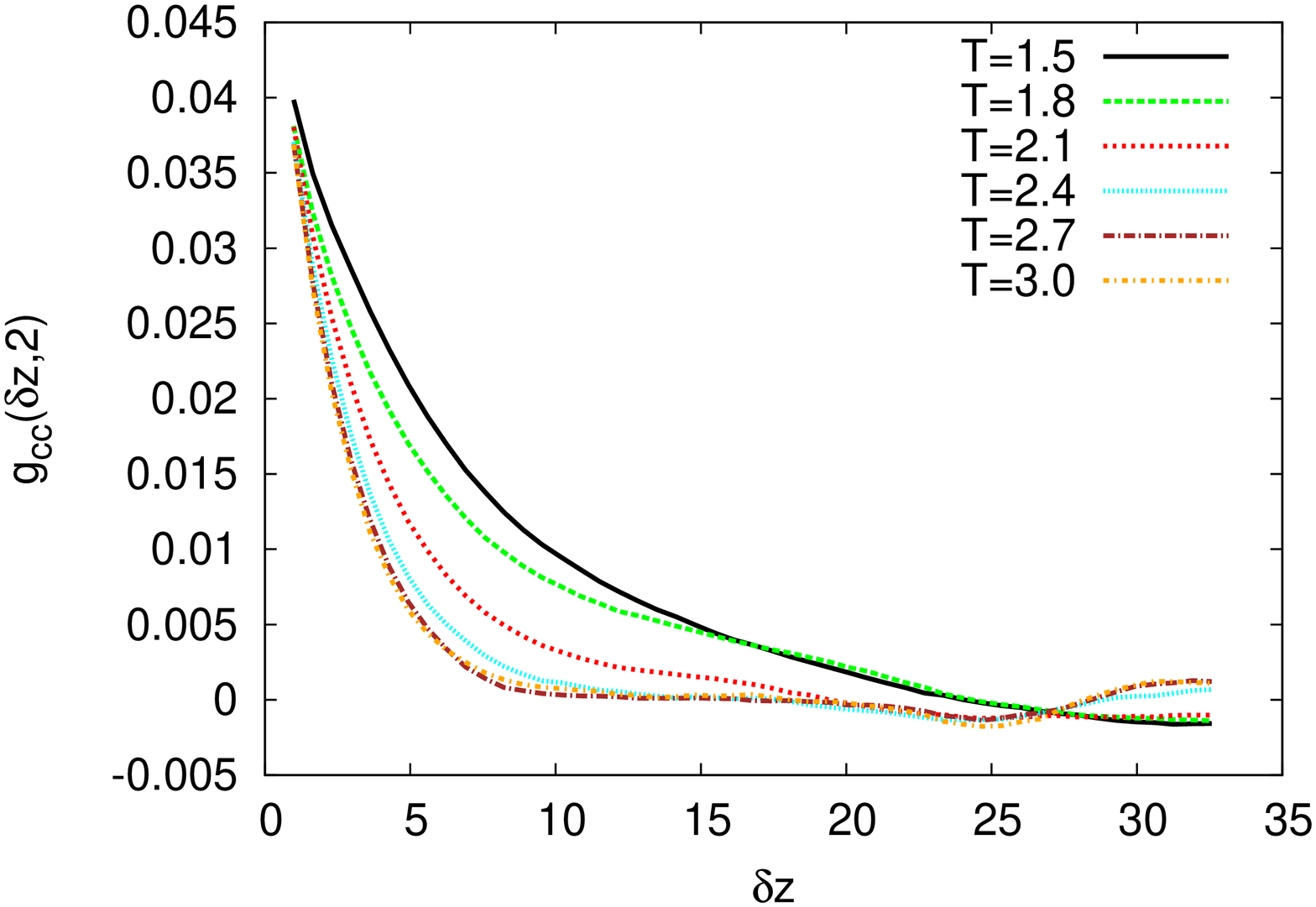}}}
      }
\caption{\label{fig15} (Color online) Same as Fig.~9, but for the case $\epsilon_{AB}=1/2,\; N = 35,\; \sigma = 0.76$.}
\end{figure}

\begin{figure}
\mbox{
      \subfloat[]{\scalebox{1.0}{\includegraphics[height=7.0cm,angle=270]{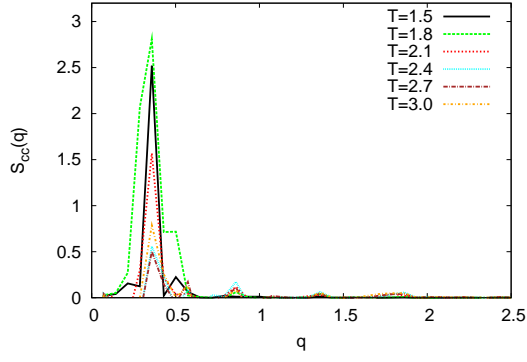}}}
      }
\mbox{
      \subfloat[]{\scalebox{1.0}{\includegraphics[height=7.0cm,angle=270]{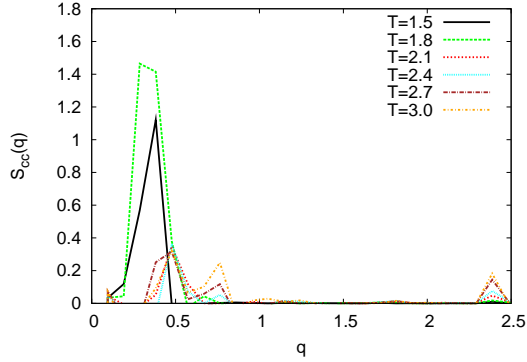}}}
      }
\mbox{
      \subfloat[]{\scalebox{1.0}{\includegraphics[height=7.0cm,angle=270]{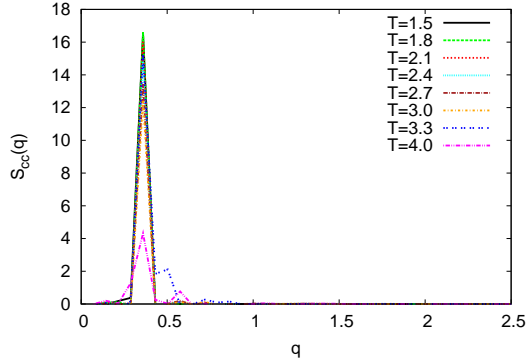}}}
      }
\mbox{
      \subfloat[]{\scalebox{1.0}{\includegraphics[height=7.0cm,angle=270]{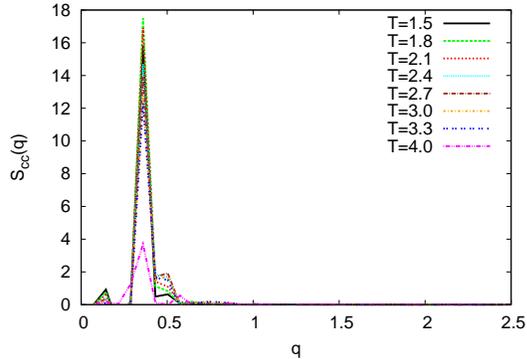}}}
      }
\caption{\label{fig16} (Color online) Structure factor $S_{cc}(q)$ plotted vs. $q$ for $\sigma = 0.57$, N = 35, $\epsilon_{AB}=1/2$ (a), $\sigma = 0.76$, N = 35, $\epsilon_{AB}=1/2$ (b) $\sigma = 1.14$, N = 35, $\epsilon_{AB}=1/2$ (c), and $\sigma = 1.14$, N = 35, $\epsilon_{AB}=7/8$ (d). The positions of the main peaks occur
near $q_{\textrm{max}} \approx 0.3$ in all cases, corresponding to a characteristic wavelength $\lambda \approx 2 \pi/q_{\textrm{max}} \approx 21$.}
\end{figure}

\begin{figure}
\mbox{
      \subfloat[]{\scalebox{1.0}{\includegraphics[height=7.0cm,angle=270]{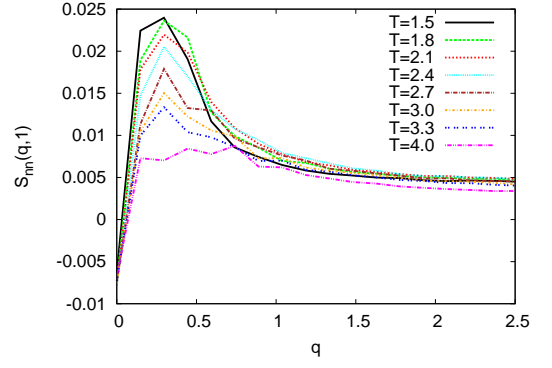}}}
      }
\mbox{
      \subfloat[]{\scalebox{1.0}{\includegraphics[height=7.0cm,angle=270]{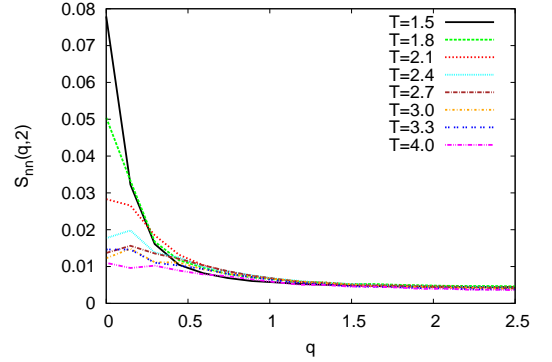}}}
      }
\mbox{
      \subfloat[]{\scalebox{1.0}{\includegraphics[height=7.0cm,angle=270]{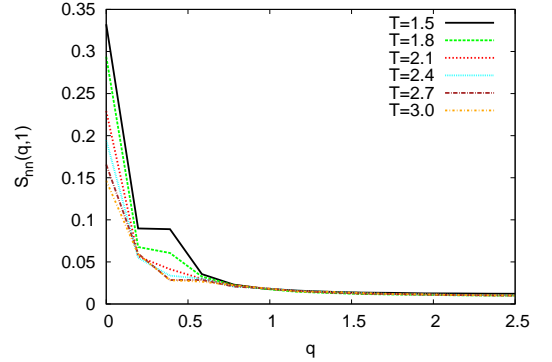}}}
      }
\mbox{
      \subfloat[]{\scalebox{1.0}{\includegraphics[height=7.0cm,angle=270]{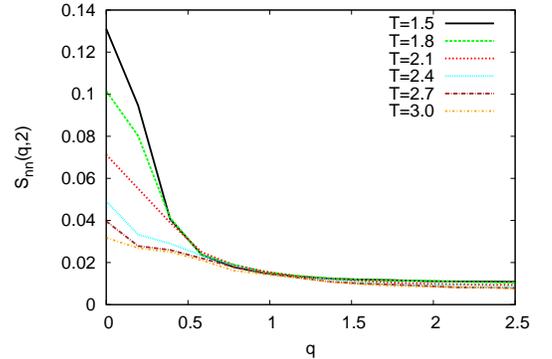}}}
      }
\caption{\label{fig17} (Color online) Fourier transforms of density fluctuation correlation functions, $S_{nn}(q,1)$ and $S_{nn}(q,2)$, as defined in the text, for the case $N=35$,
$\sigma=1.14$, $\epsilon_{AB}=3/4$ (a, b) and for the case $N=35$, $\sigma=0.76$, $\epsilon_{AB}=1/2$ (c, d).}
\end{figure}

\begin{figure}
\mbox{
      \subfloat[]{\scalebox{1.0}{\includegraphics[height=8.0cm,angle=270]{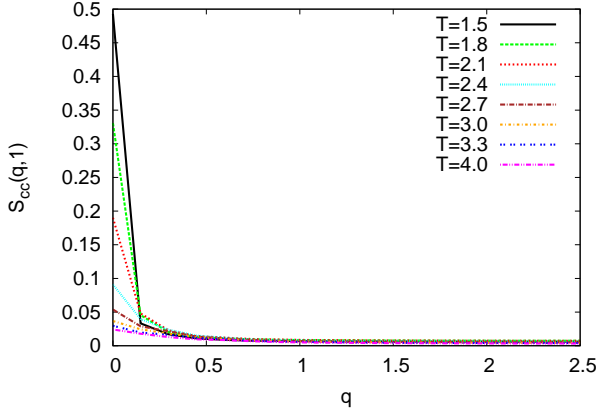}}}
      }
\mbox{
      \subfloat[]{\scalebox{1.0}{\includegraphics[height=8.0cm,angle=270]{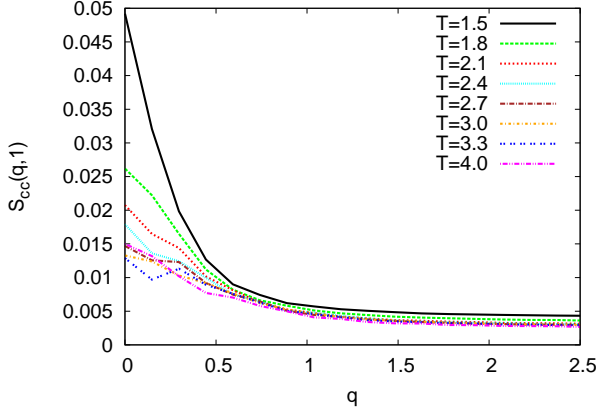}}}
      }
\mbox{
      \subfloat[]{\scalebox{1.0}{\includegraphics[height=8.0cm,angle=270]{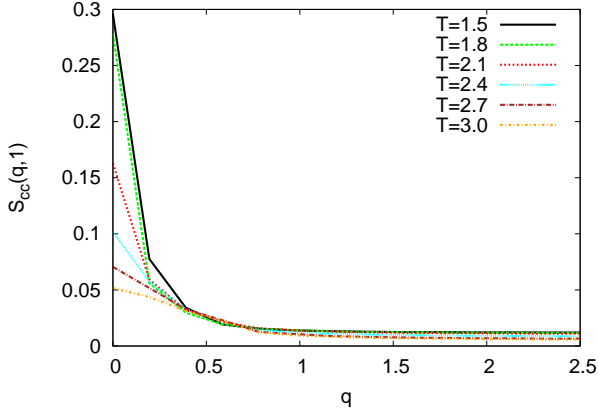}}}
      }
\caption{\label{fig18} (Color online) Fourier transforms of concentration fluctuation correlation functions $S_{cc}(q,1)$ for the cases $N=35$, $\sigma=1.14$, $\epsilon_{AB}=3/4$ (a), $N=35$, $\sigma=1.14$, $\epsilon_{AB}=15/16$ (b) and $N=35$, $\sigma=0.76$, $\epsilon_{AB}=1/2$ (c).}
\end{figure}

\begin{figure}
\mbox{
      \subfloat[]{\scalebox{1.0}{\includegraphics[height=8.0cm,angle=270]{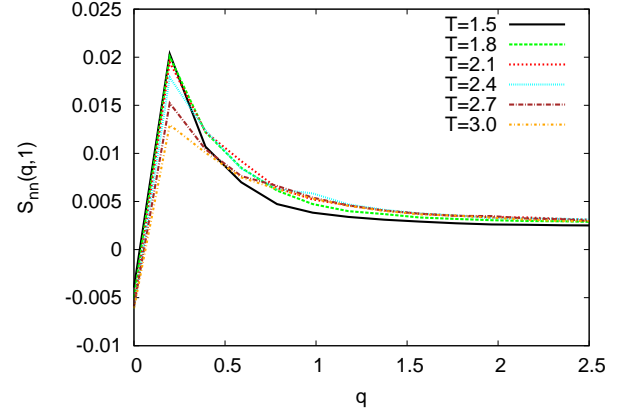}}}
      }
\mbox{
      \subfloat[]{\scalebox{1.0}{\includegraphics[height=8.0cm,angle=270]{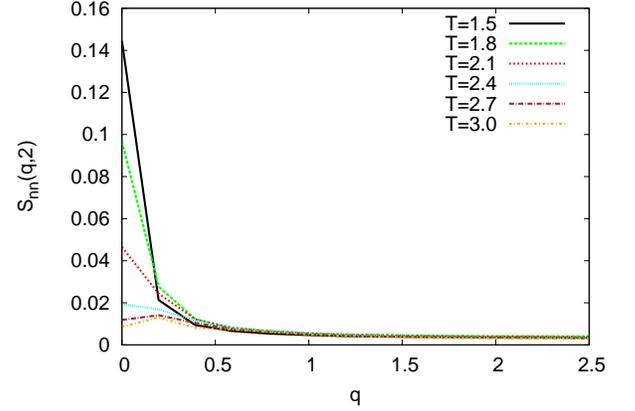}}}
      }
\mbox{
      \subfloat[]{\scalebox{1.0}{\includegraphics[height=8.0cm,angle=270]{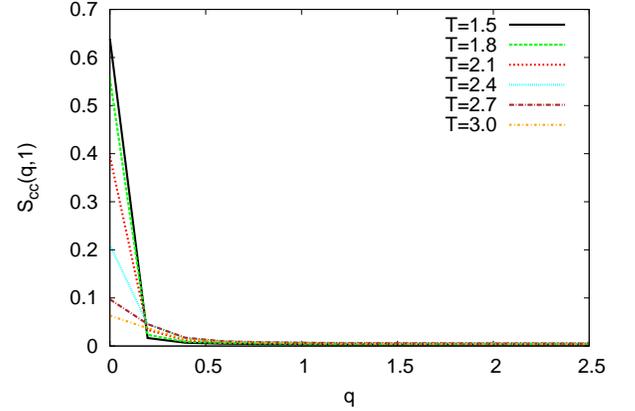}}}
      }
\caption{\label{fig19} (Color online) Fourier transforms of density fluctuation correlation
functions $S_{nn}(q,1)$ (a) and  $S_{nn}(q,2)$ (b), and concentration fluctuation correlation functions $S_{cc}(q,1)$ (c) for the case $N=35$, $\sigma=1.51$, ,and $\epsilon_{AB}=3/4$.}
\end{figure}

\end{document}